\documentclass[
 aps,
 pra,
 reprint,
 amsmath,amssymb,
 floatfix,
 nofootinbib,
 superscriptaddress,
]{revtex4-2}

\usepackage{graphicx}
\usepackage{dcolumn}
\usepackage{bm}
\usepackage{hyperref}
\usepackage{physics}
\usepackage{mathrsfs}
\usepackage{bbm}
\usepackage{mathtools}
\usepackage{braket}

\hypersetup{colorlinks = true, 
  urlcolor = blue, linkcolor = blue, citecolor = blue}

\usepackage{ifpdf}

\ifpdf

\usepackage{tikz}
\usetikzlibrary{arrows.meta}

\newcommand{\VVcorrD}{
  \tikz[baseline=1.7ex, scale=0.2]{
    \draw[line width=0.5pt] (0,3.5) -- (1.2,3.5);
    \draw[line width=0.5pt] (0,0) -- (1.2,0);
    \draw[dashed, line width=0.5pt] (0.3,0) -- (0.3,3.5);
    \draw[dashed, line width=0.5pt] (0.9,0) -- (0.9,3.5);
  }
}

\newcommand{\VVcorrC}{
  \tikz[baseline=1.7ex, scale=0.2]{
    \draw[line width=0.5pt] (0,3.5) -- (1.2,3.5);
    \draw[line width=0.5pt] (0,0) -- (1.2,0);
    \draw[dash pattern=on 2.2pt off 1.6pt, line width=0.5pt] (0.3,0) -- (0.9,3.5);
    \draw[dash pattern=on 2.2pt off 1.6pt, line width=0.5pt] (0.9,0) -- (0.3,3.5);
  }
}

\newcommand{\dummy}{%
  \mathrel{%
    \hspace{-1.5mm}\tikz[baseline=1.7ex, scale=0.2]{
      \path (0,0) rectangle (1.5,3.5);
    }
  }
}

\newcommand{\GGarrowD}{
  \tikz[baseline=1.7ex, scale=0.2]{
    \draw[line width=0.5pt] (0,3.5) -- (3,3.5);
    \draw[-{Latex[length=1.8mm]}, dashed, line width=0.5pt] (1.01,3.5) -- (1,3.5);
    \draw[dash pattern=on 2.2pt off 1.6pt, line width=0.5pt] (0,0) -- (3,0);
    \draw[-{Latex[length=1.8mm]}, dashed, line width=0.5pt] (2,0) -- (2.01,0);
  }
}

\newcommand{\GGarrowC}{
  \tikz[baseline=1.7ex, scale=0.2]{
    \draw[line width=0.5pt] (0,3.5) -- (3,3.5);
    \draw[-{Latex[length=1.8mm]}, dashed, line width=0.5pt] (1.01,3.5) -- (1,3.5);
    \draw[dash pattern=on 2.2pt off 1.6pt, line width=0.5pt] (0,0) -- (3,0);
    \draw[-{Latex[length=1.8mm]}, dashed, line width=0.5pt] (1.01,0) -- (1,0);
  }
}
\newcommand{\Diffuson}{
  \tikz[baseline=1.7ex, scale=0.2]{
    \draw[line width=0.5pt] (-0.3,3.5) -- (3.5,3.5);
    \draw[line width=0.5pt] (-0.3,0) -- (3.5,0);
    \draw[line width=0.5pt] (0,0) -- (0,3.5);
    \draw[line width=0.5pt] (3.2,0) -- (3.2,3.5);
    \path (1.6,1.8) node[inner sep=0, outer sep=0] {\scriptsize $D$};
  }
}

\newcommand{\CooperonX}{
  \tikz[baseline=1.7ex, scale=0.2]{
    \draw[line width=0.5pt] (-0.3,3.5) -- (3.5,3.5);
    \draw[line width=0.5pt] (-0.3,0) -- (3.5,0);
    \draw[line width=0.5pt] (0,0) -- (3.2,3.5);
    \draw[line width=0.5pt] (3.2,0) -- (0,3.5);
    \path (1.6,2.8) node[inner sep=0, outer sep=0] {\scriptsize $C$};
  }
}

\newcommand{\Cooperon}{
  \tikz[baseline=1.7ex, scale=0.2]{
    \draw[line width=0.5pt] (-0.3,3.5) -- (3.5,3.5);
    \draw[line width=0.5pt] (-0.3,0) -- (3.5,0);
    \draw[line width=0.5pt] (0,0) -- (0,3.5);
    \draw[line width=0.5pt] (3.2,0) -- (3.2,3.5);
    \path (1.6,1.8) node[inner sep=0, outer sep=0] {\scriptsize $C$};
  }
}

\fi

\begin{document}

\title{Mesoscopic scattering dynamics under generic uniform SU(2) gauge fields: Spin--momentum relaxation and coherent backscattering}

\author{Masataka Kakoi}
\email{kakoi@presto.phys.sci.osaka-u.ac.jp}
\affiliation{Department of Physics, The University of Osaka, Toyonaka, Osaka 560-0043, Japan}

\author{Christian Miniatura}
\email{christian.miniatura@cnrs.fr}
\affiliation{Universit{\'e} C{\^o}te d'Azur, CNRS, INPHYNI, 17, rue Julien Lauprêtre, 06200 Nice, France}
\affiliation{Centre for Quantum Technologies, National University of Singapore, Singapore}

\author{Keith Slevin}
\email{slevin.keith.sci@osaka-u.ac.jp}
\affiliation{Department of Physics, The University of Osaka, Toyonaka, Osaka 560-0043, Japan}

\date{\today}

\begin{abstract}
   We investigate the time- and momentum-resolved dynamics of matter waves undergoing elastic scattering from a disordered potential in the presence of spatially uniform SU(2) gauge fields.
   We derive the disorder-averaged density matrix as a function of time and momentum within the weak-localization regime.
   By accurately approximating the frequency dependence of the ladder and maximally crossed diagram series beyond the diffusion approximation, we describe short-time spin--momentum dynamics on timescales comparable to the scattering mean free time, for arbitrary strengths of the SU(2) gauge fields and disorder.
   We also present a cubic equation that determines the spin isotropization time, which gives accurate asymptotic forms in the limits where the spin--orbit length is much longer (Dyakonov--Perel spin relaxation regime) or much shorter than the scattering mean free path, as well as in the SU(2)-symmetric (persistent spin helix) limit. 
   Comparison with numerical simulations over a broad range of spin--orbit coupling strengths shows quantitative agreement for the predicted spin isotropization time. The same framework also reproduces both the relaxation of the momentum distribution and the transient backscattering peak with a momentum offset coexisting with the robust coherent backscattering dip.
\end{abstract}

\maketitle

\section{Introduction\label{sec:introduction}}

Wave transport in disordered systems is a fundamental problem across electronic, optical, acoustic, and atomic platforms. Multiple scattering causes waves to lose memory of their initial momentum and phase, leading to diffusion. While it is classically described as a random walk, interference effects modify this behavior.
In particular, the interference of partial waves counter-propagating along the same geometrical scattering paths manifests as coherent backscattering (CBS) and weak localization~\cite{Bergmann_1984_review,Akkermans_2007_book}. In electronic systems, this phenomenon is revealed by negative magnetoresistance measurements~\cite{Kawaguchi_1978,*Kawaguchi_1980,Dolan_1979,Kobayashi_1980}, and has been directly observed for various types of linear waves~\cite{Kuga_1984,VanAlbada_1985,*Wolf_1985,Labeyrie_1999,Bayer_1993,Sakai_1997,Larose_2004,Jendrzejewski_2012a,Labeyrie_2012,Hainaut_2017}.

For systems subject to spin--orbit coupling (SOC), spin is not a conserved quantity due to the breaking of spin-rotation symmetry, implying that any initial spin polarization decays under scattering. In noncentrosymmetric systems, this spin relaxation is dominated by the Dyakonov--Perel (DP) mechanism~\cite{Dyakonov_1972}, where spin precesses around a momentum-dependent effective field and is randomized by disorder. Similar DP-type relaxation can apply to orbital angular momentum and phonon polarization in centrosymmetric systems~\cite{Sohn_2024,Suzuki_2025}.
SOC also modifies quantum interference phenomena: destructive interference between different spin channels produces a negative quantum correction to the conductivity~\cite{Hikami_1980}, leading to a crossover from negative to positive magnetoresistance~\cite{Bergman_1982,Knap_1996}. A similar crossover occurs in Dirac-like systems, such as graphene and the surface states of topological insulators~\cite{Ando_1998a,*Ando_1998b,Suzuura_2002,McCann_2006,Y-Ando_2013_review}, whose Hamiltonians share the same structure as those of SOC systems.

When strong SOC is present, spin precession becomes rapid, leading to efficient spin relaxation. However, in special cases where the SOC can be removed through a non-Abelian gauge transformation, SU(2) symmetry is recovered, and helical spin polarization modes are preserved~\cite{Schliemann_2003,Bernevig_2006,Schliemann_2017_review}. This phenomenon, known as the persistent spin helix (PSH), has been observed in semiconductor quantum wells~\cite{Koralek_2009,Walser_2012,Kohda_2012,Altmann_2016}. Recently, symmetry-protected PSHs around specific high-symmetry points in the Brillouin zone in noncentrosymmetric bulk materials have been proposed~\cite{Tao_2018,XZ-Lu_2020,Zhao_2020,J-Ji_2022,Kilic_2025}, with potential applications in spintronics.
Strong SOC can enhance SOC-related phenomena, making it relevant for spintronics applications. Most studies with disordered potential, however, have focused on the weak SOC regime, where the spin--orbit length exceeds the scattering mean free path. Addressing the strong SOC regime (or equivalently, the high-mobility limit) requires going beyond the diffusion approximation~\cite{Golub_2005,Glazov_2006,Glazov_2009,X-Liu_2011,X-Liu_2012,Hijano_2024}. The real-time dynamics of spin and momentum relaxation, as well as the associated interference effects, in this regime therefore remain largely unexplored~\cite{Kakoi_2024,Kakoi_2026,Dornelas_arxiv}. This calls for a unified theoretical framework that can treat real-time scattering dynamics for arbitrary types and strengths of SOC.

It is worth highlighting that studies of out-of-equilibrium matter-wave scattering dynamics have largely been developed in the context of cold atomic systems. In particular, the real-time evolution of interference effects such as CBS and coherent forward scattering (CFS) has been investigated in systems belonging to the AI symmetry class~\cite{Cherroret_2012,Karpiuk_2012,Micklitz_2014,Ghosh_2014,Ghosh_2017,Thomas_2025,Cherroret_2021_review}, the AII class~\cite{Arabahmadi_2024,Kakoi_2024,Kakoi_2026}, and in the presence of many-body interactions~\cite{Scoquart_2020}. These phenomena have been directly observed in cold atom experiments using time- and momentum-resolved measurements~\cite{Jendrzejewski_2012a,Labeyrie_2012,Hainaut_2018,Arrouas_2026}. Moreover, the quantum boomerang effect has been recently predicted and observed~\cite{Prat_2019,Tessieri_2021,Janarek_2022,Sajjad_2022,X-Hou_2026,Dornelas_arxiv}. These interference signatures can serve as powerful probes of Anderson localization~\cite{Anderson_1958}. Although CBS in the AII class has not yet been observed, it is expected to become accessible using experimentally realized synthetic SOC~\cite{Galitski_2013_review,Ruseckas_2005,XJ-Liu_2009,Campbell_2011,YJ-Lin_2011,P-Wang_2012,*Cheuk_2012,Huang_2016,Z-Wu_2016,Leroux_2018,Hasan_2022,ZY-Wang_2021,Q-Liang_2024,Madasu_2025}.
Cold-atom systems are well suited for exploring short-time dynamics comparable to the scattering time, since the scattering time can be tuned to the millisecond range~\cite{Richard_2019}. In contrast, in semiconductor quantum wells the scattering time is typically on the order of picoseconds~\cite{Tschirky_2017}, which is difficult to access in most transport measurements. Recently, however, a method to probe interference echoes using terahertz pulses has been theoretically proposed~\cite{ZL-Li_2024}. This approach can be regarded as the electronic analogue of echo spectroscopy developed in cold atom systems~\cite{Micklitz_2015,Muller_2015}.

In this paper, we derive the disorder-averaged density matrix that captures the full time-resolved dynamics within a perturbative framework for multiple scattering.
This allows us to predict the spin isotropization (relaxation) time via a cubic equation~\eqref{eq:cubic_eq_transport_mean_free_time} for any uniform SU(2) gauge field, i.e., from the limit where the spin--orbit length is much longer than the scattering mean free path (weak SOC) to the opposite limit (strong SOC), and from the isotropic SOC (e.g., pure Rashba SOC) limit to the SU(2) symmetric (i.e., PSH) limit.
Similarly, it describes the dynamics of interference effects in the vicinity of the backscattering direction.
We also provide an analytical expression of a transient backscattering peak~\cite{Kakoi_2026} that appears at a position far from the exact backscattering direction.

This paper is organized as follows.
In Sec.~\ref{sec:model}, we present a model describing spin-$1/2$ particles in a two-dimensional disordered potential under an arbitrary uniform SU(2) gauge field, and discuss its basic scattering properties.
In Sec.~\ref{sec:preparation}, as a preparation for the explicit calculations, we review the disorder-averaged Green’s function for the model of interest and the general diagrammatic perturbation framework for the density matrix, including the Diffuson and Cooperon.
In Sec.~\ref{sec:analytic_results}, we apply the diagrammatic perturbation theory to the model introduced in Sec.~\ref{sec:model}. A brief overview of subsections~\ref{sec:Cooperon_Diffuson} through \ref{sec:transient_peak} is provided at the beginning of Sec.~\ref{sec:analytic_results}.
In Sec.~\ref{sec:numerical_results}, we compare the analytical results obtained in Sec.~\ref{sec:analytic_results} with numerical simulations and verify the validity of the approximations employed in our analysis.

\section{Model\label{sec:model}}

\subsection{Clean Hamiltonian for general uniform SU(2) gauge fields}

We consider a one-particle two-dimensional (2D) Hamiltonian that describes the propagation of a (pseudo) spin-$1/2$ particle confined to the 
$xy$-plane and subject to a uniform SU(2) gauge field $\hat{\bm{A}}$,
\begin{equation}\label{eq:SOC_Ham}
    \hat{H}_0 = \frac{\big(\hat{\bm{p}}+\!\hat{\bm{A}}\big)^2}{2m} = \frac{(\hat{p}_x+ \hat{A}_x)^2}{2m} + \frac{(\hat{p}_y+ \hat{A}_y)^2}{2m}.
\end{equation}
Very generally, each component of $\hat{\bm{A}}$ can be expressed as a linear combination of Pauli operators:
\begin{equation}\label{eq:vec_opt_general}
    \hat{\bm{A}} = M\hat{\bm{\sigma}},
\end{equation}
where $M$ is a $2\times3$ matrix and $\hat{\bm{\sigma}}=(\hat{\sigma}_x,\hat{\sigma}_y,\hat{\sigma}_z)$ denotes the set of Pauli operators.
For $\hat{H}_0$ to be Hermitian, all elements of $M$ must be real.
Since
\begin{equation}
    \hat{\bm{A}}^2=\Tr(MM^\top)\mathbbm{1}_2,
\end{equation}
where $\top$ denotes matrix transpose, $\hat{\bm{A}}$ determines a natural momentum and energy scales through 
\begin{equation}
\begin{aligned}
    \hbar\kappa &= \sqrt{\Tr(MM^\top)},\\
    E_{\kappa} & = \frac{\Tr(MM^\top)}{2m} = \frac{\hbar^2\kappa^2}{2m}.
\end{aligned}
\end{equation}
In turn, the corresponding length and timescales are $L_{\kappa}=\kappa^{-1}$ and $\tau_{\kappa}=\hbar/E_{\kappa}$.

It can be shown~\cite{Kakoi_2026} that, for any $M\in\mathbb{R}^{2\times3}$, a suitable singular value decomposition (SVD) of $M$ combining rotations of the spatial and spin axes map the uniform SU(2) gauge field $\hat{\bm{A}}$ to 
\begin{equation}\label{eq:def_vector_pot}
\begin{aligned}
    \hat{A}_x &= \hbar\kappa\cos\eta\,\hat{\sigma}_x = \hbar\kappa_x\hat{\sigma}_x,\\
    \hat{A}_y &= \hbar\kappa\sin\eta\,\hat{\sigma}_y = \hbar\kappa_y\hat{\sigma}_y,
\end{aligned}    
\end{equation}
which is thus characterized by only two parameters,\footnote{More generally, for $d$ spatial dimensions, $\min(d,3)$ parameters are sufficient; see Ref.~\cite{Kakoi_2026}.} the field strength $\hbar\kappa$ and the in-plane angle $\eta\in[0,\pi/4]$.\footnote{For a concrete example in solids, $\eta=\pi/4$ corresponds to pure Rashba SOC~\cite{Bychkov_1984} or Dresselhaus SOC~\cite{Dresselhaus_1955} (without the cubic term), whereas $\eta=0$ corresponds to equal strengths of Rashba and Dresselhaus couplings, which host a persistent spin texture~\cite{Schliemann_2003,Bernevig_2006}. Also, quantum dots~\cite{Aleiner_2001} and synthetic SOC~\cite{Ruseckas_2005,Leroux_2018,Hasan_2022} in cold atoms can be described in this framework.} 
Without any loss of generality, we now use Eq.~\eqref{eq:def_vector_pot} in the rest of the Paper. Indeed, the dynamics under the gauge field in Eq.~\eqref{eq:vec_opt_general} can be easily reconstructed from that governed by Eq.~\eqref{eq:def_vector_pot} once the SVD of $M$ is known.

The tuning of $\eta$ in 2D electron systems is achieved, e.g., by controlling the Rashba SOC via gate voltages or asymmetric doping~\cite{Nitta_1997,Koralek_2009}. In cold atom systems, the synthetic SOC generated by the tripod scheme~\cite{Ruseckas_2005,Leroux_2018,Hasan_2022} allows access to the entire range of $\eta$~\cite{Kakoi_2026}.

\begin{figure}[t]
\begin{center}
    \includegraphics[width=\columnwidth]{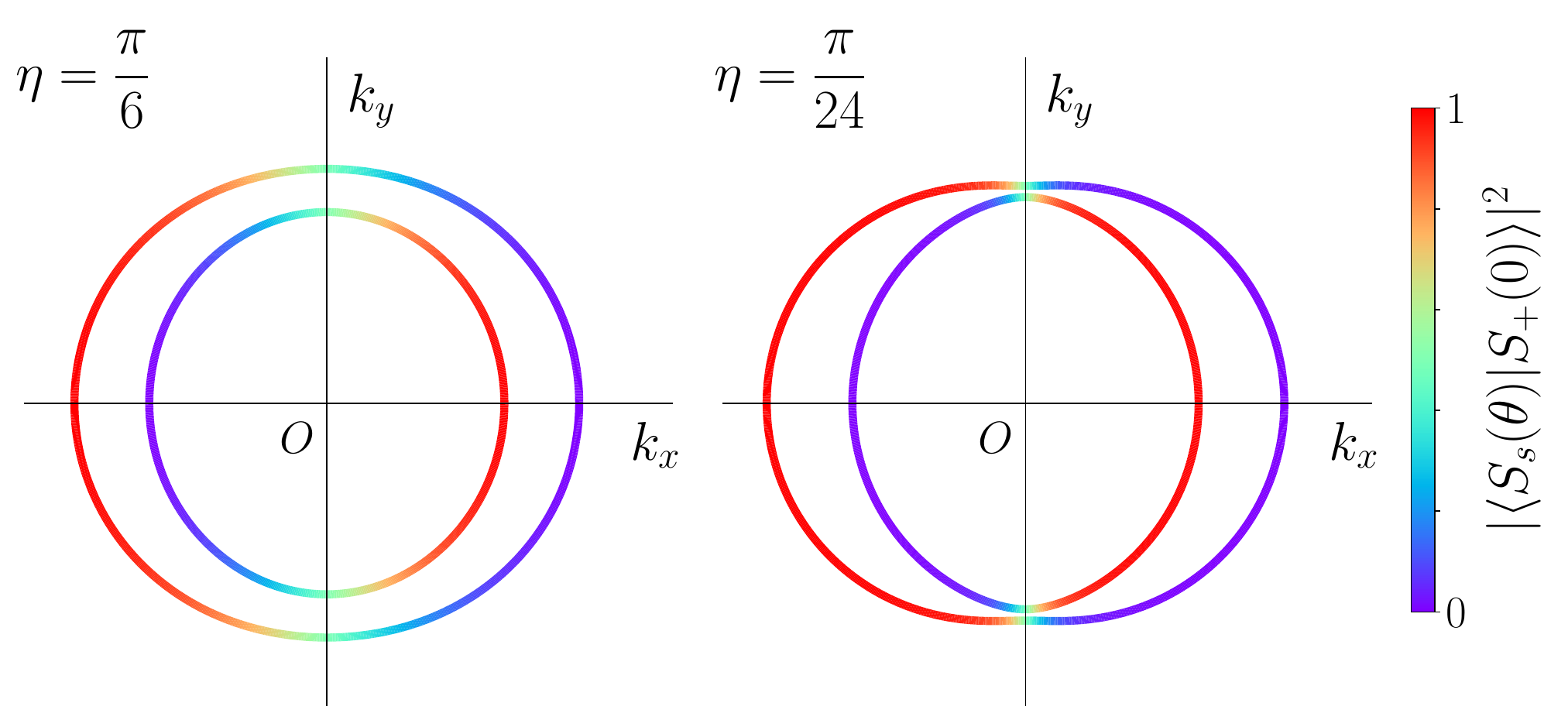}
    \caption{
    {Cut of the energy branches (Fermi surface) at a fixed energy $E_0=25E_{\kappa}$ for $\eta=\pi/6$ and $\eta=\pi/24$. The color scale represents the overlap of the spin states with the $\ket{S_+(\theta\!=\!0)}$ state, see Eq.~\eqref{eq:scattering_prob}.}
    } 
    \label{fig:Fermi_surface}
\end{center}
\end{figure}

The clean Hamiltonian $\hat{H}_0$ in Eq.~\eqref{eq:SOC_Ham} is block-diagonal in momentum space. Diagonalizing it for each momentum $\bm{p}:=\hbar\bm{k}$ yields two ($\pm$) energy branches:
\begin{equation}\label{eq:eigvals}
    \mathcal{E}_{\pm}(\bm{k}) = \frac{\hbar^2k^2}{2m} \pm \frac{\hbar^2\kappa k}{m} \sqrt{\frac{1 + \cos2\eta\cos2\theta}{2}} + E_{\kappa},
\end{equation}
where $\bm{k}$ is the wave vector of the particle.
Introducing the polar coordinates $k_x = k\cos\theta$, $k_y = k\sin\theta$, the corresponding spin-momentum eigenstates are $\ket{\bm{k},\pm}:=\ket{\bm{k}}\otimes\ket{S_{\pm}(\theta)}$ with
\begin{equation}\label{eq:eigvecs}
\begin{aligned}
    \ket{S_{+}(\theta)} &= \frac{|\!\uparrow\rangle + e^{i\phi_{\eta}(\theta)}|\!\downarrow\rangle}{\sqrt{2}},\\
    \ket{S_{-}(\theta)} &= \frac{-e^{-i\phi_{\eta}(\theta)}|\!\uparrow\rangle +|\!\downarrow\rangle}{\sqrt{2}},
\end{aligned}
\end{equation}
where
\begin{equation}\label{eq:def_phi_eta}
    \phi_{\eta}(\theta) = {\rm Arg}\left(\cos\eta\cos\theta + i\sin\eta\sin\theta\right).
\end{equation}
One may note that 
\begin{equation}
    \sum_{s=\pm} \, \ket{S_{s}(\theta)}\!\bra{S_{s} (\theta)} = \sum_{a=\uparrow,\downarrow} \, \ket{\alpha}\!\bra{\alpha} = \mathbbm{1}_2 
\end{equation}
is independent of $\bm{k}$. In Fig.~\ref{fig:Fermi_surface}, we show two typical examples of the Fermi surface obtained when cutting the energy branches at a fixed energy.

\subsection{Disordered potential}

We consider an additional spin-independent and spatially $\delta$-correlated disordered potential $V(\hat{\bm{r}})$.
The strength of the disordered potential is characterized by the correlation amplitude $\gamma_0$ given by
\begin{equation}\label{eq:correlation_potential}
    \overline{V(\bm{r})V(\bm{r}')} = \gamma_0\delta(\bm{r}\!-\!\bm{r}'),
\end{equation}
where $\overline{(\cdot\cdot\cdot)}$ denotes disorder average. This potential is spin-independent, but transitions between different spin branches ($+\leftrightarrow-$) are allowed due to the geometry of the wavefunction.
Indeed, the transition probability for a single scattering event between clean eigenstates from $(\theta',s')$ to $(\theta,s)$ 
($s=\pm$ denotes the branch index) contains a factor
\begin{align}\label{eq:scattering_prob}
    &P_{ss'}(\theta,\theta', \eta) =
    |\langle S_{s}(\theta)|S_{s'}(\theta')\rangle|^2  \notag\\[1mm]
    &\quad= \cos^2\frac{\Delta\phi_\eta}{2}\, \delta_{s,s'}
    + \sin^2\frac{\Delta\phi_\eta}{2} \, (1\!-\!\delta_{s,s'}),
\end{align}
where $\Delta\phi_{\eta}=\phi_{\eta}(\theta)-\phi_{\eta}(\theta')$~\cite{Kakoi_2026}.

When the scattering is anisotropic, we can define several timescales that characterize the transport dynamics.
Specifically, these include the scattering mean free time $\tau$ and the spin isotropization time (spin relaxation time) $\tau_{\rm iso}$. 
The spin relaxation in such a system, caused by momentum randomization and spin precession, occurs via the Dyakonov--Perel (DP) mechanism~\cite{Dyakonov_1972}.
For illustration, consider the case $\eta=\pi/4$, which corresponds, for example, to pure Rashba SOC. In this case, the SOC term $\hat{\bm{p}}\cdot\hat{\bm{A}}/m$ in $\hat{H}_0$ can be rewritten as $\hbar\bm{\Omega}(\bm{k})\cdot\bm{\sigma}/2$ in spin-momentum space where $\bm{\Omega}(\bm{k})=\sqrt{2}\,(\hbar\kappa/m) \, \bm{k}$.
When the spin precession time $\tau_{\mathrm{p}} \sim \Omega^{-1}$ is much larger than $\tau$, the typical angle $\Omega\tau$ of the precession of the spin around the precession axis $\bm{\Omega}(\bm{k})$ between scattering events is small, i.e. $\Omega\tau\ll1$.
At each scattering event, the wave vector $\bm{k}$, and the corresponding spin precession axis $\bm{\Omega}(\bm{k})$, change randomly.
As a net effect, the spin polarization is lost over a timescale $\tau_{\rm iso}\propto\tau_{\mathrm{p}}^2/\tau$ where $\tau_{\rm iso}\gg \tau_{\mathrm{p}} \gg \tau$. The DP mechanism implies that stronger momentum scattering (smaller $\tau$) leads to a longer spin polarization lifetime (larger $\tau_{\rm iso}$).
In the opposite limit, $\tau \gg \tau_{\mathrm{p}}$, a few scattering events are sufficient to randomize the spin, and there is
a crossover to $\tau_{\rm iso}\sim\tau$~\cite{Dyakonov_2008_book,Boross2013}.
This crossover has been investigated quantitatively~\cite{Burkov_2004,Szolnoki_2017,Kakoi_2024}.
For small $\eta$, the situation is different.
At $\eta=0$, the gauge field can be removed by a non-Abelian gauge transformation and the system reduces to two uncoupled systems belonging to the orthogonal symmetry class~\cite{Bernevig_2006}.
In this case, the initial spin polarization cannot be scrambled, regardless of the strength of the SOC and the disorder~\cite{Schliemann_2003}.
In this paper, we provide a unified description of the spin--momentum dynamics, consistently connecting all limiting cases.

\section{Diagrammatic calculations}

\subsection{Self-energy and intensity propagator in a nutshell\label{sec:preparation}}

\subsubsection{Disorder-averaged Green's function}

Defining $\hat{H}=\hat{H}_0 + V(\hat{\bm{r}})$, the corresponding Green's operator $\hat{G}(E) = (E-\hat{H}+i0^+)^{-1}$ at energy $E$ obeys the recursion equation $\hat{G}=\hat{G}_0+\hat{G}_0\hat{V}\hat{G}$, where $\hat{G}_0(E) = (E-\hat{H}_0+i0^+)^{-1}$ is the clean Green's operator at energy $E$. It can be shown that the disorder-averaged Green's operator $\overline{\hat{G}}(E)$ satisfies the Dyson equation $\overline{\hat{G}} = \hat{G}_0+\hat{G}_0\hat{\Sigma}\overline{\hat{G}}$ which involves the self-energy operator $\Sigma(E)$ at energy $E$. The latter is given by the infinite sum of one-particle irreducible scattering diagrams~\cite{Akkermans_2007_book}. One finds $\overline{\hat{G}}(E)= [E-\hat{H}_0-\hat{\Sigma} (E)]^{-1}$. Since disorder average restores translation invariance, $\overline{\hat{G}}(E)$ is diagonal in the clean eigenstates basis
\begin{equation}
    \bra{\bm{k},s}\overline{\hat{G}}(E)\ket{\bm{k}',s'}= (2\pi)^2 \, \delta (\bm{k}-\bm{k}') \, \delta_{ss'} \ g_s(\bm{k},E),
\end{equation}
where $s=\pm$ and $g_s(\bm{k},E)= [E-\mathcal{E}_s(\bm{k}) -\Sigma_s(\bm{k},E)]^{-1}$. For sufficiently weak disorder, the self-energy can be obtained within the Born approximation and one finds $\Sigma_s(\bm{k},E) \approx -i \pi \nu(E)\gamma_0/2$, where $\nu(E)$ is the disorder-averaged density of states (DoS) per unit area~\cite{Kakoi_2026}. We thus get
\begin{equation}\label{eq:elements_dis-av-G}
\begin{aligned}
    g_{\pm}(\bm{k},E) &= \frac{1}{E - \mathcal{E}_{\pm}(\bm{k}) + i\frac{\hbar}{2\tau(E)}},
\end{aligned}
\end{equation}
featuring the scattering time
\begin{equation}\label{eq:def_tau}
    \tau(E) = \frac{\hbar}{\pi \nu(E) \gamma_0}.
\end{equation}
In the $\ket{\uparrow\downarrow}$ basis, the Green's function reads
\begin{align}\label{eq:GF_spin_basis}
    \overline{G}(\bm{k},E) &= \frac{g_+(\bm{k},E)+g_-(\bm{k},E)}{2} \, \mathbbm{1}_2 \notag\\
    &+ \frac{g_+(\bm{k},E)-g_-(\bm{k},E)}{2} \ \bm{u}(\phi_\eta)\cdot\bm{\sigma},
\end{align}
where $\bm{u}(\phi_\eta):=(\cos\phi_\eta, \sin\phi_\eta,0)$, $\phi_\eta$ being given by Eq.~\eqref{eq:def_phi_eta}, and where $g_\pm$ are given by Eq.~\eqref{eq:elements_dis-av-G}.

\subsubsection{Diffuson and Cooperon approximation for the disorder-averaged density matrix}

The aim of this work is to obtain an appropriate approximate form of the disorder-averaged density matrix operator $\hat{\varrho}(t) = \overline{\ket{\psi(t)}\bra{\psi(t)}}$ 
as a function of time and momentum. We first start by presenting a general discussion of the approximations commonly employed in the context of weak localization. Further details can be found in Appendix~\ref{app:diagrammatic_calc} and, e.g., in Refs.~\cite{Bergmann_1984_review,Muller_2005,Kuhn_2007,Akkermans_2007_book}.

Introducing the momenta and energy entries
\begin{equation}
\bm{k}_{\pm} = \bm{k}\pm\frac{\bm{q}}{2},\quad
    \bm{k}_{\pm}' = \bm{k}'\pm\frac{\bm{q}}{2},\quad
    E_{\pm} = E\pm\frac{\hbar\omega}{2}
\end{equation}
and Greek indexes $\alpha$, $\beta$, $\gamma$ and $\delta$ to describe any internal degrees of freedom, be it the spin $\uparrow\downarrow$-indexes or the clean eigenstate $\pm$-indexes, a straightforward calculation shows that the matrix elements $\varrho_{\alpha\beta}(\bm{k},\bm{q},t) = 
    \overline{
    \braket{\bm{k}_+,\alpha|\psi(t)}
    \braket{\psi(t)|\bm{k}_-,\beta}}$ satisfy
\begin{align}
\label{eq:def_disorder_av_density_op}
    \varrho_{\alpha\beta}&(\bm{k},\bm{q},t)
    = \sum_{\gamma,\delta}\int_{-\infty}^{\infty}\! \frac{\hbar \,d\omega}{2\pi}
    \int_{-\infty}^{\infty}\! \frac{dE}{2\pi}
    \int\! \frac{d^d\bm{k}'}{(2\pi)^d}\notag\\
    &\times
    e^{-i\omega t}\,
    \Phi_{\alpha\beta,\gamma\delta}(\bm{k},\bm{k}',\bm{q},E,\omega) \, 
    \varrho_{\gamma\delta}(\bm{k}',\bm{q},t=0),
\end{align}
where
\begin{align}\label{eq:def_Phi}
    &\Phi_{\alpha\beta,\gamma\delta}(\bm{k},\bm{k}',\bm{q},E,\omega)\notag\\
    &= \overline{
    \braket{\bm{k}_+,\alpha|
    \hat{G}\!\left(E_+\right)
    |\bm{k}_+',\gamma}
    \braket{\bm{k}_-',\delta|
    \hat{G}^\dag\!\left(E_-\right)
    |\bm{k}_-,\beta}}
\end{align}
are the matrix elements of the disorder-averaged intensity propagator $\hat{\Phi} := \overline{\hat{G}(E_+)\otimes \hat{G}^\dag(E_-)}$. It can be shown that the latter satisfy the Bethe-Salpeter equation~\cite{Akkermans_2007_book} $\hat{\Phi}= \hat{\Phi}_0 + \hat{\Phi}_0 \hat{\mathcal{I}} \hat{\Phi}$ where $\hat{\Phi}_0 := \overline{\hat{G}}(E_+)\otimes \overline{\hat{G}}{}^\dag(E_-)$
corresponds to the Drude--Boltzmann approximation and where the intensity vertex $\hat{\mathcal{I}}$ is given by the infinite sum of two-particle irreducible scattering diagrams. The Bethe-Salpeter equation can be recast under the form $\hat{\Phi}=\hat{\Phi}_0 + \hat{\Phi}_0 \hat{\Gamma} \hat{\Phi}_0$, where $\hat{\Gamma}=\hat{\mathcal{I}}\,(\mathbbm{1} - \hat{\Phi}_0\hat{\mathcal{I}})^{-1}$.
For sufficiently weak disorder, the interference between scattering amplitudes propagating along different scattering paths fluctuates a lot from one disorder realization to another and is thus washed out by the disorder average. This means that only amplitudes that co-propagate or counter-propagate along identical scattering paths can contribute significantly to $\Gamma$. As a consequence, we have $\Gamma \approx \Gamma^{\mathrm{D}}+\Gamma^{\mathrm{C}}$ where $\Gamma^{\mathrm{D}}$ subsumes the contribution of co-propagating amplitudes (ladder diagram series, {\it aka} the Diffuson) while $\Gamma^{\mathrm{C}}$ subsumes the contribution of counter-propagating amplitudes (maximally-crossed diagram series, {\it aka} the Cooperon). The intensity propagator entries Eq.~\eqref{eq:def_Phi} are then given by 
\begin{align}\label{eq:Phi_Diffuson_Cooperon_expantion}
    &\Phi(\bm{k},\bm{k}',\bm{q},E,\omega) \notag\\
    &\simeq \overline{G}(\bm{k}_+,E_+) \otimes
    \overline{G}^*(\bm{k}_-,E_-) 
    \Big\{(2\pi)^d\delta(\bm{k}-\bm{k}')\notag\\
    &+
    \Gamma^{\mathrm{D}}(\bm{k},\bm{k}',\bm{q},E,\omega) \,
    \overline{G}(\bm{k}'_+,E_+) \otimes
    \overline{G}^*(\bm{k}'_-,E_-) \notag\\
    &+
    \Gamma^{\mathrm{C}}(\bm{k},\bm{k}',\bm{q},E,\omega) \,
    \overline{G}(\bm{k}'_+,E_+) \otimes
    \overline{G}^*(\bm{k}'_-,E_-)
    \Big\}.
\end{align}
For spin-independent and $\delta$-correlated disorder, the single impurity line becomes simple in the $\ket{\uparrow\downarrow}$ basis (see Appendix~\ref{app:diagonal_delta-correlated}) and we get 
\begin{align}
    \label{eq:Bethe-Salpeter_Eq_Diffuson}
    \Gamma^{\mathrm{D}}_{\alpha\beta,\gamma\delta}(\bm{k},\bm{k}',\bm{q},E,\omega) 
    &= \gamma_0\left[
    \mathbbm{1}
    -\gamma_0\Pi^{\mathrm{D}}(\bm{q},E,\omega)\right]^{-1}_{\alpha\beta,\gamma\delta} \\[1mm]
    \label{eq:Bethe-Salpeter_Eq_Cooperon}
    \Gamma^{\mathrm{C}}_{\alpha\beta,\gamma\delta}(\bm{k},\bm{k}',\bm{q},E,\omega) 
    &= \gamma_0\left[
    \mathbbm{1}
    -\gamma_0\Pi^{\mathrm{C}}(\bm{Q},E,\omega)\right]^{-1}_{\alpha\delta,\gamma\beta} \notag\\
    &\quad -\gamma_0\delta_{\alpha,\gamma}\delta_{\beta,\delta}
\end{align}
where $\alpha,\beta,\gamma,\delta\in\{\uparrow,\downarrow\}$, $\bm{Q}=\bm{k}+\bm{k}'$ and
\begin{align}
    \label{eq:def_PiD}
    \Pi^{\mathrm{D}}_{\alpha\beta,\gamma\delta}(\bm{q},E,\omega) &= \int\!\frac{d^d\bm{k}}{(2\pi)^d}\,\overline{G}_{\alpha\gamma}(\bm{k}_+,E_+) 
    \overline{G}_{\beta\delta}^*(\bm{k}_-,E_-), \displaybreak[1]\\
    \label{eq:def_PiC}
    \Pi^{\mathrm{C}}_{\alpha\beta,\gamma\delta}(\bm{q},E,\omega) &= \int\!\frac{d^d\bm{k}}{(2\pi)^d}\,\overline{G}_{\alpha\gamma}(\bm{k}_+,E_+) 
    \overline{G}_{\delta\beta}^*(-\bm{k}_-,E_-),
\end{align}
see Appendix~\ref{app:diagrammatic_calc} for details. The disorder-averaged Green's function in the $\ket{\uparrow\downarrow}$ basis is given by Eq.~\eqref{eq:GF_spin_basis}. For the spinless case, do note that reciprocity enforces the equality $\Pi^{\mathrm{D}}(\bm{q}) = \Pi^{\mathrm{C}}(\bm{q})$~\cite{Akkermans_2007_book}.

\subsection{Spin--momentum dynamics in systems under uniform SU(2) gauge fields\label{sec:analytic_results}}

Hereafter, we focus on the Hamiltonian~\eqref{eq:SOC_Ham} (characterized by $\kappa$ and $\eta$) and analyze its dynamics in the presence of a disordered potential (characterized by $\gamma_0$).
The disordered potential gives an energy scale
\begin{equation}
    E_{\ell} = \frac{\hbar^2}{2m\ell^2} = \frac{E_{\kappa}}{(\kappa\ell)^2},
\end{equation}
where $\ell$ is the scattering mean free path ($\ell:=v_\mathrm{F}\tau\simeq\hbar^2 k_{\mathrm{F}}/(\pi\nu\gamma_0m)$ for $k_\mathrm{F}/\kappa\gg1$).
Note that, in the following analyses, we assume the initial energy $E_0$ to be the largest energy scale, i.e., $E_0\gg E_{\ell},E_{\kappa}$, or equivalently, $k_\mathrm{F}\ell\gg1$ and $k_\mathrm{F}/\kappa\gg1$ ($k_\mathrm{F}$ denotes the Fermi wave number).
Under this assumption, our aim is to provide a framework that consistently describes the time- and momentum-dependent density matrix for arbitrary $\eta$, from the weak SOC (or equivalently, dirty) regime ($E_{\kappa}\ll E_{\ell}$, $\kappa\ell\ll1$) to the strong SOC (or equivalently, clean)  regime ($E_{\kappa}\gg E_{\ell}$, $\kappa\ell\gg1$).
In our calculations, we assume that the system size is sufficiently larger than the mean free path; therefore, the results are independent of the system size. Note that, this assumption does not apply, e.g., to ballistic quantum dots or ballistic wires whose width is smaller than the mean free path~\cite{Zaitsev_2005a,*Zaitsev_2005b,Kettemann_2007,Wenk_2010}, where the system size becomes the relevant length scale rather than $\ell$.

In Sec.~\ref{sec:Cooperon_Diffuson}, we derive concrete forms of the Diffuson $\Gamma^{\mathrm{D}}$ and Cooperon $\Gamma^{\mathrm{C}}$ that are valid for small $\bm{q}$ and $\bm{Q}$.
In Sec.~\ref{sec:approx_Pi}, by employing an appropriate approximation for the propagator, we derive the explicit $\bm{q}$- and $\omega$-dependencies of $\Gamma^{\mathrm{D}}$ and $\Gamma^{\mathrm{C}}$.
Using these quantities, in Sec.~\ref{sec:moment_distrib} we demonstrate that the momentum (spin) distribution can be computed and confirm consistency with previous studies in several limiting cases.
These analyses extend Ref.~\cite{Kakoi_2024} to the case of arbitrary uniform SU(2) gauge fields.
In Sec.~\ref{sec:tau_iso}, we derive a cubic equation that determines the spin isotropization time $\tau_{\rm iso}$ for any $\kappa\ell$ and $\eta$.
In Sec.~\ref{sec:transient_peak}, we discuss how to treat the Cooperon at momenta offset from the exact backscattering (large $\bm{Q}$) and introduce an approximation that describes the transient peak~\cite{Kakoi_2026}.

\subsubsection{Diffuson and Cooperon in the $\ket{\uparrow\downarrow}$ basis \label{sec:Cooperon_Diffuson}}

Substituting Eq.~\eqref{eq:GF_spin_basis} into Eqs.~\eqref{eq:def_PiD} and \eqref{eq:def_PiC}, and using $\phi_{\eta}(\theta\!+\!\pi) = \phi_{\eta}(\theta)+\pi$, we get 
\begin{align}
    \label{eq:Pi^DC_4by4}
    &\Pi^{\mathrm{D(C)}}(\bm{q},E,\omega) \notag\\
    &= \frac14 
    \sum_{s,s'=\pm}\int\!\frac{d^2\bm{k}}{(2\pi)^2} 
    \,\mathcal{P}_{ss'}(\bm{k},\bm{q},E,\omega)
    \,F^{\mathrm{D(C)}}_{ss'}(\bm{k},\bm{q}),
\end{align}
where
\begin{align}\label{eq:def_P_ss'}
    \mathcal{P}_{ss'}(\bm{k},\bm{q},E,\omega) = g_{s}\!\left(\bm{k}_+,E_+\right)\,
    g_{s'}^*\!\left(\bm{k}_-,E_-\right),
\end{align}
and
\begin{align}
    F^{\mathrm{D}}_{ss'}(\bm{k},\bm{q}) 
    &= \left[\mathbbm{1}_2 + s\cos\phi_{\eta}(\theta_+)\sigma_x + s\sin\phi_{\eta}(\theta_+)\sigma_y\right] \notag\\
    &\otimes
    \left[\mathbbm{1}_2 + s'\cos\phi_{\eta}(\theta_-)\sigma_x - s'\sin\phi_{\eta}(\theta_-)\sigma_y\right],\displaybreak[1]\\[2mm]
    F^{\mathrm{C}}_{ss'}(\bm{k},\bm{q}) 
    &= 
    \left[\mathbbm{1}_2 + s\cos\phi_{\eta}(\theta_+)\sigma_x + s\sin\phi_{\eta}(\theta_+)\sigma_y\right] \notag\\
    &\otimes
    \left[\mathbbm{1}_2 - s'\cos\phi_{\eta}(\theta_-)\sigma_x - s'\sin\phi_{\eta}(\theta_-)\sigma_y\right],
\end{align}
where $\theta_{\pm}$ denotes the polar angle of $\bm{k}_{\pm}$.
For $q\ll k_\mathrm{F}$, we can approximate $\theta_+\simeq\theta_-\simeq\theta$. Under this approximation, several terms cancel upon performing the integral in Eq.~\eqref{eq:Pi^DC_4by4} and we obtain
\begin{widetext}
\begin{align}
    \Pi^{\mathrm{D}}(\bm{q},E,\omega)
    &\simeq \frac14 
    \left(\begin{matrix}
        A_0\!+\!B_0 & 0 & 0 & A_0\!-\!B_0\\
        0 & A_0\!+\!B_0 & A_{\mathrm{c}}\!-\!A_{\mathrm{s}}\!-\!B_{\mathrm{c}}\!+\!B_{\mathrm{s}} & 0 \\
        0 & A_{\mathrm{c}}\!-\!A_{\mathrm{s}}\!-\!B_{\mathrm{c}}\!+\!B_{\mathrm{s}} & A_0\!+\!B_0 & 0 \\
        A_0\!-\!B_0 & 0 & 0 & A_0\!+\!B_0
    \end{matrix}\right) \displaybreak[1]\\[2mm]
    \label{eq:approx_PiC_for_small_lq}
    \Pi^{\mathrm{C}}(\bm{q},E,\omega)
    &\simeq \frac14 
    \left(\begin{matrix}
        A_0\!+\!B_0 & 0 & 0 & -(A_{\mathrm{c}}\!-\!A_{\mathrm{s}}\!-\!B_{\mathrm{c}}\!+\!B_{\mathrm{s}})\\
        0 & A_0\!+\!B_0 & -(A_0\!-\!B_0)& 0 \\
        0 & -(A_0\!-\!B_0)& A_0\!+\!B_0 & 0 \\
        -(A_{\mathrm{c}}\!-\!A_{\mathrm{s}}\!-\!B_{\mathrm{c}}\!+\!B_{\mathrm{s}}) & 0 & 0 & A_0\!+\!B_0
    \end{matrix}\right)
\end{align}
\end{widetext}
where
\begin{align}
    \label{eq:def_Al}
    A_l(\bm{q},E,\omega) &= \int\!\frac{d^2\bm{k}}{(2\pi)^2}
    \chi_l(\theta,\eta)\sum_{s=\pm} 
    \mathcal{P}_{ss}(\bm{k},\bm{q},\omega,E), \displaybreak[1]\\[1mm]
    \label{eq:def_Bl}
    B_l(\bm{q},E,\omega) &= \int\!\frac{d^2\bm{k}}{(2\pi)^2}
    \chi_l(\theta,\eta)\sum_{s=\pm} 
    \mathcal{P}_{s\bar{s}}(\bm{k},\bm{q},\omega,E),
\end{align}
where $\bar{s}=-s$ and
\begin{equation}\label{eq:def_chi}
    \chi_0(\theta,\eta) = 1,\ \ 
    \chi_{\mathrm{c}}(\theta,\eta) = \cos^2\phi_{\eta},\ \  
    \chi_{\mathrm{s}}(\theta,\eta) = \sin^2\phi_{\eta}.
\end{equation}
This approximation is exact when $\bm{q}=\bm{0}$.
Hereafter, for simplicity, we do not explicitly indicate the $E$ dependence, e.g., we will write $\Pi^{\mathrm{D}}(\bm{q},E,\omega)$ simply as $\Pi^{\mathrm{D}}(\bm{q},\omega)$.
Both $\Pi^{\mathrm{D}}$ and $\Pi^{\mathrm{C}}$ can be diagonalized by the basis transformation
\begin{align}\label{eq:spin2singlet-triplet}
    \left(\begin{matrix}
        \ket{\lambda_1}\\
        \ket{\lambda_2}\\
        \ket{\lambda_3}\\
        \ket{\lambda_4}
    \end{matrix}\right) = \frac{1}{\sqrt{2}}\left(\begin{matrix}
        0 & 1 & -1 & 0\\
        0 & 1 &  1 & 0\\
        1 & 0 & 0  & 1\\
        1 & 0 & 0  & -1
    \end{matrix}\right)\left(\begin{matrix}
        \ket{\uparrow\uparrow}\\
        \ket{\uparrow\downarrow}\\
        \ket{\downarrow\uparrow}\\
        \ket{\downarrow\downarrow}
    \end{matrix}\right),
\end{align}
resulting in 
\begin{equation}
\begin{aligned}
    \Pi^{\mathrm{D}}(\bm{q},\omega) = {\rm diag}(\Pi_4, \Pi_3, \Pi_2, \Pi_1), \\
    \Pi^{\mathrm{C}}(\bm{q},\omega) = {\rm diag}(\Pi_1, \Pi_2, \Pi_3, \Pi_4),
\end{aligned}
\end{equation}
where
\begin{equation}\label{eq:elemetns_Pi_diag}
\begin{aligned}
    \Pi_1(\bm{q},\omega)&=
    \frac12 A_0(\bm{q},\omega),\\[1mm]
    \Pi_2(\bm{q},\omega)&=
    \frac12 B_0(\bm{q},\omega),\\[1mm]
    \Pi_3(\bm{q},\omega)
    &= \frac12 A_{\rm s}(\bm{q},\omega) 
    + \frac12 B_{\rm c}(\bm{q},\omega),\\[1mm]
    \Pi_4(\bm{q},\omega)
    &=\frac12 A_{\rm c}(\bm{q},\omega)
    + \frac12 B_{\rm s}(\bm{q},\omega).
\end{aligned}
\end{equation}
Here, $A_{\mathrm{c}}(\bm{q},\omega)+A_{\mathrm{s}}(\bm{q},\omega)=A_0(\bm{q},\omega)$ and $B_{\mathrm{c}}(\bm{q},\omega)+B_{\mathrm{s}}(\bm{q},\omega)=B_0(\bm{q},\omega)$ are used.

By solving the Bethe--Salpeter equations~\eqref{eq:Bethe-Salpeter_Eq_Diffuson} and \eqref{eq:Bethe-Salpeter_Eq_Cooperon},\footnote{Note that the Cooperon must be “twisted” at the end; see subscripts in Eq.~\eqref{eq:Bethe-Salpeter_Eq_Cooperon} and Eq.~\eqref{eq:BS_Cooperon} in Appendix~\ref{app:diagrammatic_calc}.} we can express $\Gamma^{\mathrm{D}}$ and $\Gamma^{\mathrm{C}}$ in the $(\ket{\uparrow\uparrow},\ket{\uparrow\downarrow},\ket{\downarrow\uparrow},\ket{\downarrow\downarrow})$ basis in terms of $\Pi_n$, as
\begin{align}
    \label{eq:Diffuson_matrix_spin_rep}
    &\Gamma^{\mathrm{D}}(\bm{q},\omega) \notag\\
    &=\frac{\gamma_0}{2}\!\left(\begin{matrix}
        \Gamma_1+\Gamma_2 & 0 & 0 & \Gamma_1-\Gamma_2 \\
        0 & \Gamma_3+\Gamma_4 & -\Gamma_3+\Gamma_4 & 0 \\
        0 & -\Gamma_3+\Gamma_4 & \Gamma_3+\Gamma_4 & 0 \\
        \Gamma_1-\Gamma_2 & 0 & 0 & \Gamma_1+\Gamma_2
    \end{matrix}\right) \displaybreak[1]\\
    \label{eq:Cooperon_matrix_spin_rep}
    &\Gamma^{\mathrm{C}}(\bm{q},\omega) \notag\\
    &=\frac{\gamma_0}{2}\!\left(\begin{matrix}
        \Gamma_3+\Gamma_4 & 0 & 0 & -\Gamma_1+\Gamma_2 \\
        0 & \Gamma_1+\Gamma_2 & \Gamma_3-\Gamma_4 & 0 \\
        0 & \Gamma_3-\Gamma_4 & \Gamma_1+\Gamma_2 & 0 \\
        -\Gamma_1+\Gamma_2 & 0 & 0 & \Gamma_3+\Gamma_4
    \end{matrix}\right)\! -\gamma_0\mathbbm{1}_4,
\end{align}
where
\begin{equation}\label{eq:def_Gamma_n}
    \Gamma_n(\bm{q},\omega) = \frac{1}{1-\gamma_0\Pi_n(\bm{q},\omega)}.
\end{equation}

\subsubsection{Approximations for $\Pi_n$ and $\Gamma_n$\label{sec:approx_Pi}}

From Eqs.~\eqref{eq:Diffuson_matrix_spin_rep}--\eqref{eq:def_Gamma_n}, once the explicit form of $\Pi_n(\bm{q},\omega)$ is known, the Diffuson and Cooperon can be obtained, and consequently, the disorder-averaged density matrix can be determined by Eqs.~\eqref{eq:def_disorder_av_density_op} and \eqref{eq:Phi_Diffuson_Cooperon_expantion}.
In this subsection, we obtain a concrete expression for $\Pi_n$ as a function of $\bm{q}$ and $\omega$ using an appropriate approximation.
To simplify the problem, we assume that $A_l$ and $B_l$ can be approximated in the form of 
\begin{align}
    \label{eq:approx_A}
    \gamma_0 A_l(\bm{q},\omega)
    &\simeq -\frac{2\xi_{l}\,x_{\omega}}{x_{\omega}^2 + \frac{(q\ell)^2}{2}}
    \displaybreak[1]\\[1mm]
    \label{eq:approx_B}
    \gamma_0 B_l(\bm{q},\omega), 
    &\simeq -\frac{2\xi_l\,x_{\omega}}
    {x_{\omega}^2 + \Lambda_{l}^2+\frac{(q\ell)^2}{2}},
\end{align}
where we have introduced
\begin{equation}\label{eq:def_x_omega}
    x_{\omega} = -1 + i\omega\tau
\end{equation}
to make formulae look simpler, see Appendix~\ref{app:approx_propagator} for details.
We now define the $\eta$-dependent dimensionless functions $\xi_l$ 
\begin{align}\label{eq:def_xi}
    \xi_l(\eta) &= \int_0^{2\pi}\!\frac{d\theta}{2\pi}\,\chi_l(\theta,\eta)= \left\{\begin{array}{ll}
         1   & \text{for $l=0$},\\[1mm]
         1-r_{\eta}  & \text{for $l={\rm c}$},\\[1mm]
         r_{\eta}  & \text{for $l={\rm s}$},
    \end{array}\right.
\end{align}
where
\begin{equation}\label{eq:def_r_eta}
    r_{\eta} = \frac{|\tan\eta|}{1+|\tan\eta|},
\end{equation}
and functions  $\Lambda_l(\kappa\ell,\eta)$ satisfying
\begin{equation}\label{eq:def_Lambda}
    \frac{\xi_l(\eta)}{1+[\Lambda_{l}(\kappa\ell,\eta)]^2}
    = \int_0^{2\pi}\!\frac{d\theta}{2\pi}
    \frac{\chi_l(\theta,\eta)}{1+4(\kappa\ell)^2[f(\theta,\eta)]^2}.
\end{equation}
Here, $\chi_l$ is defined in Eq.~\eqref{eq:def_chi} and the function $f$ is given by
\begin{equation}\label{eq:def_f}
    f(\theta,\eta)=\sqrt{\frac{1+\cos2\eta\cos2\theta}{2}},
\end{equation}
which gives an angular dependence of the energy splitting, see the dispersion relation in Eq.~\eqref{eq:eigvals}.
Ansatzes~\eqref{eq:approx_A} and \eqref{eq:approx_B} are obtained by generalizing the expressions derived in the limits $\eta\to\pi/4$ (where $f(\theta,\eta)=\mathrm{const.}$) and $q\ell\to0$ under the on-shell assumption. The definitions in Eqs.~\eqref{eq:def_xi}--\eqref{eq:def_Lambda} are chosen such that Eqs.~\eqref{eq:approx_A} and \eqref{eq:approx_B} correctly reproduce the low-frequency expansion of $A_l$ and $B_l$ [Eqs.~\eqref{eq:def_Al} and \eqref{eq:def_Bl}], see Appendix~\ref{app:approx_propagator} for details. 
In the weak-SOC ($\kappa\ell\ll1$) and diffusive ($q\ell\ll1$ and $\omega\tau\ll1$) limits, Eqs.~\eqref{eq:approx_A} and \eqref{eq:approx_B} reduce to those obtained within the diffusion approximation.
As will be seen, expressions in Eqs.~\eqref{eq:approx_A} and \eqref{eq:approx_B} describe the time dependence of the momentum distribution very well for any values of $\kappa\ell$ and $\eta$. Note that, under this approximation, the equalities $A_{\mathrm{c}}+A_{\mathrm{s}}=A_0$ and $B_{\mathrm{c}}+B_{\mathrm{s}}=B_0$ no longer hold exactly while they do hold in the case of $q=0$ and $\omega=0$.
The ansatz in the weak SOC ($\kappa\ell\ll1$) and diffusive ($q\ell\ll1$ and $\omega\tau\ll1$) limit.

The functions $\Gamma_n$ defined in Eq.~\eqref{eq:def_Gamma_n} can be computed using Eqs~\eqref{eq:elemetns_Pi_diag}, \eqref{eq:approx_A} and \eqref{eq:approx_B}, and we get
\begin{align}
    \label{eq:Gamma_1}
    \Gamma_{1}(\bm{q},\omega) &= 1 - \frac{x_{\omega}}{x_{\omega}^2 + x_{\omega} + \frac{(q\ell)^2}{2}},
    \displaybreak[1]\\
    \label{eq:Gamma_2}
    \Gamma_{2}(\bm{q},\omega) &= 1 - \frac{x_{\omega}}{x_{\omega}^2 + x_{\omega} + \Lambda_0^2 + \frac{(q\ell)^2}{2}},
\end{align}
and 
\begin{equation}
    \label{eq:Gamma_34}
    \Gamma_n(\bm{q},\omega) = 
    1 - \frac{x_{\omega}(x_{\omega}^2 + C_n^{(1)})}{x_{\omega}^4 + x_{\omega}^3 
    + C_n^{(2)} x_{\omega}^2 
    + C_n^{(1)} x_{\omega} 
    + C_n^{(0)}}
\end{equation}
for $n=3,4$, where
\begin{equation}
\begin{aligned}
    C_{3}^{(2)}(\bm{q}) &= \Lambda_{\mathrm{c}}^2+(q\ell)^2,\\[1mm]
    C_{3}^{(1)}(\bm{q}) &= r_\eta\Lambda_{\mathrm{c}}^2+\frac{(q\ell)^2}{2},\\[1mm]
    C_{3}^{(0)}(\bm{q}) &= \frac{(q\ell)^2}{2} \left[
    \Lambda_{\mathrm{c}}^2+\frac{(q\ell)^2}{2}\right].
\end{aligned}
\end{equation}
and 
\begin{equation}
\begin{aligned}
    C_{4}^{(2)}(\bm{q}) &= \Lambda_{\mathrm{s}}^2+(q\ell)^2,\\[1mm]
    C_{4}^{(1)}(\bm{q}) &= (1-r_\eta)\,\Lambda_{\mathrm{s}}^2+\frac{(q\ell)^2}{2},\\[1mm]
    C_{4}^{(0)}(\bm{q}) &= \frac{(q\ell)^2}{2} \left[
    \Lambda_{\mathrm{s}}^2+\frac{(q\ell)^2}{2}\right].
\end{aligned}
\end{equation}
In the case of $\bm{q}=\bm{0}$, $\Gamma_3$ and $\Gamma_4$ simplify to
\begin{align}
    \label{eq:Gamma_3_cubic}
    \Gamma_3(\bm{0},\omega) &= 
    \frac{x_{\omega}\left(x_{\omega}^2\!+\!\Lambda_{\mathrm{c}}^2\right)}
    {x_{\omega}^3 + x_{\omega}^2 + \Lambda_{\mathrm{c}}^2\,x_{\omega} + r_{\eta}\Lambda_{\mathrm{c}}^2},
    \displaybreak[1]\\[1mm]
    \label{eq:Gamma_4_cubic}
    \Gamma_4(\bm{0},\omega) &= 
    \frac{x_{\omega}\left(x_{\omega}^2\!+\!\Lambda_{\mathrm{s}}^2\right)}
    {x_{\omega}^3 + x_{\omega}^2 + \Lambda_{\mathrm{s}}^2\,x_{\omega} + (1-r_{\eta})\Lambda_{\mathrm{s}}^2}.
\end{align}
Similarly, $\Gamma_1$  reduces to
\begin{equation}\label{eq:Gamma_linear}
    \Gamma_{1}(\bm{0},\omega) = \frac{i}{\omega\tau}+1.
\end{equation}
Since $\Gamma_{1}(\bm{0},\omega)$ has a pole at $\omega=0$, it is a dominant term in the long-time limit.

In several limiting cases, $\Lambda_l$ can be expressed in a simple form.
For $\eta=\pi/4$, where $f$ and $\chi_l$ are $\theta$-independent, we have
\begin{equation}
    \Lambda_l(\kappa\ell,\eta\!=\!\pi/4) = \left\{\begin{array}{ll}
        2\kappa\ell & \text{for $l=0$}, \\[1mm]
        \sqrt{2}\kappa\ell & \text{for $l=\mathrm{c},\mathrm{s}$}.
    \end{array}
    \right.
\end{equation} 
In this case, Eqs.~\eqref{eq:Gamma_3_cubic} and \eqref{eq:Gamma_4_cubic} coincide with those derived in Ref.~\cite{Kakoi_2024}.
Also, for $\kappa\ell\ll1$, 
\begin{align}\label{eq:approx_Lambda_small_kappal}
    [\Lambda_{l}(\kappa\ell,\eta)]^2 &\simeq
    \frac{4(\kappa\ell)^2}{\xi_l(\eta)}
    \int_0^{2\pi}\!\frac{d\theta}{2\pi}\,
    \chi_l(\theta,\eta)[f(\theta,\eta)]^2
    \displaybreak[1]\notag\\[1mm]
    &=\frac{2(\kappa\ell)^2}{\xi_l(\eta)}\times\left\{\begin{array}{ll}
        1 & \text{for $l=0$}, \\[1mm]
        \cos^2\eta & \text{for $l=\mathrm{c}$}, \\[1mm]
        \sin^2\eta & \text{for $l=\mathrm{s}$},
    \end{array}\right.
\end{align}
holds.

\subsubsection{Time evolution of the momentum distribution with an initial eigenstate of $\hat{H}_0$\label{sec:moment_distrib}}

From Eqs.~\eqref{eq:Gamma_1}--\eqref{eq:Gamma_34}, which are valid for $k_\mathrm{F}\ell\gg1$ and $k_\mathrm{F}/\kappa\gg1$, we can obtain an analytical expression for the density matrix [see also Eqs.~\eqref{eq:def_disorder_av_density_op}, \eqref{eq:Phi_Diffuson_Cooperon_expantion}, \eqref{eq:Diffuson_matrix_spin_rep}, and \eqref{eq:Cooperon_matrix_spin_rep}].
As a concrete calculation, we address the time evolution of the disorder-averaged momentum distribution, which can be directly measured in cold atom experiments~\cite{Jendrzejewski_2012a,Labeyrie_2012}. In this paper, for simplicity, we choose the initial state of the system to be some eigenstate $\ket{\bm{k}_0,s_0}$ of $\hat{H}_0$. Equivalently, we have
\begin{equation}\label{eq:init_density_op}
    \hat{\varrho}(t\!=\!0) = \ket{\bm{k}_0,s_0}\bra{\bm{k}_0,s_0}.
\end{equation}
Introducing the branch-resolved momentum distribution
\begin{equation}
    n_{s}(\bm{k},t) := \bra{\bm{k},s}\hat{\varrho}(t)\ket{\bm{k},s} \quad (s=\pm),
\end{equation}
we define the spin-unresolved momentum distribution as
\begin{equation}\label{eq:momentum_distrib}
    n(\bm{k},t) := \Tr_{\rm{int}}\braket{\bm{k}|\hat{\varrho}(t)|\bm{k}} = \sum_{s=\pm} \, n_s(\bm{k},t),
\end{equation}
where $\Tr_{\rm{int}}$ denotes the trace over the spin internal degrees of freedom. We have already obtained the explicit forms of the Diffuson $\Gamma^{\mathrm{D}}$ and Cooperon $\Gamma^{\mathrm{C}}$ in the $\ket{\uparrow\downarrow}$ basis, see Eqs.~\eqref{eq:Diffuson_matrix_spin_rep}, \eqref{eq:Cooperon_matrix_spin_rep}, and \eqref{eq:Gamma_1}--\eqref{eq:Gamma_34}, and we also know the transformation between the $\ket{\uparrow\downarrow}$ basis and the $\ket{\pm}$ basis, see Eq.~\eqref{eq:eigvecs}. Therefore, a straightforward calculation gives\footnote{The term originating from the first term in Eq.~\eqref{eq:Phi_Diffuson_Cooperon_expantion} (the Drude--Boltzmann term)
\begin{align}
n^{0}_{s}(\bm{k},t)
&=(2\pi)^2\delta(\bm{k}\!-\!\bm{k}_0)\delta_{s,s_0}\notag\\
&\times\int_{-\infty}^{\infty}\frac{dE}{2\pi} \int_{-\infty}^{\infty}\frac{\hbar \,d\omega}{2\pi} e^{-i\omega t}g_{s_0}(\bm{k}_0,E_+)g_{s_0}^*(\bm{k}_0,E_-)\notag
\end{align}
contributes only at the initial momentum and is therefore neglected in the following discussion. Note, however, that it must be taken into account when performing an integration over momentum (e.g., in calculations of the conductivity).}
\begin{align}\label{eq:branch-resolved_momentum_distrib}
    &n_{s}(\bm{k},t) = \int_{-\infty}^{\infty}\frac{dE}{2\pi} \int_{-\infty}^{\infty}\frac{\hbar \,d\omega}{2\pi} e^{-i\omega t} \notag\\
    &\quad\times\left[
    \mathit{\Gamma}^{\mathrm{D}}_{ss,s_0s_0}(\bm{k},\bm{k}_0,\bm{0},E,\omega)
    +\mathit{\Gamma}^{\mathrm{C}}_{ss,s_0s_0}(\bm{k},\bm{k}_0,\bm{0},E,\omega)\right]\notag\\
    &\quad\times g_s(\bm{k},E_+)g_s^*(\bm{k},E_-)g_{s_0}(\bm{k}_0,E_+)g_{s_0}^*(\bm{k}_0,E_-),
\end{align}
where $\mathit{\Gamma}^{\mathrm{D}}$ and $\mathit{\Gamma}^{\mathrm{C}}$ are the Diffuson and Cooperon in the $\ket{\pm}$ basis:
\begin{align}\label{eq:Diffuson_matrix_branch_rep}
    &\mathit{\Gamma}^{\mathrm{D}}_{ss,s's'}(\bm{k},\bm{k}'\!,\bm{q},E,\omega)\notag\\
    &= \frac{\gamma_0}{4} \big\{ 2\Gamma_1(\bm{q},\omega) \notag\\
    &+ ss_0 \left[ \Gamma_3(\bm{q},\omega) + \Gamma_4(\bm{q},\omega) \right] 
    \cos[\phi_{\eta}(\theta) - \phi_{\eta}(\theta')] \notag\\
    &- ss_0 \left[ \Gamma_3(\bm{q},\omega) - \Gamma_4(\bm{q},\omega) \right] 
    \cos[\phi_{\eta}(\theta) + \phi_{\eta}(\theta')] \big\},
\end{align}
and
\begin{align}\label{eq:Cooperon_matrix_branch_rep}
    &\mathit{\Gamma}^{\mathrm{C}}_{ss,s's'}(\bm{k},\bm{k}'\!,\bm{q},E,\omega)\notag\\
    &= \frac{\gamma_0}{4} \big\{ - \Gamma_1(\bm{Q},\omega) + \Gamma_2(\bm{Q},\omega)\notag\\
    & + \Gamma_3(\bm{Q},\omega) + \Gamma_4(\bm{Q},\omega) - 2 \notag\\
    &+ ss_0 \left[ \Gamma_1(\bm{Q},\omega) + \Gamma_2(\bm{Q},\omega) - 2 \right] 
    \cos[\phi_{\eta}(\theta) - \phi_{\eta}(\theta')] \notag\\
    &+ ss_0 \left[ \Gamma_3(\bm{Q},\omega) - \Gamma_4(\bm{Q},\omega) \right] 
    \cos[\phi_{\eta}(\theta) + \phi_{\eta}(\theta')] \big\},
\end{align}
where $\bm{Q}=\bm{k}+\bm{k}_0$.
$g_s(\bm{k},E)$ is given in Eq.~\eqref{eq:elements_dis-av-G}.
Integrating over $E$ in Eq.~\eqref{eq:branch-resolved_momentum_distrib}
under on-shell ($E=E_0$) approximation for the Diffuson and Cooperon, we get
\begin{align}\label{eq:momentum_distrib_analytic}
    &n_{s}^{\mathrm{D(C)}}(\bm{k},t)\notag\\
    &= -\frac{2\tau^3}{\hbar^2}\int_{-\infty}^{\infty}\frac{d\omega}{2\pi} \, \frac{
    \mathit{\Gamma}^{\mathrm{D(C)}}_{ss,s_0s_0}(\bm{k},\bm{k}_0,\bm{0},E_0,\omega)\, e^{-i\omega t}}
    {
    x_{\omega} 
    \left\{x_{\omega}^2+\tau^2[\mathcal{E}_{s}(\bm{k})\!-\!E_0]^2/\hbar^2\right\}}.
\end{align}
$x_{\omega}$ is defined in Eq.~\eqref{eq:def_x_omega}.
The functional form of $\Gamma_n(\bm{q},E,\omega)$ is given as in Eqs.~\eqref{eq:Gamma_1}--\eqref{eq:Gamma_34}, the momentum distribution can be obtained by evaluating the integral over $\omega$ in Eq.~\eqref{eq:momentum_distrib_analytic} using the residue theorem~\cite{Scoquart_2020,Kakoi_2024}.
For the specific case where $\bm{k}=-\bm{k}_0$ and $\eta=\pi/4$, the same result was derived in Ref.~\cite{Kakoi_2024}.

To confirm the validity of the derived formulae, we demonstrate that our calculations reproduce the exact results.
From $\cos[\phi_{\eta}(\theta_0+\pi) - \phi_{\eta}(\theta_0)] = \cos\pi = -1$, it follows that
\begin{align}
    &\mathit{\Gamma}^{\mathrm{D}}_{s_0s_0,s_0s_0}(-\bm{k}_0,\bm{k}_0,\bm{0},E_0,\omega)
    \notag\\
    &+ \mathit{\Gamma}^{\mathrm{C}}_{s_0s_0,s_0s_0}(-\bm{k}_0,\bm{k}_0,\bm{0},E_0,\omega) 
    = 0,
\end{align}
which is equivalent to $n_{s_0}(-\bm{k}_0,t)=0$.
This is consistent with the fact that scattering from the initial state $\ket{\bm{k}_0,s_0}$ into its time-reversed state $\ket{-\bm{k}_0,s_0}$ is completely suppressed by destructive interference.
Next, we confirm the behavior of the Diffuson contribution in the $t\!\rightarrow\!\infty$ limit.
Since the static limit corresponds to taking the $\omega\to0$ limit, only the contribution from $\Gamma_1$ is relevant, and $\mathit{\Gamma}^{\mathrm{D}}$ in Eq.~\eqref{eq:Diffuson_matrix_branch_rep} can therefore be approximated by
\begin{equation}\label{eq:GammaD_static}
    \mathit{\Gamma}^{\mathrm{D}}_{ss,s_0s_0}(\bm{k},\bm{k}_0,\bm{0},E_0,\omega) \simeq i \frac{\gamma_0}{2\omega\tau}.
\end{equation}
Substituting Eq.~\eqref{eq:GammaD_static} into Eq.~\eqref{eq:momentum_distrib_analytic} and integrating over $\omega$, we obtain
\begin{align}\label{eq:nD_long_time_limit}
    n^{\mathrm{D}}(\bm{k},t\!\to\!\infty) 
    &= \sum_s n_s^{\mathrm{D}}(\bm{k},t\!\to\!\infty)\notag\\
    &= \sum_s \frac{n_0}{1+\tau^2[\mathcal{E}_{s}(\bm{k})\!-\!E_0]^2/\hbar^2},
\end{align}
where we define
\begin{equation}\label{eq:def_n0}
    n_0 = \frac{\tau^2\gamma_0}{\hbar^2} = \frac{1}{\pi^2\nu^2\gamma_0}.
\end{equation}
This is consistent with the known result obtained in Ref.~\cite{KL-Lee_2014b}.
Similarly, for the Cooperon contribution, only the $\Gamma_1$ term is relevant in the long-time limit, and we find
\begin{align}
    &n^{\mathrm{C}}_{s}(\bm{k},t\!\to\!\infty) = 
    \left\{\begin{array}{ll}
    -n_0 e^{-tDQ^2}
    & \text{for $s=s_0$}, \\
    0
    & \text{for $s=\bar{s}_0$},
    \end{array}
    \right.
\end{align}
where 
\begin{equation}\label{eq:diffusion_const}
    D=\frac{\ell^2}{2\tau}
\end{equation}
is the diffusion constant for 2D systems.
This behavior in the diffusive limit agrees with known results~\cite{Akkermans_2007_book,Arabahmadi_2024} and immediately shows that the FWHM of the CBS dip observed for such disordered spin--orbit systems is given by $2\sqrt{\log2/Dt}$. 
In the localized regime, the CBS width is time independent (given by the inverse localization length), and finite-time scaling allows one to extract the critical exponent of the Anderson transition~\cite{Arabahmadi_2024,Ghosh_2015}.

\subsubsection{Spin isotropization time\label{sec:tau_iso}}

We have derived analytical expressions for the time evolution of the momentum distribution.
This allows us to determine the spin isotropization time $\tau_{\rm iso}$, namely, the time it takes for the Diffuson contribution to the momentum distribution $n^{\mathrm{D}}$ to reach its steady state in Eq.~\eqref{eq:nD_long_time_limit}. Specifically, as done in Ref.~\cite{Kakoi_2024}, $\tau_{\rm iso}$ is determined from the pole of the Diffuson.

Equation~\eqref{eq:Diffuson_matrix_branch_rep} shows that the difference between the momentum distributions in the $s_0$ and $\bar{s}_0$ branches for the initial $s_0$ state arises from the contributions of $\Gamma_3(\bm{0},\omega)$ and $\Gamma_4(\bm{0},\omega)$. These contributions vanish at sufficiently long times, that is, at times much longer than the spin isotropization time.
The characteristic lifetimes of these contributions are determined by their poles, since the integral in Eq.~\eqref{eq:momentum_distrib_analytic} can be evaluated using the residue theorem. As long as the domain of $\eta$ is given by $[0,\pi/4]$, the contribution from $\Gamma_4$ is always longer lived than that from $\Gamma_3$. We therefore focus on $\Gamma_4$ in the following.

To identify the poles of $\Gamma_4(\bm{0},\omega)$, one needs to solve the following cubic equation in $\omega$:
\begin{equation}\label{eq:cubic_eq_transport_mean_free_time}
    x_{\omega}^3 + x_{\omega}^2 
    + [\Lambda_{\mathrm{s}}(\kappa\ell,\eta)]^2x_{\omega} 
    + (1-r_{\eta})[\Lambda_{\mathrm{s}}(\kappa\ell,\eta)]^2 = 0,
\end{equation}
which is obtained by setting the denominator of $\Gamma_4(\bm{0},\omega)$ equal to zero; see Eq.~\eqref{eq:Gamma_4_cubic}.\footnote{$x_\omega$, $r_{\eta}$, and $\Lambda_{\mathrm{s}}(\kappa\ell,\eta)$ are defined in Eqs.~\eqref{eq:def_x_omega}, \eqref{eq:def_r_eta}, and \eqref{eq:def_Lambda}, respectively [see also Eqs.~\eqref{eq:def_chi}, \eqref{eq:def_xi}, and \eqref{eq:def_f}].}
Note that, for the special case $\eta=\pi/4$, where $r_\eta=1/2$ and $\Lambda_{\mathrm{s}}=\sqrt{2}\kappa\ell$ hold, the cubic equation~\eqref{eq:cubic_eq_transport_mean_free_time} coincides with that derived previously; see Refs.~\cite{Szolnoki_2017,Kakoi_2024}.
We denote the solutions of Eq.~\eqref{eq:cubic_eq_transport_mean_free_time} as $\omega_m$ ($m\in\{1,2,3\}$).
All of these poles have non-positive imaginary parts. The real parts are either zero or appear in positive--negative pairs.
In two limiting cases ($\Lambda_{\mathrm{s}}\ll1$ and $\Lambda_{\mathrm{s}}\gg1$), the poles $\omega_m$ of $\Gamma_{4}(\bm{0},\omega)$ are approximated by simple expressions, and we summarize them in Table~\ref{tab:poles_Gamma}. 
Each pole $\omega_m$ contributes the exponential factors $e^{t\,\mathrm{Im}(\omega_m)}\, e^{-it\,\mathrm{Re}(\omega_m)}$ to the time dependence of the momentum distribution.
The actual time evolution is given by the sum of contributions from all poles. Since $\mathrm{Im}(\omega_m) \leq 0$, the pole with the smallest imaginary part contributes to the time decay that survives the longest.
Therefore, it is reasonable to define the spin isotropization time by
\begin{equation}\label{eq:def_tau_iso}
    \tau_{\rm iso} = \max_{m}\!\left(\frac{1}{|{\rm Im}\,\omega_m|}\right).
\end{equation}
\begin{table}[b]
    \caption{
    Summary of the asymptotic expressions for the poles $\omega_m$ of $\Gamma_{4}(\bm{0},\omega)$, nondimensionalized by $1/\tau$.
    \label{tab:poles_Gamma}}
    \begin{ruledtabular}
    \renewcommand{\arraystretch}{1.7}
    \begin{tabular}{cc}
     Condition & Poles $\omega\tau_m$\\
    \hline
    \rule{0pt}{1.7\normalbaselineskip}
    $\displaystyle \Lambda_{\mathrm{s}}^2 \ll 1$ & 
    $\displaystyle -ir_{\eta}\Lambda_{\mathrm{s}}^2$,\quad 
    $\displaystyle -i\!\pm\! \sqrt{1\!-\!r_{\eta}}\,\Lambda_{\mathrm{s}}$\\[2mm]
    $\displaystyle \Lambda_{\mathrm{s}}^2 \gg 1$ & 
    $\displaystyle -ir_{\eta}$,\quad
    $\displaystyle -i\!\left(1\!-\!\frac{r_{\eta}}{2}\right)\!\pm\Lambda_{\mathrm{s}}$
    \end{tabular}
    \end{ruledtabular}
\end{table}

\begin{figure}[t]
\begin{center}
    \includegraphics[width=\columnwidth]{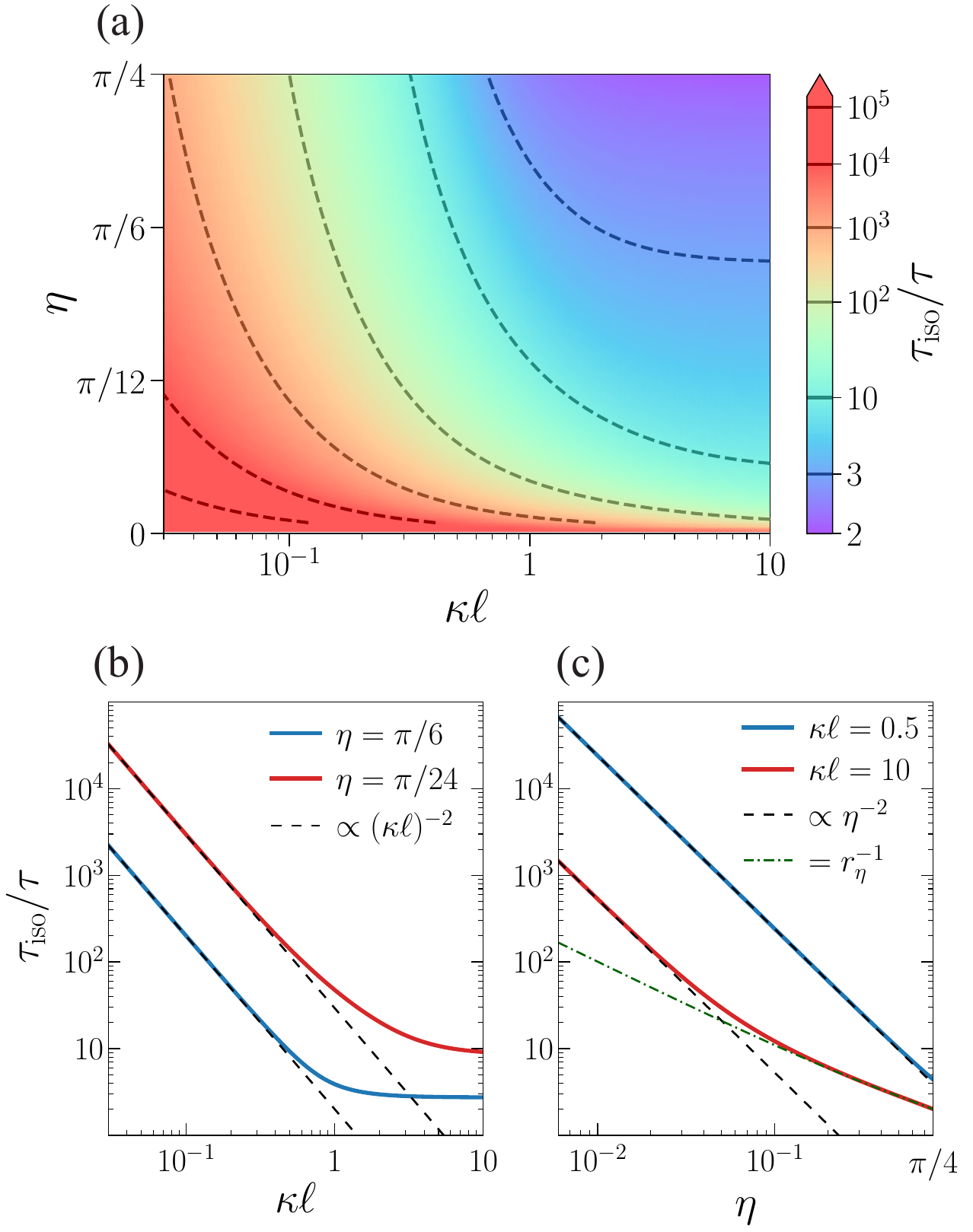}
    \caption{
    Panel (a): Spin isotropization time $\tau_{\rm iso}$ (in units of the scattering time $\tau$) in the ($\kappa\ell$, $\eta$) plane with its associated color scale values, as obtained from the prescription Eq.~\eqref{eq:def_tau_iso} applied to the solutions of Eq.~\eqref{eq:cubic_eq_transport_mean_free_time}. 
    The dashed lines mark the contours of $\tau_{\rm iso}/\tau = C$ for $C = 3, 10, 10^2, 10^3, 10^4,$ and $10^5$.
    This result holds regardless of $E_0$ as long as the conditions $k_\mathrm{F}\ell\gg1$ and $k_\mathrm{F}/\kappa\gg1$ are satisfied. Panel (b): $\kappa\ell$ dependence of $\tau_{\rm iso}$ for $\eta= \pi/6$ (blue curve) and $\eta=\pi/24$ (red curve). The dashed lines represent the predicted asymptotic behavior $(\kappa\ell)^{-2}$ in Eq.~\eqref{eq:tau_iso_small_kappal}. Panel (c): $\eta$ dependence of $\tau_{\rm iso}$ for $\kappa\ell=0.5$ (blue curve) and $\kappa\ell=10$ (red curve). The dashed lines represent the analytical asymptotic estimates given by Eq.~\eqref{eq:tau_iso_asymptotic}. $r_{\eta}$ is defined in Eq.~\eqref{eq:def_r_eta}. For the behavior $\eta^{-2}$, see also Eq.~\eqref{eq:tau_iso_small_kappal} at small $\eta$.}
    \label{fig:isotropization_time}
\end{center}
\end{figure}
In Fig.~\ref{fig:isotropization_time}, we show the $\kappa\ell$- and $\eta$-dependence of $\tau_{\rm iso}/\tau$, obtained by solving Eq.~\eqref{eq:cubic_eq_transport_mean_free_time}. 
The behavior at $\eta=\pi/4$ has been studied in detail from weak SOC regime ($\kappa \ell\ll1$) to strong SOC regime ($\kappa \ell\gg1$) in Refs.~\cite{Szolnoki_2017,Kakoi_2024}, yielding the same results. It is also well known that spin relaxation is absent ($\tau_{\rm iso}\to \infty$) at $\eta=0$, which corresponds to the case where the Rashba and Dresselhaus spin--orbit couplings have equal strengths~\cite{Schliemann_2003,Bernevig_2006} and the gauge field can be gauged away. Our results continuously interpolate between these two limiting cases.

In the limits $\Lambda_{\mathrm{s}}^2 \ll 1$ and $\Lambda_{\mathrm{s}}^2 \gg 1$, the following asymptotic forms hold [see also Table~\ref{tab:poles_Gamma}]:
\begin{equation}\label{eq:tau_iso_asymptotic}
    \frac{\tau_{\rm iso}}{\tau} \simeq 
    \begin{cases}
        \displaystyle \frac{1}{r_\eta \Lambda_{\mathrm{s}}^2}, 
        & \text{for $\Lambda_{\mathrm{s}}^2 \ll 1$}, \\[3mm]
        \displaystyle \frac{1}{r_\eta}, 
        & \text{for $\Lambda_{\mathrm{s}}^2 \gg 1$}.
    \end{cases}
\end{equation}
In particular, for $\kappa\ell \ll 1$, where the condition $\Lambda_{\mathrm{s}}^2 \ll 1$ is automatically satisfied, one further obtains
\begin{equation}\label{eq:tau_iso_small_kappal}
    \frac{\tau_{\rm iso}}{\tau} \simeq \frac{1}{2}(\kappa\ell \sin\eta)^{-2},
\end{equation}
where Eq.~\eqref{eq:approx_Lambda_small_kappal} is used.
In Fig.~\ref{fig:isotropization_time}(b) and (c), we present these asymptotic behaviors for several representative parameters.

Spin relaxation in the regime $\kappa\ell\ll1$ is well understood as a manifestation of the DP mechanism~\cite{Dyakonov_1972}. In this regime, the spin isotropization timescales as $\tau_{\rm iso}\propto\tau(\kappa\ell)^{-2}\propto1/\tau$, a behavior known as {\it motional narrowing}.
Furthermore, Ref.~\cite{Wenk_2010} extended this result to the case of an arbitrary mixture of Rashba and Dresselhaus spin--orbit couplings (arbitrary $\eta$), and the asymptotic form~\eqref{eq:tau_iso_small_kappal} obtained here is consistent with that result.
It is noteworthy that, in the context of spin relaxation, both the crossover with respect to $\kappa\ell$ and that with respect to $\eta$ can be described by a single cubic equation~\eqref{eq:cubic_eq_transport_mean_free_time}.

\subsubsection{Cooperon for the transient backscattering peak\label{sec:transient_peak}}

In Ref.~\cite{Kakoi_2026}, the present authors demonstrated that a transient peak emerges at a momentum offset from the backscattering direction under the condition $\kappa\ell\eta\lesssim1\ll \kappa\ell$. 
In the present work, we provide a quantitative analysis of this transient peak based on a perturbative calculation using the Cooperon, with the aim of complementing Ref.~\cite{Kakoi_2026}. 
The momentum offset is proportional to $\kappa$, and in the regime $\kappa\ell\ll1$ (i.e., when the momentum offset is nearly zero), the peak dynamics can be fully described within the scope of previous analyses, namely by Eq.~\eqref{eq:Cooperon_matrix_spin_rep} [or Eq.~\eqref{eq:Cooperon_matrix_branch_rep}]. 
In contrast, for $\kappa\ell\gtrsim 1$, the Cooperon analysis presented in Sec.~\ref{sec:Cooperon_Diffuson}, which relies on the assumption $q\ell \ll1$, is no longer applicable for the transient peak. Here, assuming a small $\eta$, we examine an approximate expression for the Cooperon that is valid for any $\kappa\ell$.
As discussed in Ref.~\cite{Kakoi_2026}, it is more convenient to use the $\ket{\rightleftarrows}$ basis defined as
\begin{equation}
    \ket{\rightarrow} = \frac{\ket{\uparrow}+\ket{\downarrow}}{\sqrt{2}},
    \hspace{8mm} \ket{\leftarrow} = \frac{\ket{\uparrow}-\ket{\downarrow}}{\sqrt{2}}.
\end{equation}
Here, we reconstruct the Diffuson and Cooperon using this basis.
The disorder-averaged Green's function is recast with the $\ket{\rightleftarrows}$ basis into
\begin{align}\label{eq:GF_minus_k_rotated_spin_basis}
     \overline{\hat{\mathcal{G}}}(\bm{k},E)
    =\, &\frac{g_+(\bm{k},E)}{2}\left[\mathbbm{1}_2 + \cos\phi_{\eta}(\theta)\hat{\sigma}_z' - \sin\phi_{\eta}(\theta)\hat{\sigma}_y'\right] \notag\\
    +\, &\frac{g_-(\bm{k},E)}{2}\left[\mathbbm{1}_2 - \cos\phi_{\eta}(\theta)\hat{\sigma}_z' + \sin\phi_{\eta}(\theta)\hat{\sigma}_y'\right].
\end{align}
$\hat{\sigma}_x'$, $\hat{\sigma}_y'$, and $\hat{\sigma}_z'$ are Pauli operators in the $\ket{\rightleftarrows}$ basis.\footnote{$\hat{\sigma}_x'=\ket{\rightarrow}\bra{\leftarrow}+\ket{\leftarrow}\bra{\rightarrow}$, \ $\hat{\sigma}_y'=-i\ket{\rightarrow}\bra{\leftarrow}+i\ket{\leftarrow}\bra{\rightarrow}$, and $\hat{\sigma}_z'=\ket{\rightarrow}\bra{\rightarrow}-\ket{\leftarrow}\bra{\leftarrow}$.}
Since the disorder potential is diagonal also in the $\ket{\rightleftarrows}$ basis, the discussion in Sec.~\ref{app:diagonal_delta-correlated} can be applied directly.
The propagator in the $\ket{\rightleftarrows}$ basis is
\begin{align}\label{eq:Pi^C_4by4_rotated_spin_basis}
    &\tilde{\Pi}^{\mathrm{C}}(\bm{q},\omega) \notag\\
    &:= \int\!\frac{d^2\bm{k}}{(2\pi)^2} 
    \,\mathcal{G}(\bm{k}_+,E_+)
    \otimes\mathcal{G}^{\dag}(-\bm{k}_-,E_-)\notag\\
    &= \frac14 
    \sum_{s,s'=\pm}\int\!\frac{d^2\bm{k}}{(2\pi)^2} 
    \,\mathcal{P}_{ss'}(\bm{k},\bm{q},E,\omega)
    \,\mathcal{F}_{ss'}^{\mathrm{C}}(\bm{k},\bm{q}),
\end{align}
where
\begin{align}
    \mathcal{F}_{ss'}^{\mathrm{C}}(\bm{k},\bm{q}) &= 
    \left[\mathbbm{1}_2 + s\cos\phi_{\eta}(\theta_+)\sigma_z - s\sin\phi_{\eta}(\theta_+)\sigma_y\right] \notag\\
    &\otimes
    \left[\mathbbm{1}_2 - s'\cos\phi_{\eta}(\theta_-)\sigma_z + s'\sin\phi_{\eta}(\theta_-)\sigma_y\right],
\end{align}
Remember that $\theta_{\pm}$ denotes the polar angle of $\bm{k}_{\pm}=\bm{k}\pm{\bm{q}}/2$.
Using $\tilde{\Pi}^{\mathrm{C}}$, the Cooperon in the $\ket{\rightleftarrows}$ basis is given as
\begin{align}
    \tilde{\Gamma}^{\mathrm{C}}(\bm{q},\omega) = \frac{\gamma_0}{1-\gamma_0\tilde{\Pi}^{\mathrm{C}}(\bm{q},\omega)} - \gamma_0.
\end{align}
Here, we assume that the off-diagonal elements of $\tilde{\Pi}^{\mathrm{C}}$ in Eq.~\eqref{eq:Pi^C_4by4_rotated_spin_basis} are negligible.
In this case, the $(SS,SS)$-component ($S\in\{\rightarrow,\leftarrow\}$) of the Cooperon is expressed by only the $(SS,SS)$-component of $\tilde{\Pi}^{\mathrm{C}}$:
\begin{align}\label{eq:assumption_diagonal_Pi}
    \tilde{\Gamma}^{\mathrm{C}}_{S}(\bm{q},\omega) = \frac{\gamma_0}{1-\gamma_0\tilde{\Pi}^{\mathrm{C}}_{S}(\bm{q},\omega)} - \gamma_0.
\end{align}
We express $\tilde{\Gamma}^{\mathrm{C}}_{SS,SS}$ and $\tilde{\Pi}^{\mathrm{C}}_{SS,SS}$ as $\tilde{\Gamma}^{\mathrm{C}}_{S}$ and $\tilde{\Pi}^{\mathrm{C}}_{S}$ for simplicity. 
Specifically, when $S=\rightarrow$,
we have 
\begin{align}
    \tilde{\Pi}^{\mathrm{C}}_{\rightarrow}&(\bm{q},\omega) \notag\\ 
    &= \int\!\frac{d^2\bm{k}}{(2\pi)^2} 
    \Bigg[ \notag\\
    &\quad \mathcal{P}_{++}(\bm{k},\bm{q},\omega,E)
    \cos^2\!\frac{\phi_{\eta}\!\left(\theta_+\right)}{2}
    \sin^2\!\frac{\phi_{\eta}\!\left(\theta_-\right)}{2} \displaybreak[1]\notag\\
    &+ \mathcal{P}_{--}(\bm{k},\bm{q},\omega,E)
    \sin^2\!\frac{\phi_{\eta}\!\left(\theta_+\right)}{2}
    \cos^2\!\frac{\phi_{\eta}\!\left(\theta_-\right)}{2} \displaybreak[1]\notag\\
    &+ \mathcal{P}_{+-}(\bm{k},\bm{q},\omega,E)
    \cos^2\!\frac{\phi_{\eta}\!\left(\theta_+\right)}{2}
    \cos^2\!\frac{\phi_{\eta}\!\left(\theta_-\right)}{2} \displaybreak[1]\notag\\
    &+ \mathcal{P}_{-+}(\bm{k},\bm{q},\omega,E)
    \sin^2\!\frac{\phi_{\eta}\!\left(\theta_+\right)}{2}
    \sin^2\!\frac{\phi_{\eta}\!\left(\theta_-\right)}{2} \Bigg].
\end{align}
$\mathcal{P}_{ss'}$ is defined in Eq.~\eqref{eq:def_P_ss'}.
The assumption~\eqref{eq:assumption_diagonal_Pi} holds exactly when $\eta=0$, where $\sin\phi_\eta(\theta) = 0$ holds. 
Even when this is not the case, since the contribution in $\tilde{\Gamma}^{\mathrm{C}}$ from the diagonal components of $\tilde{\Pi}^{\mathrm{C}}$ is of order $\mathcal{O}(\gamma_0^2)$, while that from the off-diagonal components is of order $\mathcal{O}(\gamma_0^3)$, assumption~\eqref{eq:assumption_diagonal_Pi} is still justified for weak disorder.

Here, we consider the expansion of the propagator $\tilde{\Pi}^{\mathrm{C}}_{S}$ around $\omega = 0$, as is commonly done in the analysis of weak localization~\cite{Akkermans_2007_book}. 
This is justified in the regime $t \gg \tau$.
As will be shown later in comparison with simulations, it also gives reasonably good results even for $t \sim \tau$.
At the same time, we also consider an expansion around the wave vector
\begin{equation}
    \tilde{\bm{Q}}_{S} = \underset{\bm{q}}{\rm argmax}\!\left[
    \mathrm{Re}\ \Pi^{\mathrm{C}}_{S}(\bm{q},0)\right],
\end{equation}
 at which the interference effect is maximized.
 As a result, we get
\begin{align}\label{eq:GammaC_approx_SigmaSigma}
    \gamma_0\tilde{\Pi}^{\mathrm{C}}_{S}(\bm{q},\omega) &\simeq 
    \gamma_0\tilde{\Pi}^{\mathrm{C}}_{S}(\tilde{\bm{Q}}_{S},0) + i\tilde{\tau}\omega \notag\\
    &+ \frac{1}{2}(\bm{q}\!-\!\tilde{\bm{Q}}_{S})^\top 
    \mathrm{H}^{\mathrm{C}}_{S}
    (\tilde{\bm{Q}}_{S})(\bm{q}\!-\!\tilde{\bm{Q}}_{S}),
\end{align}
where 
\begin{equation}
    \tilde{\tau} := -i\gamma_0\left.\frac{d}{d\omega}\tilde{\Pi}^{\mathrm{C}}(\tilde{\bm{Q}}_{S},\omega)\right|_{\omega=0},
\end{equation}
and $\mathrm{H}^{\mathrm{C}}(\tilde{\bm{Q}})$ denotes the Hessian matrix,
\begin{equation}
    \mathrm{H}^{\mathrm{C}}_{S}(\tilde{\bm{Q}}_{S}) := \gamma_0\nabla_{\!\bm{q}}\left[\nabla_{\!\bm{q}}\, \tilde{\Pi}^{\mathrm{C}}_{S}(\bm{q},0)\right]^\top \Big|_{\bm{q}=\tilde{\bm{Q}}_{S}}.
\end{equation}

For $\eta=0$, an exact gauge transformation exists that eliminates the vector potential from the Hamiltonian~\eqref{eq:SOC_Ham} entirely~\cite{Bernevig_2006,Kakoi_2026}. 
Consequently, the off-diagonal elements of the Green’s function~\eqref{eq:GF_minus_k_rotated_spin_basis} become zero, and the diagonal elements take~\cite{Kakoi_2026}
\begin{equation}\label{eq:GF_for_eta=0}
    \overline{\mathcal{G}}_{SS}(\bm{k},E) = \Big[E-\frac{\hbar^2(\bm{k}\!-\!\tilde{\bm{Q}}_{S})^2}{2m}+i\frac{\hbar}{2\tau}\Big]^{-1},
\end{equation}
where
\begin{equation}
    \tilde{\bm{Q}}_{S}=\left\{
    \begin{array}{ll}
        -2\kappa_x\va{e}_x=-2\kappa\cos\eta\,\va{e}_x & {\rm for}\ \,S=\rightarrow \\[1mm]
        +2\kappa_x\va{e}_x=+2\kappa\cos\eta\,\va{e}_x & {\rm for}\ \,S=\leftarrow.
    \end{array}
    \right.
\end{equation}
Since $\overline{\mathcal{G}}$ is diagonal in the $\ket{\rightleftarrows}$ basis, the propagator $\tilde{\Pi}^{\mathrm{C}}$ is also diagonal.
Moreover, Eq.~\eqref{eq:GF_for_eta=0} coincides with the disorder-averaged Green’s function for spinless particles, apart from the momentum shift $\tilde{\bm{Q}}_{S}$.
We can therefore make use of the known result for the propagator of spinless particles~\cite{Akkermans_2007_book},
\begin{equation}\label{eq:approx_Gamma_C_w_absorption_time}
    \gamma_0\tilde{\Pi}^{\mathrm{C}}_{S}(\bm{q},\omega) = 1 + i\omega\tau-\frac{\ell^2}{2}(\bm{q}\!-\!\tilde{\bm{Q}}_{S})^2.
\end{equation}
From the comparison with Eq.~\eqref{eq:GammaC_approx_SigmaSigma}, we obtain
\begin{equation}\label{eq:Pi^C_eta=0}
    \gamma_0\tilde{\Pi}^{\mathrm{C}}_{S}(\tilde{\bm{Q}}_{S},0) = 1,
\end{equation}
and
\begin{equation}\label{eq:tau_Hesse_eta=0}
    \tilde{\tau} = \tau\quad{\rm and}\quad
    \mathrm{H}^{\mathrm{C}}_{S}(\tilde{\bm{Q}}_{S}) = -\frac{\ell^2}{2}\mathbbm{1}_2.
\end{equation}
For nonzero $\eta$, Eqs.~\eqref{eq:Pi^C_eta=0} and \eqref{eq:tau_Hesse_eta=0} are no longer exact; however, when $\eta$ is small, they are still expected to provide a good approximation. 
From the correction to Eq.~\eqref{eq:Pi^C_eta=0}, we define the {\it dephasing time} $\tau_{\gamma}$ by
\begin{equation}\label{eq:dephasing_time}
    \frac{\tau}{\tau_{\gamma}} = 
    1 - \gamma_0\tilde{\Pi}^{\mathrm{C}}_{S}(\tilde{\bm{Q}}_{S},0),
\end{equation}
With this definition, the expansion of $\tilde{\Pi}^C$ becomes
\begin{equation}\label{eq:approx_Gamma_C_w_dephasing_time}
    \gamma_0\tilde{\Pi}^{\mathrm{C}}_{S}(\bm{q},\omega) \simeq 1 + i\omega\tau-\frac{\ell^2}{2}(\bm{q}\!-\!\tilde{\bm{Q}}_{S})^2 - \frac{\tau}{\tau_{\gamma}}.
\end{equation}
By substituting Eq.~\eqref{eq:approx_Gamma_C_w_dephasing_time} into Eq.~\eqref{eq:assumption_diagonal_Pi}, we find that the pole of $\tilde{\Gamma}^C_{S}$ is given by 
\begin{equation}
    \omega_{\bm{q}} = -i\left(
    \frac{1}{\tau_{\gamma}} + D(\bm{q}\!-\!\tilde{\bm{Q}}_{S})^2
    \right),
\end{equation}
where $D$ denotes the diffusion constant~\eqref{eq:diffusion_const}.
Therefore, for the initial $\ket{\bm{k}_0,S}$ state, the contribution of the Cooperon to the momentum distribution becomes~\cite{Kakoi_2026}
\begin{equation}
    n^C(\bm{k},t) = 2n_0 e^{-\omega_{\bm{k+\bm{k}_0}}t} = 2n_0 e^{-tD(\bm{k}+\bm{k}_0+\tilde{\bm{Q}}_{S})^2}e^{-t/\tau_{\gamma}},
\end{equation}
at $\bm{k}+\bm{k}_0\simeq\tilde{\bm{Q}}_{S}$ and $t\gg\tau$, which accounts for the transient peak with the momentum offset $\tilde{\bm{Q}}_{S}$ and the lifetime $\tau_{\gamma}$.

\section{Comparison with numerical simulations\label{sec:numerical_results}}

In the previous section, we analytically derived the disorder-averaged density matrix, as well as the time- and momentum-dependent dynamics of an initial plane-wave state, by combining several approximations. In this section, we confirm the validity of these approximations by comparing the analytical results with numerical simulations.

We compute, using the split-step method~\cite{Weideman_1986}, the time evolution of the momentum distribution starting from an initial plane-wave state $\ket{\bm{k}_0,+}$, i.e, we choose $s_0=+$.
We average over 4000 disorder realizations.
The computational conditions are the same as those described in Ref.~\cite{Kakoi_2026}.
We use a $\delta$-correlated disordered potential and fix the disorder strength and initial wave vector to $\gamma_0 = 2E_{\kappa}^2L_{\kappa}^2$ and $\bm{k}_0=\kappa_x(4,1)$,\footnote{We use $\kappa_x$ rather than $\kappa$ solely for numerical convenience, so that the peak maximum lies on a grid point.} respectively.
With these parameters, $\tau/\tau_{\kappa}\approx1$ and $\kappa\ell\approx10$.\footnote{We performed calculations for different values of $\eta$; however, within the scope of this work, $\kappa\ell$ is nearly independent of $\eta$.}

\begin{figure}[t]
\begin{center}
    \includegraphics[width=\columnwidth]{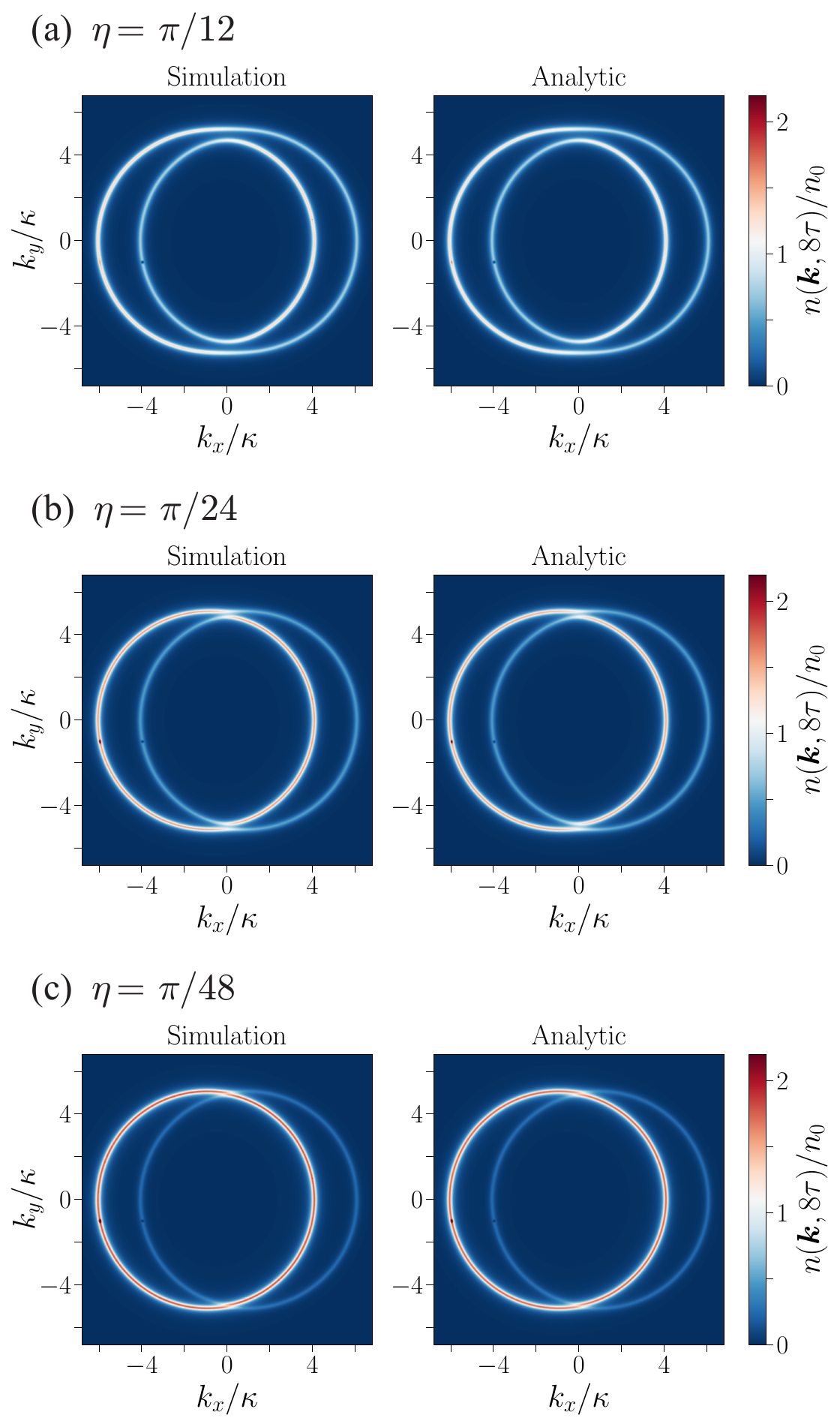}
    \caption{
    Disorder-averaged momentum distribution $n(\bm{k},t)$ in Eq.~\eqref{eq:momentum_distrib} at $t=8\tau$ (in units of $n_0$, see Eq.~\eqref{eq:def_n0}) as a function of $\bm{k}$. Panel (a): $\eta=\pi/12$. Panel (b): $\eta=\pi/24$. Panel (c): $\eta=\pi/48$.
    The left panels are obtained from numerical simulations, while the right panels are derived analytically using Eqs.~\eqref{eq:Diffuson_matrix_branch_rep}--\eqref{eq:momentum_distrib_analytic}, Eqs.~\eqref{eq:Gamma_1}--\eqref{eq:Gamma_34}, and Eqs.~\eqref{eq:assumption_diagonal_Pi} and \eqref{eq:approx_Gamma_C_w_dephasing_time} for the transient peak. As one can see, the analytics reproduces well the numerical simulations.
    } 
    \label{fig:num_vs_anal_distrib}
\end{center}
\end{figure}

In Fig.~\ref{fig:num_vs_anal_distrib}, we compare the disorder-averaged momentum distributions $n(\bm{k},t)$ obtained from numerical simulations with the analytically derived ones at a fixed time $t=8\tau$ for various values of $\eta\in\{\pi/12,\pi/24,\pi/48\}$.
At this timescale, the diffusive background is still building up, and the spin isotropization time increases as $\eta$ decreases. The spin isotropization times obtained from the cubic equation~\eqref{eq:cubic_eq_transport_mean_free_time} are $\tau_{\rm iso}=4.8\tau$, $9.1\tau$, and $20.3\tau$ for $\eta=\pi/12$, $\pi/24$, and $\pi/48$, respectively. In all cases, the analytical results are in good agreement with the numerical simulations. We emphasize that the agreement between the numerical simulations and the analytical calculations is obtained without the use of any adjustable fitting parameters.

\begin{figure*}[t]
\begin{center}
    \includegraphics[width=\linewidth]{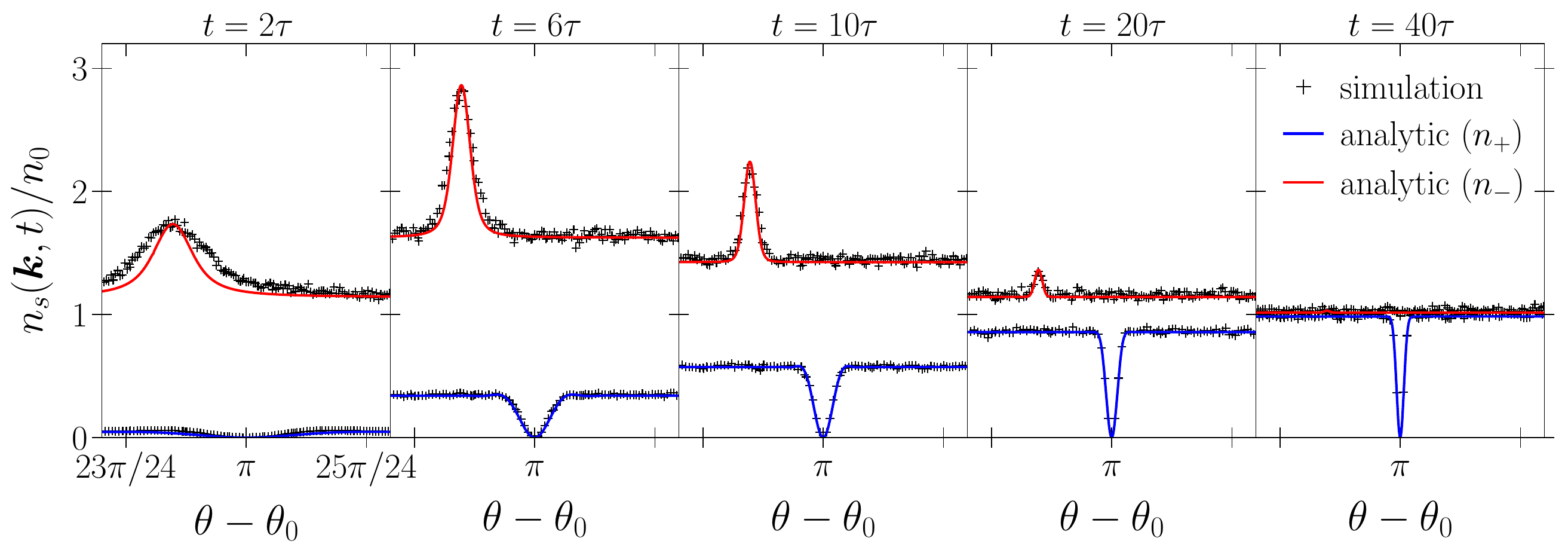}
    \caption{
    Momentum distributions obtained at different (small) times along each branch of the Fermi surface around the backscattering direction. The parameter $\eta$ is fixed at $\pi/24$. The blue and red solid lines represent the analytical results for the $(+)$-branch and $(-)$-branch respectively. They reproduce well the numerical simulation data without any adjustable parameters. The disagreement when $t\approx\tau$ likely arises because Eq.~\eqref{eq:approx_Gamma_C_w_dephasing_time} for the transient peak assumes $\omega\tau\ll1$ (corresponding to $t\gg\tau$) and $(\bm{q}-\tilde{\bm{Q}}_{S})^2\ell^2\ll1$.}
    \label{fig:full_num_vs_anal}
\end{center}
\end{figure*}

\begin{figure}[t]
\begin{center}
    \includegraphics[width=\columnwidth]{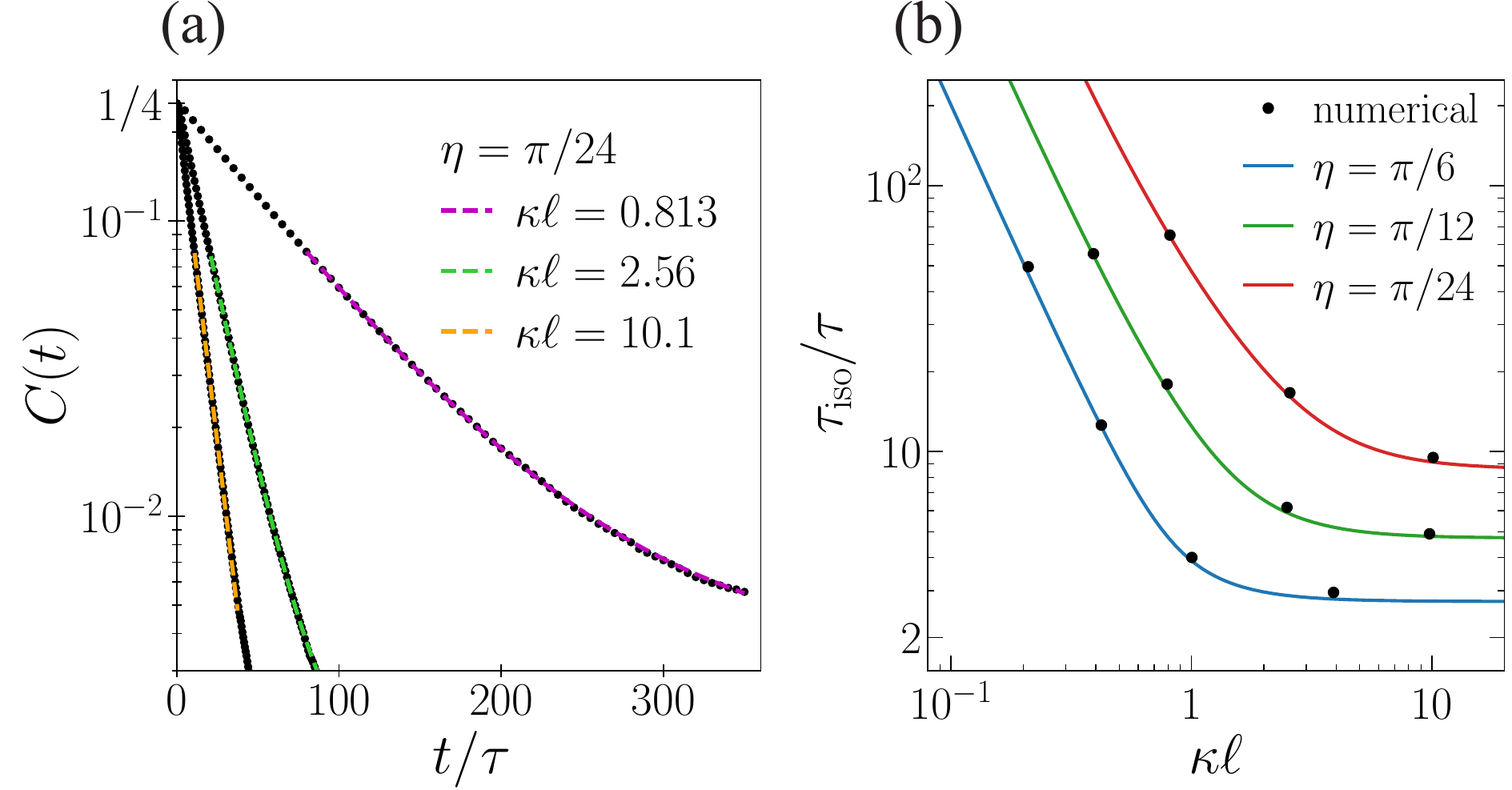}
    \caption{
    (a) Disorder-averaged spin autocorrelation function $C(t):=\overline{\braket{\bm{S}(t)\cdot\bm{S}(0)}}$ for several values of $\kappa\ell$. The spin isotropization times $\tau_{\rm iso}$ are extracted by fitting the data to Eq.~\eqref{eq:fit} using data points satisfying $C(t)<0.08$ (dashed lines). In all cases, $A\approx0.25$ and $B<0.01$. 
    We perform a goodness-of-fit test for each fit. If the $p$-value is below $0.05$, the fit is rejected, and a single-exponential fit ($B=0$) is instead performed over the range $0.005<C(t)<0.08$ (dashed yellow line).
    (b)~Comparison of $\tau_{\rm iso}$ obtained analytically from Eqs.~\eqref{eq:cubic_eq_transport_mean_free_time} and \eqref{eq:def_tau_iso} (solid lines) with numerical results for several values of $\eta$ and $\kappa\ell$.
    The uncertainty $\delta \tau_{\rm iso}$ in the numerically extracted isotropization time, estimated using the synthetic data method, satisfies $\delta \tau_{\rm iso}/\tau_{\rm iso} < 0.013$ for all data points.
    } 
    \label{fig:tau_iso_num_vs_anal}
\end{center}
\end{figure}

Next, to verify that our Cooperon analysis accurately captures the 
effect of interference on the time evolution, we examine in detail the momentum distribution around the backscattering direction. In Fig.~\ref{fig:full_num_vs_anal}, we show the short-time behavior of $n_s(\bm{k}, t)$ along each branch $s=\pm$ of the Fermi surface at $\eta=\pi/24$.
The analytical calculation accurately captures the time evolution of the diffusive background, the CBS dip, and the transient peak.
The dephasing time of the transient peak calculated by Eq.~\eqref{eq:dephasing_time} is $\tau_{\gamma}=7.6\tau$.
The disagreement when $t\approx\tau$ can be attributed to the fact that Eq.~\eqref{eq:approx_Gamma_C_w_dephasing_time} is an expansion around $\omega=0$ and $\bm{q}=\tilde{\bm{Q}}_{S}$, which assumes $t\gg\tau$ and focuses on the immediate vicinity of the peak. Nevertheless, the analytical expression still provides an accurate prediction for the peak maximum even in this short-time regime.

Finally, we compare the spin isotropization time obtained from the cubic equation~\eqref{eq:cubic_eq_transport_mean_free_time} with numerical simulations for various values of $\kappa\ell$ and $\eta$.
Starting from the initial plane-wave state $\ket{k_0\va{e}_x,+}$, we simulate its time evolution for various values of $\eta$ and $\kappa\ell$ (controlled by $\gamma_0$ and $k_0$) and calculate the spin autocorrelation function.
The results are averaged over $1000$ disorder realizations, and a $256\times256$ grid is used.
Although the initial energy $E_0$ varies with different $\eta$ and $\kappa\ell$, we note that the value of $\tau_{\rm iso}/\tau$ is insensitive to $E_0$ as long as the condition $E_0\gg E_{\ell},E_{\kappa}$ is satisfied.
In Fig.~\ref{fig:tau_iso_num_vs_anal}(a), we show the spin autocorrelation function $C(t):=\overline{\braket{\bm{S}(t)\cdot\bm{S}(0)}}$ for several values of $\kappa\ell$ at the representative value $\eta=\pi/24$.
To extract the spin isotropization time $\tau_{\rm iso}$, we analyze the decay of $C(t)$ using exponential fits.
The numerically obtained $C(t)$ cannot be adequately described by a single exponential decay and is instead approximated by the following two-component form:
\begin{equation}\label{eq:fit}
    C(t) = A e^{-t/\tau_{\rm iso}} + B e^{-t/\tau_{\rm slow}}
\end{equation}
The dominant component is associated with the spin isotropization time $\tau_{\rm iso}$.
The slower component is likely to originate from higher-order diagrams~\cite{Hikami_1981,Karpiuk_2012,Ghosh_2014}, including those responsible for coherent forward scattering (CFS), which makes the two-component form of Eq.~\eqref{eq:fit} physically plausible.
For all parameter regimes investigated, we find $A\approx0.25$ and $B<0.01$, indicating that the contribution of the slow component is small.
We perform a goodness-of-fit test for each fit. If the $p$-value is below $0.05$, a single-exponential fit ($B=0$) is instead performed\footnote{We performed single-exponential fits for two data points at $(\kappa\ell,\eta)=(10.1,\pi/24)$ and $(3.90,\pi/6)$. Note that, for the latter, the goodness-of-fit $p$-value is below $0.05$.} using the data points satisfying $0.005<C(t)<0.08$.
In Fig.~\ref{fig:tau_iso_num_vs_anal}(b), we compare the values of $\tau_{\rm iso}/\tau$ extracted from the numerical simulations with those obtained analytically from Eqs.~\eqref{eq:cubic_eq_transport_mean_free_time} and \eqref{eq:def_tau_iso}.
Numerical results show excellent agreement with analytical predictions throughout the entire parameter range considered, demonstrating that the cubic equation~\eqref{eq:cubic_eq_transport_mean_free_time} provides a unified description of the spin isotropization time for arbitrary values of $\eta$ and $\kappa\ell$.

\section{Discussion\label{sec:discussion}}

We presented a theoretical framework for describing the dynamics of spin-$1/2$ particles weakly scattered by disorder under general uniform SU(2) gauge fields. 
Using diagrammatic perturbation theory, we evaluated the Diffuson and Cooperon contributions, with careful consideration of their $\omega$-dependence, and derived the disorder-averaged density matrix as a function of time and momentum. The results obtained here are applicable not only in the $\kappa\ell \ll 1$ regime but also in the $\kappa\ell \gg 1$ regime.
By examining the pole of the Diffuson, one can obtain the characteristic timescale of spin--momentum relaxation. The spin isotropization time $\tau_{\rm iso}$ obtained from the cubic equation~\eqref{eq:cubic_eq_transport_mean_free_time} consistently interpolates from $\eta=\pi/4$ (corresponding to pure Rashba SOC) to $\eta=0$, where spin relaxation is absent.
Also, the time evolution of the Cooperon contribution can be analyzed in the same manner. 
For $\kappa\ell\eta\lesssim1\lesssim\kappa\ell$, a transient peak of constructive interference appears at a direction offset from exact backscattering. 
The dephasing time of the peak can be derived by applying an appropriate approximation to the Cooperon that takes the momentum shift into account.
We summarize the characteristic timescales in the momentum dynamics in Table~\ref{tab:timescales}.
We confirmed the validity of our calculations by comparing the analytically derived momentum dynamics with numerical simulations.

\begin{table}[b]
    \caption{
    Summary of important timescales
    \label{tab:timescales}}
    \begin{ruledtabular}
    \renewcommand{\arraystretch}{1.2}
    \begin{tabular}{cll}
    Symbol & Name & Formula\\
    \hline
    $\tau$ &  Scattering mean free time & Eq.~\eqref{eq:def_tau}\\
    $\tau_{\rm iso}$ & Spin isotropization time  & Eq.~\eqref{eq:def_tau_iso}\\
    $\tau_{\gamma}$ & Dephasing time  & Eq.~\eqref{eq:dephasing_time}
    \end{tabular}
    \end{ruledtabular}
\end{table}

The momentum distribution studied here could be directly measured by time-of-flight experiments in cold atom systems, similar to experiments in the AI symmetry class~\cite{Jendrzejewski_2012a,Labeyrie_2012}.
The Hamiltonian~\eqref{eq:SOC_Ham} can be implemented using synthetic SOC generated by a tripod scheme~\cite{Ruseckas_2005,Leroux_2018,Hasan_2022}, which in principle allows access to the full range of $\eta$ (see Ref.~\cite{Kakoi_2026} for details).
In addition, synthetic SOC has been realized in other physical platforms~\cite{Bliokh_2015_review,Y-Yang_2024_review,Ma_2016,Rechcinska_2019,Y-Chen_2019,Polimeno_2021,Y-Li_2022,J-Wu2022,Muszynski_2022}, suggesting that our results can be relevant for those systems.
Although we focused on uniform SU(2) gauge fields, the method presented here relies only on the knowledge of the dispersion relation and the wave functions, and could therefore be extended to nonuniform SU(2) gauge fields.

In cold-atom systems, disorder is often realized using optical speckle potentials.
These potentials are spatially correlated, the effects of which have been discussed, e.g., in Refs.~\cite{Kuhn_2005,Kuhn_2007}.
In this case, the momentum change imparted by a single scattering event 
is restricted to a range given by the inverse of the speckle correlation length ($1/\zeta$).
When the splitting of the Fermi surface is much larger than $1/\zeta$, transitions between the two branches are strongly suppressed and the momentum distribution approaches a steady state more slowly. 
The effect of spatial correlations can be investigated by evaluating the Bethe--Salpeter equation in its integral form, as presented in Appendix~\ref{app:ladder_and_crossed_diagrams}, instead of using Eqs.~\eqref{eq:Bethe-Salpeter_Eq_Diffuson} and \eqref{eq:Bethe-Salpeter_Eq_Cooperon}.
A detailed analysis is left for the future.

Another natural extension is to consider systems with random SOC~\cite{Arabahmadi_2024,Sherman_2003,Sherman_2005,Glazov_2005,Glazov_2010_review}. In semiconductor quantum wells, electric fields from ionized donors generate a random Rashba field, and fluctuations of the well width can also lead to random SOC~\cite{Glazov_2010_review}. Exploring such effects within the present framework is also left for future work.

\begin{acknowledgments}
    This work was supported by JSPS KAKENHI, Grant No.~JP25KJ1758 and by the French government, through the UCA$^\text{JEDI}$ ``Investissements d'Avenir" project, managed by the National Research Agency (ANR) with the Reference No.~ANR-15-IDEX-01. 
    M.K. thanks the Program for Leading Graduate Schools:``Interactive Materials Science Cadet Program" and the Research Fellowship for Young Scientists (Grant No.~JP25KJ1758) for support. M.K. also thanks UMR 7010 INPHYNI (UniCA and CNRS) for support and kind hospitality.
\end{acknowledgments}

\appendix

\section{General formalism for Diagrammatic perturbation theory\label{app:diagrammatic_calc}}
\subsection{Ladder (Diffuson) and maximally-crossed (Cooperon) diagrams\label{app:ladder_and_crossed_diagrams}}

We review the perturbation expansion for the intensity propagator
\begin{equation}
    \hat{\Phi}(E,\omega) = \overline{\hat{G}(E_+) \otimes\hat{G}^\dag(E_-)},
\end{equation}
which is valid in the weak-disorder regime~\cite{Bergmann_1984_review,Muller_2005,Kuhn_2007,Akkermans_2007_book}.
The simplest approximation is obtained by factorizing the disorder average as
\begin{equation}
    \hat{\Phi}(E,\omega) \simeq \hat{\Phi}_0(E,\omega) 
    := \overline{\hat{G}}(E_+) \otimes
    \overline{\hat{G}}{}^\dag(E_-),
\end{equation}
which corresponds to the Drude--Boltzmann approximation.
The full intensity propagator satisfies a Bethe--Salpeter equation
\begin{align}
    \hat{\Phi}(E,\omega) &= 
    \hat{\Phi}_0(E,\omega) + 
    \hat{\Phi}_0(E,\omega) \hat{\mathcal{I}}(E,\omega) \hat{\Phi}(E,\omega) \notag\\
    &= \hat{\Phi}_0(E,\omega) 
    \left[1 - \hat{\Phi}_0(E,\omega) \hat{\mathcal{I}}(E,\omega)\right]^{-1}
\end{align}
where $\hat{\mathcal{I}}$ denotes the irreducible impurity-scattering vertex, which generates multiple-scattering processes. Approximating $\hat{\mathcal{I}}$ by a single impurity line,
\begin{equation}
    \hat{\mathcal{I}} \simeq \overline{\hat{V}\otimes\hat{V}^\dag},
\end{equation}
generates the ladder-type series (Diffuson), which recovers the classical diffusion picture.
Contributions not included in this approximation give rise to quantum corrections. In the following, we focus on the quantum corrections arising from maximally-crossed diagrams (Cooperon), which correspond to the interference of wave amplitudes traveling along the same scattering paths but in opposite directions. Higher-order contributions beyond these are described by Hikami boxes~\cite{Hikami_1981}.

To simplify the formulae, we introduce the ``diagrams" defined as
\ifpdf
\begin{widetext}
\begin{align}
    \label{eq:diagram_VVD}
    \dummy^{\raisebox{0.3ex}{$\scriptstyle \ \ \ \bm{k},\alpha\!$}}_{\raisebox{-0.3ex}{$\scriptstyle \bm{k}-\bm{q},\beta\,$}}\hspace{-1.0mm}
    \VVcorrD
    \hspace{-2.2mm}\dummy^{\raisebox{0.3ex}{$\scriptstyle \bm{k}'\!,\gamma$}}_{\raisebox{-0.3ex}{$\scriptstyle \bm{k}'\!-\bm{q},\delta$}}
    &:= 
    \overline{
    \bra{\bm{k},\alpha}\hat{V}\ket{\bm{k}',\gamma}
    \bra{\bm{k}',\delta}\hat{V}^{\dag}\ket{\bm{k},\beta}
    }
    = \left[\overline{
    V(\bm{k}\!-\!\bm{k}') \otimes V^*(\bm{k}'\!-\!\bm{k})
    }\right]_{\alpha\beta,\gamma\delta},\\[3mm]
    \label{eq:diagram_VVC}
    \dummy^{\raisebox{0.3ex}{$\scriptstyle \ \ \ \bm{k},\alpha\,$}}_{\raisebox{-0.3ex}{$\scriptstyle \bm{k}'\!-\bm{q},\beta\,$}}\hspace{-1.0mm}
    \VVcorrC
    \hspace{-2.2mm}\dummy^{\raisebox{0.3ex}{$\scriptstyle \bm{k}'\!,\gamma$}}_{\raisebox{-0.3ex}{$\scriptstyle \bm{k}-\bm{q},\delta$}}
    &:= \dummy^{\raisebox{0.3ex}{$\scriptstyle \ \ \ \bm{k},\alpha\!$}}_{\raisebox{-0.3ex}{$\scriptstyle \bm{k}-\bm{q},\delta\,$}}\hspace{-1.0mm}
    \VVcorrD
    \hspace{-2.2mm}\dummy^{\raisebox{0.3ex}{$\scriptstyle \bm{k}'\!,\gamma$}}_{\raisebox{-0.3ex}{$\scriptstyle \bm{k}'\!-\bm{q},\beta$}}
    = \left[\overline{
    V(\bm{k}\!-\!\bm{k}') \otimes V(\bm{k}'\!-\!\bm{k})
    }\right]_{\alpha\beta,\gamma\delta},
\end{align}
and
\begin{align}
    \dummy^{\raisebox{0.3ex}{$\scriptstyle \alpha\,$}}_{\raisebox{-0.3ex}{$\scriptstyle \beta\,$}}\hspace{-1.0mm}
    \overset{\bm{k},E}{\underset{\bm{k}',E'}{\GGarrowD}}
    \hspace{-2.2mm}\dummy^{\raisebox{0.3ex}{$\scriptstyle \gamma$}}_{\raisebox{-0.3ex}{$\scriptstyle \delta$}}
    &:= \overline{G}_{\alpha\gamma}\!\left(\bm{k},E\right)
    \overline{G}^*_{\beta\delta}\!\left(\bm{k}',E'\right) 
    = \left[
    \overline{G}\!\left(\bm{k},E\right) \otimes
    \overline{G}^*\!\left(\bm{k}',E'\right)
    \right]_{\alpha\beta,\gamma\delta},\\[2mm]
    \dummy^{\raisebox{0.3ex}{$\scriptstyle \alpha\,$}}_{\raisebox{-0.3ex}{$\scriptstyle \beta\,$}}\hspace{-1.0mm}
    \overset{\bm{k},E}{\underset{\bm{k}',E'}{\GGarrowC}}
    \hspace{-2.2mm}\dummy^{\raisebox{0.3ex}{$\scriptstyle \gamma$}}_{\raisebox{-0.3ex}{$\scriptstyle \delta$}}
    &:= \dummy^{\raisebox{0.3ex}{$\scriptstyle \alpha\,$}}_{\raisebox{-0.3ex}{$\scriptstyle \delta\,$}}\hspace{-1.0mm}
    \overset{\bm{k},E}{\underset{\bm{k}',E'}{\GGarrowD}}
    \hspace{-2.2mm}\dummy^{\raisebox{0.3ex}{$\scriptstyle \gamma$}}_{\raisebox{-0.3ex}{$\scriptstyle \beta$}}
    = \left[
    \overline{G}\!\left(\bm{k},E\right) \otimes
    \overline{G}^\dag\!\left(\bm{k}',E'\right)
    \right]_{\alpha\beta,\gamma\delta}.
\end{align}
Equations~\eqref{eq:diagram_VVD} and \eqref{eq:diagram_VVC} are given in the same form as, e.g., in Ref.~\cite{Muller_2005}.
Applying the Diffuson and Cooperon approximation, which retains only the contributions from ladder and maximally crossed diagrams, we can approximate $\Phi$ given in Eq.~\eqref{eq:def_Phi} by
\begin{align}\label{eq:Phi_diagrams}
    \Phi(\bm{k},\bm{k}',\bm{q},E,\omega) 
    = (2\pi)^d\,\delta(\bm{k}\!-\!\bm{k}')\,\overset{\bm{k}_+,E_+}{\underset{\bm{k}_-,E_-}{\GGarrowD}}
    +\dummy^{\raisebox{0.3ex}{$\scriptstyle \bm{k}_+\,$}}_{\raisebox{-0.3ex}{$\scriptstyle \bm{k}_-\,$}}\hspace{-1.0mm}
    \overset{E_+}{\underset{E_-}{\GGarrowD \Diffuson \GGarrowD}}
    \hspace{-2.2mm}\dummy^{\raisebox{0.3ex}{$\scriptstyle \bm{k}_+'\!$}}_{\raisebox{-0.3ex}{$\scriptstyle \bm{k}_-'$}}
    + \dummy^{\raisebox{0.3ex}{$\scriptstyle \bm{k}_+\,$}}_{\raisebox{-0.3ex}{$\scriptstyle \bm{k}_-\,$}}\hspace{-1.0mm}
    \overset{E_+}{\underset{E_-}{\GGarrowD \CooperonX \GGarrowD}}
    \hspace{-2.2mm}\dummy^{\raisebox{0.3ex}{$\scriptstyle \bm{k}_+'\!$}}_{\raisebox{-0.3ex}{$\scriptstyle \bm{k}_-$}}.
\end{align}
This corresponds to Eq.~\eqref{eq:Phi_Diffuson_Cooperon_expantion} in the main text.
Here, the Diffuson and Cooperon satisfies
\begin{align}\label{eq:BS_Diffuson}
    \Gamma^{\mathrm{D}}_{\alpha\beta,\gamma\delta}(\bm{k},\bm{k}',\bm{q},E,\omega) :=
    \dummy^{\raisebox{0.3ex}{$\scriptstyle \bm{k}_+,\alpha\,$}}_{\raisebox{-0.3ex}{$\scriptstyle \bm{k}_-,\beta\,$}}\hspace{-1.0mm}
    \Diffuson
    \hspace{-2.2mm}\dummy^{\raisebox{0.3ex}{$\scriptstyle \bm{k}_+',\gamma$}}_{\raisebox{-0.3ex}{$\scriptstyle \bm{k}_-',\delta$}}
    &= \dummy^{\raisebox{0.3ex}{$\scriptstyle \bm{k}_+,\alpha\,$}}_{\raisebox{-0.3ex}{$\scriptstyle \bm{k}_-,\beta\,$}}\hspace{-1.0mm}
    \VVcorrD
    \hspace{-2.2mm}\dummy^{\raisebox{0.3ex}{$\scriptstyle \bm{k}_+',\gamma$}}_{\raisebox{-0.3ex}{$\scriptstyle \bm{k}_-',\delta$}}
    + \dummy^{\raisebox{0.3ex}{$\scriptstyle \bm{k}_+,\alpha\,$}}_{\raisebox{-0.3ex}{$\scriptstyle \bm{k}_-,\beta\,$}}\hspace{-1.0mm}
    \VVcorrD \GGarrowD \VVcorrD
    \hspace{-2.2mm}\dummy^{\raisebox{0.3ex}{$\scriptstyle \bm{k}_+',\gamma$}}_{\raisebox{-0.3ex}{$\scriptstyle \bm{k}_-',\delta$}}
    + \dummy^{\raisebox{0.3ex}{$\scriptstyle \bm{k}_+,\alpha\,$}}_{\raisebox{-0.3ex}{$\scriptstyle \bm{k}_-,\beta\,$}}\hspace{-1.0mm}
    \VVcorrD \GGarrowD \VVcorrD \GGarrowD \VVcorrD
    \hspace{-2.2mm}\dummy^{\raisebox{0.3ex}{$\scriptstyle \bm{k}_+',\gamma$}}_{\raisebox{-0.3ex}{$\scriptstyle \bm{k}_-',\delta$}}
    + \cdots \notag\\[2mm]
    &= \dummy^{\raisebox{0.3ex}{$\scriptstyle \bm{k}_+,\alpha\,$}}_{\raisebox{-0.3ex}{$\scriptstyle \bm{k}_-,\beta\,$}}\hspace{-1.0mm}
    \VVcorrD
    \hspace{-2.2mm}\dummy^{\raisebox{0.3ex}{$\scriptstyle \bm{k}_+',\gamma$}}_{\raisebox{-0.3ex}{$\scriptstyle \bm{k}_-',\delta$}}
    + \dummy^{\raisebox{0.3ex}{$\scriptstyle \bm{k}_+,\alpha\,$}}_{\raisebox{-0.3ex}{$\scriptstyle \bm{k}_-,\beta\,$}}\hspace{-1.0mm}
    \VVcorrD \GGarrowD \Diffuson
    \hspace{-2.2mm}\dummy^{\raisebox{0.3ex}{$\scriptstyle \bm{k}_+',\gamma$}}_{\raisebox{-0.3ex}{$\scriptstyle \bm{k}_-',\delta$}},
\end{align}
and
\begin{align}\label{eq:BS_Cooperon}
    \Gamma^{\mathrm{C}}_{\alpha\beta,\gamma\delta}(\bm{k},\bm{k}',\bm{q},E,\omega) :=
    \dummy^{\raisebox{0.3ex}{$\scriptstyle \bm{k}_+,\alpha\,$}}_{\raisebox{-0.3ex}{$\scriptstyle \bm{k}_-,\beta\,$}}\hspace{-1.0mm}
    \CooperonX
    \hspace{-2.2mm}\dummy^{\raisebox{0.3ex}{$\scriptstyle \bm{k}_+',\gamma$}}_{\raisebox{-0.3ex}{$\scriptstyle \bm{k}_-',\delta$}}
    &= \dummy^{\raisebox{0.3ex}{$\scriptstyle \bm{k}_+,\alpha\,$}}_{\raisebox{-0.3ex}{$\scriptstyle \bm{k}_-',\delta\,$}}\hspace{-1.0mm}
    \Cooperon
    \hspace{-2.2mm}\dummy^{\raisebox{0.3ex}{$\scriptstyle \bm{k}_+',\gamma$}}_{\raisebox{-0.3ex}{$\scriptstyle \bm{k}_-,\beta$}} \notag\\[2mm]
    &= \dummy^{\raisebox{0.3ex}{$\scriptstyle \bm{k}_+,\alpha\,$}}_{\raisebox{-0.3ex}{$\scriptstyle \bm{k}_-',\delta\,$}}\hspace{-1.0mm}
    \VVcorrC \GGarrowC \VVcorrC
    \hspace{-2.2mm}\dummy^{\raisebox{0.3ex}{$\scriptstyle \bm{k}_+',\gamma$}}_{\raisebox{-0.3ex}{$\scriptstyle \bm{k}_-,\beta$}}
    + \dummy^{\raisebox{0.3ex}{$\scriptstyle \bm{k}_+,\alpha\,$}}_{\raisebox{-0.3ex}{$\scriptstyle \bm{k}_-',\delta\,$}}\hspace{-1.0mm}
    \VVcorrC \GGarrowC \VVcorrC \GGarrowC \VVcorrC
    \hspace{-2.2mm}\dummy^{\raisebox{0.3ex}{$\scriptstyle \bm{k}_+',\gamma$}}_{\raisebox{-0.3ex}{$\scriptstyle \bm{k}_-,\beta$}}
    + \cdots \notag\\[2mm]
    &= \dummy^{\raisebox{0.3ex}{$\scriptstyle \bm{k}_+,\alpha\,$}}_{\raisebox{-0.3ex}{$\scriptstyle \bm{k}_-',\delta\,$}}\hspace{-1.0mm}
    \VVcorrC \GGarrowC \VVcorrC
    \hspace{-2.2mm}\dummy^{\raisebox{0.3ex}{$\scriptstyle \bm{k}_+',\gamma$}}_{\raisebox{-0.3ex}{$\scriptstyle \bm{k}_-,\beta$}}
    + \dummy^{\raisebox{0.3ex}{$\scriptstyle \bm{k}_+,\alpha\,$}}_{\raisebox{-0.3ex}{$\scriptstyle \bm{k}_-',\delta\,$}}\hspace{-1.0mm}
    \VVcorrC \GGarrowC \Cooperon
    \hspace{-2.2mm}\dummy^{\raisebox{0.3ex}{$\scriptstyle \bm{k}_+',\gamma$}}_{\raisebox{-0.3ex}{$\scriptstyle \bm{k}_-,\beta$}}.
\end{align}
To ease the notation, we omit the subscripts $E_+$ and $E_-$. Do note however that the top horizontal solid lines with arrows correspond to the disorder-averaged retarded Green's functions computed at energy $E_+$ while the bottom dashed lines with arrows correspond to the advanced disorder-averaged Green's function computed at energy $E_-$.
Connected diagrams are calculated as follows,
\begin{align}
    \dummy^{\raisebox{0.3ex}{$\scriptstyle \bm{k}_+,\alpha\,$}}_{\raisebox{-0.3ex}{$\scriptstyle \bm{k}_-,\beta\,$}}\hspace{-1.0mm}
    \VVcorrD \GGarrowD \Diffuson
    \hspace{-2.2mm}\dummy^{\raisebox{0.3ex}{$\scriptstyle \bm{k}_+',\gamma$}}_{\raisebox{-0.3ex}{$\scriptstyle \bm{k}_-',\delta$}}
    &= \sum_{\mu,\nu,\mu'\!,\nu'}
    \int\!\frac{d^d\bm{k}''}{(2\pi)^d}
    \dummy^{\raisebox{0.3ex}{$\scriptstyle \bm{k}_+,\alpha\,$}}_{\raisebox{-0.3ex}{$\scriptstyle \bm{k}_-,\beta\,$}}\hspace{-1.0mm}
    \VVcorrD
    \hspace{-2.2mm}\dummy^{\raisebox{0.3ex}{$\scriptstyle \bm{k}''\!+\bm{q}/2,\mu\!$}}_{\raisebox{-0.3ex}{$\scriptstyle \bm{k}''\!-\bm{q}/2,\nu$}}
    \times
    \dummy^{\raisebox{0.3ex}{$\scriptstyle \mu\,$}}_{\raisebox{-0.3ex}{$\scriptstyle \nu\,$}}\hspace{-1.0mm}
    \overset{\bm{k}''\!+\bm{q}/2}{\underset{\bm{k}''\!-\bm{q}/2}{\GGarrowD}}
    \hspace{-2.2mm}\dummy^{\raisebox{0.3ex}{$\scriptstyle \mu'$}}_{\raisebox{-0.3ex}{$\scriptstyle \nu'$}}
    \times
    \dummy^{\raisebox{0.3ex}{$\scriptstyle \bm{k}''\!+\bm{q}/2,\mu'$}}_{\raisebox{-0.3ex}{$\scriptstyle \bm{k}''\!-\bm{q}/2,\nu'$}}\hspace{-1.0mm}
    \Diffuson
    \hspace{-2.2mm}\dummy^{\raisebox{0.3ex}{$\scriptstyle \bm{k}_+',\gamma$}}_{\raisebox{-0.3ex}{$\scriptstyle \bm{k}_-',\delta$}}, \\[2mm]
    \dummy^{\raisebox{0.3ex}{$\scriptstyle \bm{k}_+,\alpha\,$}}_{\raisebox{-0.3ex}{$\scriptstyle \bm{k}_-',\delta\,$}}\hspace{-1.0mm}
    \VVcorrC \GGarrowC \Cooperon
    \hspace{-2.2mm}\dummy^{\raisebox{0.3ex}{$\scriptstyle \bm{k}_+',\gamma$}}_{\raisebox{-0.3ex}{$\scriptstyle \bm{k}_-,\beta$}}
    &= \sum_{\mu,\nu,\mu'\!,\nu'}
    \int\!\frac{d^d\bm{k}''}{(2\pi)^d}
    \dummy^{\raisebox{0.3ex}{$\scriptstyle \bm{k}_+,\alpha\,$}}_{\raisebox{-0.3ex}{$\scriptstyle \bm{k}_-',\delta\,$}}\hspace{-1.0mm}
    \VVcorrC
    \hspace{-2.2mm}\dummy^{\raisebox{0.3ex}{$\scriptstyle \bm{k}''\!+\bm{Q}/2,\mu\!$}}_{\raisebox{-0.3ex}{$\scriptstyle \bm{Q}/2-\bm{k}''\!\!,\nu$}}
    \times
    \dummy^{\raisebox{0.3ex}{$\scriptstyle \mu\,$}}_{\raisebox{-0.3ex}{$\scriptstyle \nu\,$}}\hspace{-1.0mm}
    \overset{\bm{k}''\!+\bm{Q}/2}{\underset{\bm{Q}/2-\bm{k}''}{\GGarrowC}}
    \hspace{-2.2mm}\dummy^{\raisebox{0.3ex}{$\scriptstyle \mu'$}}_{\raisebox{-0.3ex}{$\scriptstyle \nu'$}}
    \times
    \dummy^{\raisebox{0.3ex}{$\scriptstyle \bm{k}''\!+\bm{Q}/2,\mu'\,$}}_{\raisebox{-0.3ex}{$\scriptstyle \bm{Q}/2-\bm{k}''\!\!,\nu'\!$}}\hspace{-1.0mm}
    \Cooperon
    \hspace{-2.2mm}\dummy^{\raisebox{0.3ex}{$\scriptstyle \bm{k}_+',\gamma$}}_{\raisebox{-0.3ex}{$\scriptstyle \bm{k}_-,\beta$}}.
\end{align}
To simplify the expression, we define $\bm{Q} = \bm{k}+\bm{k}'$.
\end{widetext}
\else
\begin{align}
    \label{eq:diagram_VVD}
    \text{(See the PDF)}\\[3mm]
    \label{eq:diagram_VVC}
    \text{(See the PDF)}
\end{align}
and
\begin{align}
    \text{(See the PDF)}
\end{align}
Equations~\eqref{eq:diagram_VVD} and \eqref{eq:diagram_VVC} are given in the same form as, e.g., in Ref.~\cite{Muller_2005}.
Applying the Diffuson and Cooperon approximation, which retains only the contributions from ladder and maximally crossed diagrams, we can approximate $\Phi$ given in Eq.~\eqref{eq:def_Phi} by
\begin{align}\label{eq:Phi_diagrams}
    \text{(See the PDF)}
\end{align}
This corresponds to Eq.~\eqref{eq:Phi_Diffuson_Cooperon_expantion} in the main text.
Here, the Diffuson and Cooperon satisfies
\begin{align}\label{eq:BS_Diffuson}
    \text{(See the PDF)}
\end{align}
and
\begin{align}\label{eq:BS_Cooperon}
    \text{(See the PDF)}
\end{align}
To ease the notation, we omit the subscripts $E_+$ and $E_-$. Do note however that the top horizontal solid lines with arrows correspond to the disorder-averaged retarded Green's functions computed at energy $E_+$ while the bottom dashed lines with arrows correspond to the advanced disorder-averaged Green's function computed at energy $E_-$.
Connected diagrams are calculated as follows,
\begin{align}
    \text{(See the PDF)}
\end{align}
To simplify the expression, we define $\bm{Q} = \bm{k}+\bm{k}'$.
\fi

\subsection{Simplification for (pseudo)spin-independent and spatially $\delta$-correlated disorder\label{app:diagonal_delta-correlated}}

When the disordered potential is diagonal in the internal degrees of freedom such as spin and is spatially $\delta$-correlated,
\begin{equation}
    \overline{V_{\alpha\gamma}(\bm{r})V_{\beta\delta}(\bm{r')}} = \gamma_0\,\delta(\bm{r}\!-\!\bm{r}')\, \delta_{\alpha\gamma}\delta_{\beta\delta},
\end{equation}
In this case, Eqs.~\eqref{eq:diagram_VVD} and \eqref{eq:diagram_VVC} are independent of the wave vector and diagonal with respect to the (pseudo)spin,
\ifpdf
\begin{equation}
    \VVcorrC = \VVcorrD
    = \gamma_0 \,\delta_{\alpha\gamma} \delta_{\beta\delta} = \gamma_0\, \mathbbm{1}.
\end{equation}
Now, Eqs.~\eqref{eq:BS_Diffuson} and \eqref{eq:BS_Cooperon} can be expressed as infinite geometric series of matrix products, and the Diffuson and Cooperon are finally obtained as
\begin{align}
    \label{eq:BS_simplified_Diffuson}
    \Diffuson &= \gamma_0
    \left(\mathbbm{1} - \gamma_0\int\!\frac{d^d\bm{k}}{(2\pi)^d}
    \overset{\bm{k}+\bm{q}/2,E_+}{\underset{\bm{k}-\bm{q}/2,E_-}{\GGarrowD}}
    \right)^{\!-1},\displaybreak[1]\\[2mm]
    \label{eq:BS_simplified_Cooperon}
    \Cooperon &= \gamma_0
    \left(\mathbbm{1} - \gamma_0\int\!\frac{d^d\bm{k}}{(2\pi)^d}
    \overset{\bm{k}+\bm{Q}/2,E_+}{\underset{\bm{Q}/2-\bm{k},E_-}{\GGarrowC}}
    \right)^{\!-1} \hspace{-2mm}- \gamma_0\mathbbm{1},
\end{align}
respectively.
Equations~\eqref{eq:BS_simplified_Diffuson} and \eqref{eq:BS_simplified_Cooperon} are the same as Eqs.~\eqref{eq:Bethe-Salpeter_Eq_Diffuson} and \eqref{eq:Bethe-Salpeter_Eq_Cooperon} in the main text.
\else
\begin{equation}
    \text{(See the PDF)}
\end{equation}
Now, Eqs.~\eqref{eq:BS_Diffuson} and \eqref{eq:BS_Cooperon} can be expressed as infinite geometric series of matrix products, and the Diffuson and Cooperon are finally obtained as
\begin{align}
    \label{eq:BS_simplified_Diffuson}
    \text{(See the PDF)}\\
    \label{eq:BS_simplified_Cooperon}
    \text{(See the PDF)}
\end{align}
respectively.
Equations~\eqref{eq:BS_simplified_Diffuson} and \eqref{eq:BS_simplified_Cooperon} are the same as Eqs.~\eqref{eq:Bethe-Salpeter_Eq_Diffuson} and \eqref{eq:Bethe-Salpeter_Eq_Cooperon} in the main text.
\fi

\section{Details for approximations~\eqref{eq:approx_A} and \eqref{eq:approx_B}\label{app:approx_propagator}}

Here, we give some details about approximations~\eqref{eq:approx_A} and \eqref{eq:approx_B} introduced in Sec.~\ref{sec:approx_Pi}. We start by the following relations:
\begin{align}\label{eq:approx_gRgA} 
    &\mathcal{P}_{ss'}(\bm{k},\bm{q},\omega,E) + \mathcal{P}_{s's}(-\bm{k},\bm{q},\omega,E)\notag\\
    &= -i\frac{\tau}{\hbar}\left\{\frac{g_{s}\left(\bm{k}_+,E_+\right)-g_{s'}^*\left(\bm{k}_-,E_-\right)}
    {x_{\omega} +\frac{i\tau}{\hbar} [\mathcal{E}_{s}(\bm{k}_+) - \mathcal{E}_{s'}(\bm{k}_-)]} \right. \notag\\
    &\left.\qquad+ \frac{g_{s'}\left(\bm{k}_-,E_+\right)-g_{s}^*\left(\bm{k}_+,E_-\right)}
    {x_{\omega} +\frac{i\tau}{\hbar} [\mathcal{E}_{s'}(\bm{k}_-) - \mathcal{E}_{s}(\bm{k}_+)]} \right\}
    \displaybreak[1]\notag\\
    &\simeq -\frac{2\pi\tau}{\hbar}\frac{x_{\omega}[\mathcal{A}_{s}\left(\bm{k}_+,E\right)+\mathcal{A}_{s'}\left(\bm{k}_-,E\right)]}
    {x_{\omega}^2 +\frac{\tau^2}{\hbar^2} [\mathcal{E}_{s}(\bm{k}_+) - \mathcal{E}_{s'}(\bm{k}_-)]^2}.
\end{align}
where $\mathcal{A}_s(\bm{k},E) = -(1/\pi)\,\mathrm{Im}\, g_s(\bm{k},E)$ is the $s$ component of the spectral function.
Using Eq.~\eqref{eq:approx_gRgA}, $A_l$ and $B_l$ given in Eqs.~\eqref{eq:def_Al} and \eqref{eq:def_Bl} are approximated as
\begin{align}
    \label{eq:approx_Al_k_integral}
    &A_l(\bm{q},\omega) \notag\\
    &= \int\!\frac{d^2\bm{k}}{(2\pi)^2}
    \chi_l(\theta,\eta)\sum_{s}\frac{\mathcal{P}_{ss}(\bm{k},\bm{q},\omega,E) + \mathcal{P}_{ss}(-\bm{k},\bm{q},\omega,E)}{2}\displaybreak[1]\notag\\
    &\simeq -\frac{\pi\tau}{\hbar}\int\!\frac{d^2\bm{k}}{(2\pi)^2}
    \chi_l(\theta,\eta) \sum_{s} \frac{x_{\omega}[\mathcal{A}_{s}\left(\bm{k}_+,E\right)+\mathcal{A}_{s}\left(\bm{k}_-,E\right)]}
    {x_{\omega}^2 +\frac{\tau^2}{\hbar^2} [\mathcal{E}_{s}(\bm{k}_+) - \mathcal{E}_{s}(\bm{k}_-)]^2}\displaybreak[1]\notag\\
    &\simeq -\frac{2\pi\tau}{\hbar}\int\!\frac{d^2\bm{k}}{(2\pi)^2}
    \chi_l(\theta,\eta) \frac{x_{\omega}\mathcal{A}\left(\bm{k},E\right)}
    {x_{\omega}^2 + (\tau\bm{v}\cdot\bm{q})^2},\displaybreak[1]\\[2mm]
    \label{eq:approx_Bl_k_integral}
    &B_l(\bm{q},\omega) \notag\\
    &= \int\!\frac{d^2\bm{k}}{(2\pi)^2}
    \chi_l(\theta,\eta)\sum_{s}\frac{\mathcal{P}_{s\bar{s}}(\bm{k},\bm{q},\omega,E) + \mathcal{P}_{\bar{s}s}(-\bm{k},\bm{q},\omega,E)}{2}\displaybreak[1]\notag\\
    &\simeq -\frac{\tau}{\hbar}\int\!\frac{d^2\bm{k}}{(2\pi)^2}
    \chi_l(\theta,\eta)\sum_{s} \frac{x_{\omega}[\mathcal{A}_{s}\left(\bm{k}_+,E\right)+\mathcal{A}_{\bar{s}}\left(\bm{k}_-,E\right)]}
    {x_{\omega}^2 +\frac{\tau^2}{\hbar^2} [\mathcal{E}_{s}(\bm{k}_+) - \mathcal{E}_{\bar{s}}(\bm{k}_-)]^2}\displaybreak[1]\notag\\
    &\simeq -\frac{2\pi\tau}{\hbar}\int\!\frac{d^2\bm{k}}{(2\pi)^2}
    \chi_l(\theta,\eta) \frac{x_{\omega}\mathcal{A}\left(\bm{k},E\right)}
    {x_{\omega}^2 + [\Delta(\bm{k})+\tau\bm{v}\cdot\bm{q}]^2},
\end{align}
where $\mathcal{A}(\bm{k},E) = \mathcal{A}_+(\bm{k},E)+\mathcal{A}_-(\bm{k},E)$, where $\bm{v}$ denotes the group velocity, and where
\begin{equation}
    \Delta(\bm{k}) = \frac{\tau}{\hbar}[\mathcal{E}_{+}(\bm{k}) - \mathcal{E}_{-}(\bm{k})] = \frac{2\tau\hbar\kappa k}{m}f(\theta,\eta).
\end{equation}
Under the weak disorder condition $k_\mathrm{F}\ell\gg1$, the spectral weight is concentrated on the Fermi surface.
Assuming the spectral function to be
\begin{equation}
    \mathcal{A}(\bm{k},E) = \frac{2\pi}{k_\mathrm{F}} \nu(E_0)\,\delta(k-k_\mathrm{F}),
\end{equation}
the integral in Eqs.~\eqref{eq:approx_Al_k_integral} and \eqref{eq:approx_Bl_k_integral} can be replaced by an angular integral on the Fermi surface (on-shell approximation):
\begin{align}
    \label{eq:approx_Al_theta_integral}
    A_l(\bm{q},\omega) &= -\frac{2}{\gamma_0}\int_0^{2\pi}\!\frac{d\theta}{2\pi}
    \frac{\chi_l(\theta,\eta)\, x_{\omega}}
    {x_{\omega}^2 +(\tau\bm{v}_\mathrm{F}\cdot\bm{q})^2},\\
    \label{eq:approx_Bl_theta_integral}
    B_l(\bm{q},\omega) &= -\frac{2}{\gamma_0}\int_0^{2\pi}\!\frac{d\theta}{2\pi}
    \frac{\chi_l(\theta,\eta)\, x_{\omega}}
    {x_{\omega}^2 + [2\kappa\ell f(\theta,\eta)+\tau\bm{v}_\mathrm{F}\cdot\bm{q}]^2}.
\end{align}
Here, the relation $\tau = \hbar/\pi\nu\gamma_0$ [Eq.~\eqref{eq:def_tau}] is used.
Strictly speaking, the Fermi wave number $k_\mathrm{F}$ and the Fermi velocity $v_\mathrm{F}$ depends on $\theta$, but under condition $k_\mathrm{F}/\kappa\gg1$, this dependence is negligible.
With this condition, $\hbar k_\mathrm{F}/m\approx v_\mathrm{F}$.
Assuming that the $A_l$ and $B_l$ follow the functional forms
\begin{align}
    \gamma_0A_l(\bm{q},\omega) = \frac{c_1 x_{\omega}}{x_{\omega}^2 +c_2},\quad
    \gamma_0B_l(\bm{q},\omega) = \frac{c_1 x_{\omega}}{x_{\omega}^2 +c_3},
\end{align}
and determining $c_1$, $c_2$, and $c_3$ so that the expansion of Eqs.~\eqref{eq:approx_Al_theta_integral} and \eqref{eq:approx_Bl_theta_integral} around $q=0$ at $\omega=0$ ($x_{\omega}=-1$) remains consistent, we obtain Eqs.~\eqref{eq:approx_A} and \eqref{eq:approx_B}.
Here, we used
\begin{equation}
    \int_0^{2\pi}\!\frac{d\theta}{2\pi} (\tau\bm{v}_\mathrm{F}\cdot\bm{q})^2 = \frac{(\tau v_{\mathrm{F}} q)^2}{2} = \frac{(q\ell)^2}{2},
\end{equation}
and the term linear in $q$ in $B_l$ is negligibly small since it cancels upon integration.

\bibliography{refs}

@book{Akkermans_2007_book,
   author = {Akkermans, E. and Montambaux, G.},
   title = {{Mesoscopic Physics of Electrons and Photons}},
   publisher = {Cambridge University Press},
   address = {London},
   year = {2007}
}

@article{Y-Ando_2013_review,
author = {Ando ,Yoichi},
title = {{Topological Insulator Materials}},
journal = {J. Phys. Soc. Jpn.},
volume = {82},
number = {10},
pages = {102001},
year = {2013},
doi = {10.7566/JPSJ.82.102001},
URL = {https://doi.org/10.7566/JPSJ.82.102001},
}

@article{Altmann_2016,
  title = {{Current-Controlled Spin Precession of Quasistationary Electrons in a Cubic Spin-Orbit Field}},
  author = {Altmann, P. and Hernandez, F. G. G. and Ferreira, G. J. and Kohda, M. and Reichl, C. and Wegscheider, W. and Salis, G.},
  journal = {Phys. Rev. Lett.},
  volume = {116},
  issue = {19},
  pages = {196802},
  numpages = {5},
  year = {2016},
  month = {May},
  publisher = {American Physical Society},
  doi = {10.1103/PhysRevLett.116.196802},
  url = {https://link.aps.org/doi/10.1103/PhysRevLett.116.196802}
}

@article{Cherroret_2021_review,
   author = {Cherroret, N. and Scoquart, T. and Delande, D.},
   title = {{Coherent multiple scattering of out-of-equilibrium interacting Bose gases}},
   journal = {Ann. Phys.},
   volume = {435},
   pages = {168543},
   ISSN = {0003-4916},
   DOI = {10.1016/j.aop.2021.168543},
   url = {<Go to ISI>://WOS:000735455600027},
   year = {2021},
   type = {Journal Article}
}

@misc{Dornelas_arxiv,
      title={{Jittery Quantum Boomerang Effect}}, 
      author={Pedro Dornelas and Gerson J. Ferreira},
      eprint={2606.10067},
      archivePrefix={arXiv},
}

@incollection{Dyakonov_2008_book,
  author    = {Dyakonov, M. I.},
  title     = {{Basics of Semiconductor and Spin Physics}},
  booktitle = {{Spin Physics in Semiconductors}},
  editor    = {Dyakonov, M. I.},
  publisher = {Springer},
  address   = {Berlin},
  year      = {2008}
}

@article{Bergmann_1984_review,
   author = {Bergmann, G.},
   title = {{Weak localization in thin films: a time-of-flight experiment with conduction electrons}},
   journal = {Phys. Rep.},
   volume = {107},
   number = {1},
   pages = {1-58},
   ISSN = {0370-1573},
   DOI = {10.1016/0370-1573(84)90103-0},
   url = {<Go to ISI>://WOS:A1984TA69300001},
   year = {1984},
   type = {Journal Article}
}

@article{Glazov_2010_review,
title = {{Two-dimensional electron gas with spin--orbit coupling disorder}},
journal = {Physica E},
volume = {42},
number = {9},
pages = {2157-2177},
year = {2010},
issn = {1386-9477},
doi = {https://doi.org/10.1016/j.physe.2010.04.021},
url = {https://www.sciencedirect.com/science/article/pii/S1386947710002213},
author = {M.M. Glazov and E.Ya. Sherman and V.K. Dugaev},
}

@article{Schliemann_2017_review,
  title = {{Colloquium: Persistent spin textures in semiconductor nanostructures}},
  author = {Schliemann, John},
  journal = {Rev. Mod. Phys.},
  volume = {89},
  issue = {1},
  pages = {011001},
  numpages = {17},
  year = {2017},
  month = {Jan},
  publisher = {American Physical Society},
  doi = {10.1103/RevModPhys.89.011001},
  url = {https://link.aps.org/doi/10.1103/RevModPhys.89.011001}
}

@article{Aleiner_2001,
  title = {{Spin-Orbit Coupling Effects on Quantum Transport in Lateral Semiconductor Dots}},
  author = {Aleiner, I. L. and Fal'ko, Vladimir I.},
  journal = {Phys. Rev. Lett.},
  volume = {87},
  issue = {25},
  pages = {256801},
  numpages = {4},
  year = {2001},
  month = {Nov},
  publisher = {American Physical Society},
  doi = {10.1103/PhysRevLett.87.256801},
  url = {https://link.aps.org/doi/10.1103/PhysRevLett.87.256801}
}

@article{Anderson_1958,
  title = {{Absence of Diffusion in Certain Random Lattices}},
  author = {Anderson, P. W.},
  journal = {Phys. Rev.},
  volume = {109},
  issue = {5},
  pages = {1492--1505},
  numpages = {0},
  year = {1958},
  month = {Mar},
  publisher = {American Physical Society},
  doi = {10.1103/PhysRev.109.1492},
  url = {https://link.aps.org/doi/10.1103/PhysRev.109.1492}
}

@article{Ando_1998a,
author = {Ando ,Tsuneya and Nakanishi ,Takeshi},
title = {{Impurity Scattering in Carbon Nanotubes    – Absence of Back Scattering –}},
journal = {J. Phys. Soc. Jpn.},
volume = {67},
number = {5},
pages = {1704-1713},
year = {1998},
doi = {10.1143/JPSJ.67.1704},
URL = {https://doi.org/10.1143/JPSJ.67.1704},
}

@article{Ando_1998b,
author = {Ando ,Tsuneya and Nakanishi ,Takeshi and Saito ,Riichiro},
title = {{Berry's Phase and Absence of Back Scattering in Carbon Nanotubes}},
journal = {J. Phys. Soc. Jpn.},
volume = {67},
number = {8},
pages = {2857-2862},
year = {1998},
doi = {10.1143/JPSJ.67.2857},
URL = {https://doi.org/10.1143/JPSJ.67.2857},
}

@article{Arabahmadi_2024,
  title = {{Momentum-space signatures of the Anderson transition in a symplectic, two-dimensional, disordered ultracold gas}},
  author = {Arabahmadi, Ehsan and Schumayer, Daniel and Gr\'emaud, Beno\^{\i}t and Miniatura, Christian and Hutchinson, David A. W.},
  journal = {Phys. Rev. Res.},
  volume = {6},
  issue = {1},
  pages = {L012021},
  numpages = {6},
  year = {2024},
  month = {Jan},
  publisher = {American Physical Society},
  doi = {10.1103/PhysRevResearch.6.L012021},
  url = {https://link.aps.org/doi/10.1103/PhysRevResearch.6.L012021}
}

@Article{Arrouas_2026,
author={Arrouas, F. and H{\'e}braud, J. and Ombredane, N. and Flament, E. and Ronco, D. and Dupont, N. and Lemari{\'e}, G. and Georgeot, B. and Miniatura, Ch. and Billy, J. and Peaudecerf, B. and Gu{\'e}ry-Odelin, D.},
title={Probing non-ergodicity and symmetry via direct measurement of coherent scattering in a shaken rotor},
journal={Nat. Commun.},
year={2026},
volume = {},
issue = {},
pages = {},
doi={10.1038/s41467-026-74302-7},
url={https://doi.org/10.1038/s41467-026-74302-7}
}

@article{Bayer_1993,
  title = {{Weak localization of acoustic waves in strongly scattering media}},
  author = {Bayer, G. and Niederdr\"ank, T.},
  journal = {Phys. Rev. Lett.},
  volume = {70},
  issue = {25},
  pages = {3884--3887},
  numpages = {0},
  year = {1993},
  month = {Jun},
  publisher = {American Physical Society},
  doi = {10.1103/PhysRevLett.70.3884},
  url = {https://link.aps.org/doi/10.1103/PhysRevLett.70.3884}
}

@article{Bergman_1982,
  title = {{Influence of Spin-Orbit Coupling on Weak Localization}},
  author = {Bergman, Gerd},
  journal = {Phys. Rev. Lett.},
  volume = {48},
  issue = {15},
  pages = {1046--1049},
  numpages = {0},
  year = {1982},
  month = {Apr},
  publisher = {American Physical Society},
  doi = {10.1103/PhysRevLett.48.1046},
  url = {https://link.aps.org/doi/10.1103/PhysRevLett.48.1046}
}

@article{Bernevig_2006,
  title = {{Exact SU(2) Symmetry and Persistent Spin Helix in a Spin-Orbit Coupled System}},
  author = {Bernevig, B. Andrei and Orenstein, J. and Zhang, Shou-Cheng},
  journal = {Phys. Rev. Lett.},
  volume = {97},
  issue = {23},
  pages = {236601},
  numpages = {4},
  year = {2006},
  month = {Dec},
  publisher = {American Physical Society},
  doi = {10.1103/PhysRevLett.97.236601},
  url = {https://link.aps.org/doi/10.1103/PhysRevLett.97.236601}
}

@Article{Boross2013,
author={Boross, P{\'e}ter and D{\'o}ra, Bal{\'a}zs and Kiss, Annam{\'a}ria and Simon, Ferenc},
title={{A unified theory of spin-relaxation due to spin-orbit coupling in metals and semiconductors}},
journal={Sci. Rep.},
year={2013},
month={Nov},
day={20},
volume={3},
number={1},
pages={3233},
issn={2045-2322},
doi={10.1038/srep03233},
url={https://doi.org/10.1038/srep03233}
}

@article{Burkov_2004,
  title = {{Spin relaxation in a two-dimensional electron gas in a perpendicular magnetic field}},
  author = {Burkov, A. A. and Balents, Leon},
  journal = {Phys. Rev. B},
  volume = {69},
  issue = {24},
  pages = {245312},
  numpages = {6},
  year = {2004},
  month = {Jun},
  publisher = {American Physical Society},
  doi = {10.1103/PhysRevB.69.245312},
  url = {https://link.aps.org/doi/10.1103/PhysRevB.69.245312}
}

@article{Bychkov_1984,
doi = {10.1088/0022-3719/17/33/015},
url = {https://dx.doi.org/10.1088/0022-3719/17/33/015},
year = {1984},
month = {nov},
publisher = {},
volume = {17},
number = {33},
pages = {6039},
author = {Yu A Bychkov and E I Rashba},
title = {{Oscillatory effects and the magnetic susceptibility of carriers in inversion layers}},
journal = {J. Phys. C},
}

@article{Campbell_2011,
  title = {{Realistic Rashba and Dresselhaus spin-orbit coupling for neutral atoms}},
  author = {Campbell, D. L. and Juzeli\ifmmode \bar{u}\else \={u}\fi{}nas, G. and Spielman, I. B.},
  journal = {Phys. Rev. A},
  volume = {84},
  issue = {2},
  pages = {025602},
  numpages = {4},
  year = {2011},
  month = {Aug},
  publisher = {American Physical Society},
  doi = {10.1103/PhysRevA.84.025602},
  url = {https://link.aps.org/doi/10.1103/PhysRevA.84.025602}
}

@article{Cherroret_2012,
  title = {{Coherent backscattering of ultracold matter waves: Momentum space signatures}},
  author = {Cherroret, Nicolas and Karpiuk, Tomasz and M\"uller, Cord A. and Gr\'emaud, Beno\^{\i}t and Miniatura, Christian},
  journal = {Phys. Rev. A},
  volume = {85},
  issue = {1},
  pages = {011604},
  numpages = {4},
  year = {2012},
  month = {Jan},
  publisher = {American Physical Society},
  doi = {10.1103/PhysRevA.85.011604},
  url = {https://link.aps.org/doi/10.1103/PhysRevA.85.011604}
}

@article{Dolan_1979,
  title = {{Nonmetallic Conduction in Thin Metal Films at Low Temperatures}},
  author = {Dolan, G. J. and Osheroff, D. D.},
  journal = {Phys. Rev. Lett.},
  volume = {43},
  issue = {10},
  pages = {721--724},
  numpages = {0},
  year = {1979},
  month = {Sep},
  publisher = {American Physical Society},
  doi = {10.1103/PhysRevLett.43.721},
  url = {https://link.aps.org/doi/10.1103/PhysRevLett.43.721}
}

@article{Dresselhaus_1955,
  title = {{Spin-Orbit Coupling Effects in Zinc Blende Structures}},
  author = {Dresselhaus, G.},
  journal = {Phys. Rev.},
  volume = {100},
  issue = {2},
  pages = {580--586},
  numpages = {0},
  year = {1955},
  month = {Oct},
  publisher = {American Physical Society},
  doi = {10.1103/PhysRev.100.580},
  url = {https://link.aps.org/doi/10.1103/PhysRev.100.580}
}

@article{Dyakonov_1972,
   author = {D'yakonov, M. I. and Perel', V. I.},
   title = {{Spin relaxation of conduction electrons in noncentrosymmetric semiconductors}},
   journal = {Fiz. Tverd. Tela (Leningrad)},
   volume = {13},
   pages = {3581},
   year = {1971},
   note={[Sov. Phys. Solid State \textbf{13}, 3023 (1972)]},
   type = {Journal Article}
}

@article{Ghosh_2014,
  title = {{Coherent forward scattering in two-dimensional disordered systems}},
  author = {Ghosh, S. and Cherroret, N. and Gr\'emaud, B. and Miniatura, C. and Delande, D.},
  journal = {Phys. Rev. A},
  volume = {90},
  issue = {6},
  pages = {063602},
  numpages = {12},
  year = {2014},
  month = {Dec},
  publisher = {American Physical Society},
  doi = {10.1103/PhysRevA.90.063602},
  url = {https://link.aps.org/doi/10.1103/PhysRevA.90.063602}
}

@article{Ghosh_2015,
  title = {{Coherent Backscattering Reveals the Anderson Transition}},
  author = {Ghosh, S. and Delande, D. and Miniatura, C. and Cherroret, N.},
  journal = {Phys. Rev. Lett.},
  volume = {115},
  issue = {20},
  pages = {200602},
  numpages = {5},
  year = {2015},
  month = {Nov},
  publisher = {American Physical Society},
  doi = {10.1103/PhysRevLett.115.200602},
  url = {https://link.aps.org/doi/10.1103/PhysRevLett.115.200602}
}

@article{Ghosh_2017,
  title = {{Coherent forward scattering as a signature of Anderson metal-insulator transitions}},
  author = {Ghosh, Sanjib and Miniatura, Christian and Cherroret, Nicolas and Delande, Dominique},
  journal = {Phys. Rev. A},
  volume = {95},
  issue = {4},
  pages = {041602},
  numpages = {5},
  year = {2017},
  month = {Apr},
  publisher = {American Physical Society},
  doi = {10.1103/PhysRevA.95.041602},
  url = {https://link.aps.org/doi/10.1103/PhysRevA.95.041602}
}

@article{Glazov_2005,
  title = {Nonexponential spin relaxation in magnetic fields in quantum wells with random spin-orbit coupling},
  author = {Glazov, M. M. and Sherman, E. Ya.},
  journal = {Phys. Rev. B},
  volume = {71},
  issue = {24},
  pages = {241312},
  numpages = {4},
  year = {2005},
  month = {Jun},
  publisher = {American Physical Society},
  doi = {10.1103/PhysRevB.71.241312},
  url = {https://link.aps.org/doi/10.1103/PhysRevB.71.241312}
}

@Article{Glazov_2006,
author={Glazov, M. M.
and Golub, L. E.},
title={Nondiffusive weak localization in two-dimensional systems with spin-orbit splitting of the spectrum},
journal={Semiconductors},
year={2006},
month={Oct},
day={01},
volume={40},
number={10},
pages={1209-1217},
issn={1090-6479},
doi={10.1134/S1063782606100150},
url={https://doi.org/10.1134/S1063782606100150}
}

@article{Glazov_2009,
doi = {10.1088/0268-1242/24/6/064007},
url = {https://doi.org/10.1088/0268-1242/24/6/064007},
year = {2009},
month = {may},
publisher = {},
volume = {24},
number = {6},
pages = {064007},
author = {Glazov, M M and Golub, L E},
title = {{Spin--orbit interaction and weak localization in heterostructures}},
journal = {Semicond. Sci. Technol.},
}

@article{Golub_2005,
  title = {{Weak antilocalization in high-mobility two-dimensional systems}},
  author = {Golub, L. E.},
  journal = {Phys. Rev. B},
  volume = {71},
  issue = {23},
  pages = {235310},
  numpages = {6},
  year = {2005},
  month = {Jun},
  publisher = {American Physical Society},
  doi = {10.1103/PhysRevB.71.235310},
  url = {https://link.aps.org/doi/10.1103/PhysRevB.71.235310}
}

@article{Hainaut_2017,
  title = {{Return to the Origin as a Probe of Atomic Phase Coherence}},
  author = {Hainaut, Cl\'ement and Manai, Isam and Chicireanu, Radu and Cl\'ement, Jean-Fran\c{c}ois and Zemmouri, Samir and Garreau, Jean Claude and Szriftgiser, Pascal and Lemari\'e, Gabriel and Cherroret, Nicolas and Delande, Dominique},
  journal = {Phys. Rev. Lett.},
  volume = {118},
  issue = {18},
  pages = {184101},
  numpages = {6},
  year = {2017},
  month = {May},
  publisher = {American Physical Society},
  doi = {10.1103/PhysRevLett.118.184101},
  url = {https://link.aps.org/doi/10.1103/PhysRevLett.118.184101}
}

@article{Hainaut_2018,
   author = {Hainaut, C. and Manai, I. and Cl\'{e}ment, J.-F. and Garreau, J. C. and Szriftgiser, P. and Lemari\'{e}, G. and Cherroret, N. and Delande, D. and Chicireanu, R.},
   title = {{Controlling symmetry and localization with an artificial gauge field in a disordered quantum system}},
   journal = {Nat. Commun.},
   volume = {9},
   pages = {1382},
   ISSN = {2041-1723},
   DOI = {10.1038/s41467-018-03481-9},
   url = {<Go to ISI>://WOS:000429689800005},
   year = {2018},
   type = {Journal Article}
}

@article{Hijano_2024,
  title = {{Weak localization at arbitrary disorder strength in systems with generic spin-dependent fields}},
  author = {Hijano, Alberto and Ili\ifmmode \acute{c}\else \'{c}\fi{}, Stefan and Bergeret, F. Sebasti\'an},
  journal = {Phys. Rev. Res.},
  volume = {6},
  issue = {2},
  pages = {023100},
  numpages = {12},
  year = {2024},
  month = {Apr},
  publisher = {American Physical Society},
  doi = {10.1103/PhysRevResearch.6.023100},
  url = {https://link.aps.org/doi/10.1103/PhysRevResearch.6.023100}
}

@article{Hikami_1980,
   author = {Hikami, S. and Larkin, A. I. and Nagaoka, Y.},
   title = {{Spin-Orbit Interaction and Magnetoresistance in the Two Dimensional Random System}},
   journal = {Prog. Theor. Phys.},
   volume = {63},
   number = {2},
   pages = {707-710},
   ISSN = {0033-068X},
   DOI = {10.1143/ptp.63.707},
   url = {<Go to ISI>://WOS:A1980JJ19900030},
   year = {1980},
   type = {Journal Article}
}

@article{Hikami_1981,
  title = {{Anderson localization in a nonlinear-$\ensuremath{\sigma}$-model representation}},
  author = {Hikami, Shinobu},
  journal = {Phys. Rev. B},
  volume = {24},
  issue = {5},
  pages = {2671--2679},
  numpages = {0},
  year = {1981},
  month = {Sep},
  publisher = {American Physical Society},
  doi = {10.1103/PhysRevB.24.2671},
  url = {https://link.aps.org/doi/10.1103/PhysRevB.24.2671}
}

@article{Janarek_2022,
  title = {Quantum boomerang effect in systems without time-reversal symmetry},
  author = {Janarek, Jakub and Gr\'emaud, Beno\^{\i}t and Zakrzewski, Jakub and Delande, Dominique},
  journal = {Phys. Rev. B},
  volume = {105},
  issue = {18},
  pages = {L180202},
  numpages = {5},
  year = {2022},
  month = {May},
  publisher = {American Physical Society},
  doi = {10.1103/PhysRevB.105.L180202},
  url = {https://link.aps.org/doi/10.1103/PhysRevB.105.L180202}
}

@article{Jendrzejewski_2012a,
  title = {{Coherent Backscattering of Ultracold Atoms}},
  author = {Jendrzejewski, F. and M\"uller, K. and Richard, J. and Date, A. and Plisson, T. and Bouyer, P. and Aspect, A. and Josse, V.},
  journal = {Phys. Rev. Lett.},
  volume = {109},
  issue = {19},
  pages = {195302},
  numpages = {5},
  year = {2012},
  month = {Nov},
  publisher = {American Physical Society},
  doi = {10.1103/PhysRevLett.109.195302},
  url = {https://link.aps.org/doi/10.1103/PhysRevLett.109.195302}
}

@article{Kakoi_2024,
  title = {{Time evolution of coherent wave propagation and spin relaxation in spin-orbit-coupled systems}},
  author = {Kakoi, Masataka and Slevin, Keith},
  journal = {Phys. Rev. A},
  volume = {109},
  issue = {3},
  pages = {033303},
  numpages = {15},
  year = {2024},
  month = {Mar},
  publisher = {American Physical Society},
  doi = {10.1103/PhysRevA.109.033303},
  url = {https://link.aps.org/doi/10.1103/PhysRevA.109.033303}
}

@article{Kakoi_2026,
  title = {{Coherent transport in two-dimensional disordered potentials under spatially uniform $\text{SU}(2)$ gauge fields}},
  author = {Kakoi, Masataka and Miniatura, Christian and Slevin, Keith},
  journal = {Phys. Rev. A},
  volume = {113},
  issue = {4},
  pages = {L041302},
  numpages = {8},
  year = {2026},
  month = {Apr},
  publisher = {American Physical Society},
  doi = {10.1103/kdcq-9q67},
  url = {https://link.aps.org/doi/10.1103/kdcq-9q67}
}

@article{Karpiuk_2012,
  title = {{Coherent Forward Scattering Peak Induced by Anderson Localization}},
  author = {Karpiuk, T. and Cherroret, N. and Lee, K. L. and Gr\'emaud, B. and M\"uller, C. A. and Miniatura, C.},
  journal = {Phys. Rev. Lett.},
  volume = {109},
  issue = {19},
  pages = {190601},
  numpages = {5},
  year = {2012},
  month = {Nov},
  publisher = {American Physical Society},
  doi = {10.1103/PhysRevLett.109.190601},
  url = {https://link.aps.org/doi/10.1103/PhysRevLett.109.190601}
}

@article{Kawaguchi_1978,
   author = {Kawaguchi, Y. and Kitahara, H. and Kawaji, S.},
   title = {{Negative magnetoresistance in a two-dimensional impurity band in cesiated p-Si(111) surface inversion layers}},
   journal = {Surf. Sci.},
   volume = {73},
   number = {1},
   pages = {520-527},
   ISSN = {0039-6028},
   DOI = {10.1016/0039-6028(78)90535-6},
   url = {<Go to ISI>://WOS:A1978FC93500071},
   year = {1978},
   type = {Journal Article}
}

@article{Kawaguchi_1980,
   author = {Kawaguchi, Y. and Kawaji, S.},
   title = {{Negative Magnetoresistance in Silicon (100) MOS Inversion Layers}},
   journal = {J. Phys. Soc. Jpn.},
   volume = {48},
   number = {2},
   pages = {699-700},
   ISSN = {0031-9015},
   DOI = {10.1143/jpsj.48.699},
   url = {<Go to ISI>://WOS:A1980JM64000055},
   year = {1980},
   type = {Journal Article}
}

@article{Kobayashi_1980,
author = {Kobayashi ,Shunichi and Komori ,Fumio and Ootuka ,Youiti and Sasaki ,Wataru},
title = {{ln T Dependence of Resistivity in Two-Dimensionally Coupled Fine Particles of Cu}},
journal = {J. Phys. Soc. Jpn.},
volume = {49},
number = {4},
pages = {1635-1636},
year = {1980},
doi = {10.1143/JPSJ.49.1635},
URL = {https://doi.org/10.1143/JPSJ.49.1635},
}

@article{Kettemann_2007,
  title = {{Dimensional Control of Antilocalization and Spin Relaxation in Quantum Wires}},
  author = {Kettemann, S.},
  journal = {Phys. Rev. Lett.},
  volume = {98},
  issue = {17},
  pages = {176808},
  numpages = {4},
  year = {2007},
  month = {Apr},
  publisher = {American Physical Society},
  doi = {10.1103/PhysRevLett.98.176808},
  url = {https://link.aps.org/doi/10.1103/PhysRevLett.98.176808}
}

@article{Knap_1996,
  title = {{Weak antilocalization and spin precession in quantum wells}},
  author = {Knap, W. and Skierbiszewski, C. and Zduniak, A. and Litwin-Staszewska, E. and Bertho, D. and Kobbi, F. and Robert, J. L. and Pikus, G. E. and Pikus, F. G. and Iordanskii, S. V. and Mosser, V. and Zekentes, K. and Lyanda-Geller, Yu. B.},
  journal = {Phys. Rev. B},
  volume = {53},
  issue = {7},
  pages = {3912--3924},
  numpages = {0},
  year = {1996},
  month = {Feb},
  publisher = {American Physical Society},
  doi = {10.1103/PhysRevB.53.3912},
  url = {https://link.aps.org/doi/10.1103/PhysRevB.53.3912}
}

@Article{Koralek_2009,
author={Koralek, J. D. and Weber, C. P. and Orenstein, J. and Bernevig, B. A. and Zhang, Shou-Cheng and Mack, S. and Awschalom, D. D.},
title={{Emergence of the persistent spin helix in semiconductor quantum wells}},
journal={Nature},
year={2009},
month={Apr},
day={01},
volume={458},
number={7238},
pages={610-613},
issn={1476-4687},
doi={10.1038/nature07871},
url={https://doi.org/10.1038/nature07871}
}

@article{Kuga_1984,
   author = {Kuga, Y. and Ishimaru, A.},
   title = {{Retroreflectance from a dense distribution of spherical particles}},
   journal = {J. Opt. Soc. Am. A},
   volume = {1},
   number = {8},
   pages = {831-835},
   ISSN = {0740-3232},
   DOI = {10.1364/josaa.1.000831},
   url = {<Go to ISI>://WOS:A1984TD14300005},
   year = {1984},
   type = {Journal Article}
}

@article{Kuhn_2005,
  title = {{Localization of Matter Waves in Two-Dimensional Disordered Optical Potentials}},
  author = {Kuhn, R. C. and Miniatura, C. and Delande, D. and Sigwarth, O. and M\"uller, C. A.},
  journal = {Phys. Rev. Lett.},
  volume = {95},
  issue = {25},
  pages = {250403},
  numpages = {4},
  year = {2005},
  month = {Dec},
  publisher = {American Physical Society},
  doi = {10.1103/PhysRevLett.95.250403},
  url = {https://link.aps.org/doi/10.1103/PhysRevLett.95.250403}
}

@article{Kuhn_2007,
   author = {Kuhn, R. C. and Sigwarth, O. and Miniatura, C. and Delande, D. and M\"{u}ller, C. A.},
   title = {{Coherent matter wave transport in speckle potentials}},
   journal = {New J. Phys.},
   volume = {9},
   pages = {161},
   ISSN = {1367-2630},
   DOI = {10.1088/1367-2630/9/6/161},
   url = {<Go to ISI>://WOS:000247182800001},
   year = {2007},
   type = {Journal Article}
}

@article{Labeyrie_1999,
  title = {{Coherent Backscattering of Light by Cold Atoms}},
  author = {Labeyrie, G. and de Tomasi, F. and Bernard, J.-C. and M\"uller, C. A. and Miniatura, C. and Kaiser, R.},
  journal = {Phys. Rev. Lett.},
  volume = {83},
  issue = {25},
  pages = {5266--5269},
  numpages = {0},
  year = {1999},
  month = {Dec},
  publisher = {American Physical Society},
  doi = {10.1103/PhysRevLett.83.5266},
  url = {https://link.aps.org/doi/10.1103/PhysRevLett.83.5266}
}

@article{Labeyrie_2012,
   author = {Labeyrie, G. and Karpiuk, T. and Schaff, J.-F. and Gr\'{e}maud, B. and Miniatura, C. and Delande, D.},
   title = {{Enhanced backscattering of a dilute Bose-Einstein condensate}},
   journal = {EPL},
   volume = {100},
   number = {6},
   pages = {66001},
   ISSN = {0295-5075},
   DOI = {10.1209/0295-5075/100/66001},
   url = {<Go to ISI>://WOS:000313894800017},
   year = {2012},
   type = {Journal Article}
}

@article{Larose_2004,
  title = {{Weak Localization of Seismic Waves}},
  author = {Larose, E. and Margerin, L. and van Tiggelen, B. A. and Campillo, M.},
  journal = {Phys. Rev. Lett.},
  volume = {93},
  issue = {4},
  pages = {048501},
  numpages = {4},
  year = {2004},
  month = {Jul},
  publisher = {American Physical Society},
  doi = {10.1103/PhysRevLett.93.048501},
  url = {https://link.aps.org/doi/10.1103/PhysRevLett.93.048501}
}

@article{KL-Lee_2014b,
  title = {{Dynamics of localized waves in one-dimensional random potentials: Statistical theory of the coherent forward scattering peak}},
  author = {Lee, Kean Loon and Gr\'emaud, Beno\^{\i}t and Miniatura, Christian},
  journal = {Phys. Rev. A},
  volume = {90},
  issue = {4},
  pages = {043605},
  numpages = {14},
  year = {2014},
  month = {Oct},
  publisher = {American Physical Society},
  doi = {10.1103/PhysRevA.90.043605},
  url = {https://link.aps.org/doi/10.1103/PhysRevA.90.043605}
}

@article{ZL-Li_2024,
  title = {{Time-domain interferometry of electron weak localization through terahertz nonlinear response}},
  author = {Li, Zi-Long and Li, Xiao-Hui and Wan, Yuan},
  journal = {Phys. Rev. Res.},
  volume = {6},
  issue = {3},
  pages = {033125},
  numpages = {12},
  year = {2024},
  month = {Aug},
  publisher = {American Physical Society},
  doi = {10.1103/PhysRevResearch.6.033125},
  url = {https://link.aps.org/doi/10.1103/PhysRevResearch.6.033125}
}

@article{McCann_2006,
  title = {{Weak-Localization Magnetoresistance and Valley Symmetry in Graphene}},
  author = {McCann, E. and Kechedzhi, K. and Fal'ko, Vladimir I. and Suzuura, H. and Ando, T. and Altshuler, B. L.},
  journal = {Phys. Rev. Lett.},
  volume = {97},
  issue = {14},
  pages = {146805},
  numpages = {4},
  year = {2006},
  month = {Oct},
  publisher = {American Physical Society},
  doi = {10.1103/PhysRevLett.97.146805},
  url = {https://link.aps.org/doi/10.1103/PhysRevLett.97.146805}
}

@article{Micklitz_2014,
  title = {{Strong Anderson Localization in Cold Atom Quantum Quenches}},
  author = {Micklitz, T. and M\"uller, C. A. and Altland, A.},
  journal = {Phys. Rev. Lett.},
  volume = {112},
  issue = {11},
  pages = {110602},
  numpages = {5},
  year = {2014},
  month = {Mar},
  publisher = {American Physical Society},
  doi = {10.1103/PhysRevLett.112.110602},
  url = {https://link.aps.org/doi/10.1103/PhysRevLett.112.110602}
}

@article{Micklitz_2015,
  title = {{Echo spectroscopy of Anderson localization}},
  author = {Micklitz, T. and M\"uller, C. A. and Altland, A.},
  journal = {Phys. Rev. B},
  volume = {91},
  issue = {6},
  pages = {064203},
  numpages = {13},
  year = {2015},
  month = {Feb},
  publisher = {American Physical Society},
  doi = {10.1103/PhysRevB.91.064203},
  url = {https://link.aps.org/doi/10.1103/PhysRevB.91.064203}
}

@article{Muller_2015,
  title = {{Suppression and Revival of Weak Localization through Control of Time-Reversal Symmetry}},
  author = {M\"uller, K. and Richard, J. and Volchkov, V. V. and Denechaud, V. and Bouyer, P. and Aspect, A. and Josse, V.},
  journal = {Phys. Rev. Lett.},
  volume = {114},
  issue = {20},
  pages = {205301},
  numpages = {5},
  year = {2015},
  month = {May},
  publisher = {American Physical Society},
  doi = {10.1103/PhysRevLett.114.205301},
  url = {https://link.aps.org/doi/10.1103/PhysRevLett.114.205301}
}

@article{Tschirky_2017,
  title = {{Scattering mechanisms of highest-mobility $\mathrm{InAs}/{\mathrm{Al}}_{x}{\mathrm{Ga}}_{1\ensuremath{-}x}\mathrm{Sb}$ quantum wells}},
  author = {Tschirky, T. and Mueller, S. and Lehner, Ch. A. and F\"alt, S. and Ihn, T. and Ensslin, K. and Wegscheider, W.},
  journal = {Phys. Rev. B},
  volume = {95},
  issue = {11},
  pages = {115304},
  numpages = {8},
  year = {2017},
  month = {Mar},
  publisher = {American Physical Society},
  doi = {10.1103/PhysRevB.95.115304},
  url = {https://link.aps.org/doi/10.1103/PhysRevB.95.115304}
}

@article{Nitta_1997,
  title = {{Gate Control of Spin-Orbit Interaction in an Inverted I${\mathrm{n}}_{0.53}$G${\mathrm{a}}_{0.47}$As/I${\mathrm{n}}_{0.52}$A${\mathrm{l}}_{0.48}$As Heterostructure}},
  author = {Nitta, Junsaku and Akazaki, Tatsushi and Takayanagi, Hideaki and Enoki, Takatomo},
  journal = {Phys. Rev. Lett.},
  volume = {78},
  issue = {7},
  pages = {1335--1338},
  numpages = {0},
  year = {1997},
  month = {Feb},
  publisher = {American Physical Society},
  doi = {10.1103/PhysRevLett.78.1335},
  url = {https://link.aps.org/doi/10.1103/PhysRevLett.78.1335}
}

@article{Prat_2019,
  title = {{Quantum boomeranglike effect of wave packets in random media}},
  author = {Prat, Tony and Delande, Dominique and Cherroret, Nicolas},
  journal = {Phys. Rev. A},
  volume = {99},
  issue = {2},
  pages = {023629},
  numpages = {5},
  year = {2019},
  month = {Feb},
  publisher = {American Physical Society},
  doi = {10.1103/PhysRevA.99.023629},
  url = {https://link.aps.org/doi/10.1103/PhysRevA.99.023629}
}

@article{Richard_2019,
  title = {{Elastic Scattering Time of Matter Waves in Disordered Potentials}},
  author = {Richard, J\'er\'emie and Lim, Lih-King and Denechaud, Vincent and Volchkov, Valentin V. and Lecoutre, Baptiste and Mukhtar, Musawwadah and Jendrzejewski, Fred and Aspect, Alain and Signoles, Adrien and Sanchez-Palencia, Laurent and Josse, Vincent},
  journal = {Phys. Rev. Lett.},
  volume = {122},
  issue = {10},
  pages = {100403},
  numpages = {6},
  year = {2019},
  month = {Mar},
  publisher = {American Physical Society},
  doi = {10.1103/PhysRevLett.122.100403},
  url = {https://link.aps.org/doi/10.1103/PhysRevLett.122.100403}
}

@article{Sajjad_2022,
  title = {{Observation of the Quantum Boomerang Effect}},
  author = {Sajjad, Roshan and Tanlimco, Jeremy L. and Mas, Hector and Cao, Alec and Nolasco-Martinez, Eber and Simmons, Ethan Q. and Santos, Fl\'avio L. N. and Vignolo, Patrizia and Macr\`{\i}, Tommaso and Weld, David M.},
  journal = {Phys. Rev. X},
  volume = {12},
  issue = {1},
  pages = {011035},
  numpages = {10},
  year = {2022},
  month = {Feb},
  publisher = {American Physical Society},
  doi = {10.1103/PhysRevX.12.011035},
  url = {https://link.aps.org/doi/10.1103/PhysRevX.12.011035}
}

@article{Schliemann_2003,
  title = {{Nonballistic Spin-Field-Effect Transistor}},
  author = {Schliemann, John and Egues, J. Carlos and Loss, Daniel},
  journal = {Phys. Rev. Lett.},
  volume = {90},
  issue = {14},
  pages = {146801},
  numpages = {4},
  year = {2003},
  month = {Apr},
  publisher = {American Physical Society},
  doi = {10.1103/PhysRevLett.90.146801},
  url = {https://link.aps.org/doi/10.1103/PhysRevLett.90.146801}
}

@article{Scoquart_2020,
  title = {{Quench dynamics of a weakly interacting disordered Bose gas in momentum space}},
  author = {Scoquart, Thibault and Wellens, Thomas and Delande, Dominique and Cherroret, Nicolas},
  journal = {Phys. Rev. Res.},
  volume = {2},
  issue = {3},
  pages = {033349},
  numpages = {15},
  year = {2020},
  month = {Sep},
  publisher = {American Physical Society},
  doi = {10.1103/PhysRevResearch.2.033349},
  url = {https://link.aps.org/doi/10.1103/PhysRevResearch.2.033349}
}

@article{Sherman_2003,
    author = {Sherman, E. Ya.},
    title = {{Random spin--orbit coupling and spin relaxation in symmetric quantum wells}},
    journal = {Appl. Phys. Lett.},
    volume = {82},
    number = {2},
    pages = {209-211},
    year = {2003},
    month = {01},
    issn = {0003-6951},
    doi = {10.1063/1.1533839},
    url = {https://doi.org/10.1063/1.1533839},
}

@article{Sherman_2005,
  title = {{Spin relaxation in quantum dots with random spin-orbit coupling}},
  author = {Sherman, E. Ya. and Lockwood, D. J.},
  journal = {Phys. Rev. B},
  volume = {72},
  issue = {12},
  pages = {125340},
  numpages = {7},
  year = {2005},
  month = {Sep},
  publisher = {American Physical Society},
  doi = {10.1103/PhysRevB.72.125340},
  url = {https://link.aps.org/doi/10.1103/PhysRevB.72.125340}
}

@article{Suzuura_2002,
  title = {{Crossover from Symplectic to Orthogonal Class in a Two-Dimensional Honeycomb Lattice}},
  author = {Suzuura, Hidekatsu and Ando, Tsuneya},
  journal = {Phys. Rev. Lett.},
  volume = {89},
  issue = {26},
  pages = {266603},
  numpages = {4},
  year = {2002},
  month = {Dec},
  publisher = {American Physical Society},
  doi = {10.1103/PhysRevLett.89.266603},
  url = {https://link.aps.org/doi/10.1103/PhysRevLett.89.266603}
}

@Article{Szolnoki_2017,
   author={Szolnoki, L{\'e}n{\'a}rd and Kiss, Annam{\'a}ria and D{\'o}ra, Bal{\'a}zs and Simon, Ferenc},
   title={{Spin-relaxation time in materials with broken inversion symmetry and large spin-orbit coupling}},
   journal={Sci. Rep.},
   year={2017},
   month={Aug},
   day={30},
   volume={7},
   number={1SN  - 2045-2322},
   pages={9949},
   doi={10.1038/s41598-017-09759-0},
   url={https://doi.org/10.1038/s41598-017-09759-0}
}

@article{Tessieri_2021,
  title = {{Quantum boomerang effect: Beyond the standard Anderson model}},
  author = {Tessieri, L. and Akdeniz, Z. and Cherroret, N. and Delande, D. and Vignolo, P.},
  journal = {Phys. Rev. A},
  volume = {103},
  issue = {6},
  pages = {063316},
  numpages = {11},
  year = {2021},
  month = {Jun},
  publisher = {American Physical Society},
  doi = {10.1103/PhysRevA.103.063316},
  url = {https://link.aps.org/doi/10.1103/PhysRevA.103.063316}
}

@article{Thomas_2025,
  title = {{Coherent backscattering and coherent forward-scattering effects in variations of the random quantum kicked rotor}},
  author = {Thomas, Hugo and H\'ebraud, Julien and Georgeot, Bertrand and Lemari\'e, Gabriel and Miniatura, Christian and Giraud, Olivier},
  journal = {Phys. Rev. A},
  volume = {111},
  issue = {6},
  pages = {063302},
  numpages = {12},
  year = {2025},
  month = {Jun},
  publisher = {American Physical Society},
  doi = {10.1103/PhysRevA.111.063302},
  url = {https://link.aps.org/doi/10.1103/PhysRevA.111.063302}
}

@article{VanAlbada_1985,
   author = {Van Albada, M. P. and Lagendijk, A.},
   title = {{Observation of Weak Localization of Light in a Random Medium}},
   journal = {Phys. Rev. Lett.},
   volume = {55},
   number = {24},
   pages = {2692-2695},
   ISSN = {0031-9007},
   DOI = {10.1103/PhysRevLett.55.2692},
   url = {<Go to ISI>://WOS:A1985AVL2200017},
   year = {1985},
   type = {Journal Article}
}

@article{Wenk_2010,
  title = {{Dimensional dependence of weak localization corrections and spin relaxation in quantum wires with Rashba spin-orbit coupling}},
  author = {Wenk, P. and Kettemann, S.},
  journal = {Phys. Rev. B},
  volume = {81},
  issue = {12},
  pages = {125309},
  numpages = {18},
  year = {2010},
  month = {Mar},
  publisher = {American Physical Society},
  doi = {10.1103/PhysRevB.81.125309},
  url = {https://link.aps.org/doi/10.1103/PhysRevB.81.125309}
}

@article{Weideman_1986,
author = {Weideman, J. A. C. and Herbst, B. M.},
title = {{Split-Step Methods for the Solution of the Nonlinear Schrödinger Equation}},
journal = {SIAM J. Numer. Anal.},
volume = {23},
number = {3},
pages = {485-507},
year = {1986},
doi = {10.1137/0723033},
}

@article{Wolf_1985,
   author = {Wolf, P.-E. and Maret, G.},
   title = {{Weak Localization and Coherent Backscattering of Photons in Disordered Media}},
   journal = {\emph{ibid.}},
   volume = {55},
   number = {24},
   pages = {2696-2699},
   ISSN = {0031-9007},
   DOI = {10.1103/PhysRevLett.55.2696},
   url = {<Go to ISI>://WOS:A1985AVL2200018},
   year = {1985},
   type = {Journal Article}
}

@article{Zaitsev_2005a,
  title = {{Role of Orbital Dynamics in Spin Relaxation and Weak Antilocalization in Quantum Dots}},
  author = {Zaitsev, Oleg and Frustaglia, Diego and Richter, Klaus},
  journal = {Phys. Rev. Lett.},
  volume = {94},
  issue = {2},
  pages = {026809},
  numpages = {4},
  year = {2005},
  month = {Jan},
  publisher = {American Physical Society},
  doi = {10.1103/PhysRevLett.94.026809},
  url = {https://link.aps.org/doi/10.1103/PhysRevLett.94.026809}
}

@article{Zaitsev_2005b,
  title = {{Semiclassical theory of weak antilocalization and spin relaxation in ballistic quantum dots}},
  author = {Zaitsev, Oleg and Frustaglia, Diego and Richter, Klaus},
  journal = {Phys. Rev. B},
  volume = {72},
  issue = {15},
  pages = {155325},
  numpages = {18},
  year = {2005},
  month = {Oct},
  publisher = {American Physical Society},
  doi = {10.1103/PhysRevB.72.155325},
  url = {https://link.aps.org/doi/10.1103/PhysRevB.72.155325}
}

@article{Ruseckas_2005,
  title = {{Non-Abelian Gauge Potentials for Ultracold Atoms with Degenerate Dark States}},
  author = {Ruseckas, J. and Juzeli\ifmmode \bar{u}\else \={u}\fi{}nas, G. and \"Ohberg, P. and Fleischhauer, M.},
  journal = {Phys. Rev. Lett.},
  volume = {95},
  issue = {1},
  pages = {010404},
  numpages = {4},
  year = {2005},
  month = {Jun},
  publisher = {American Physical Society},
  doi = {10.1103/PhysRevLett.95.010404},
  url = {https://link.aps.org/doi/10.1103/PhysRevLett.95.010404}
}

@article{Sakai_1997,
  title = {{Observation of acoustic coherent backscattering}},
  author = {Sakai, Keiji and Yamamoto, Ken and Takagi, Kenshiro},
  journal = {Phys. Rev. B},
  volume = {56},
  issue = {17},
  pages = {10930--10933},
  numpages = {0},
  year = {1997},
  month = {Nov},
  publisher = {American Physical Society},
  doi = {10.1103/PhysRevB.56.10930},
  url = {https://link.aps.org/doi/10.1103/PhysRevB.56.10930}
}

@article{Sohn_2024,
  title = {{Dyakonov-Perel-like Orbital and Spin Relaxations in Centrosymmetric Systems}},
  author = {Sohn, Jeonghun and Lee, Jongjun M. and Lee, Hyun-Woo},
  journal = {Phys. Rev. Lett.},
  volume = {132},
  issue = {24},
  pages = {246301},
  numpages = {7},
  year = {2024},
  month = {Jun},
  publisher = {American Physical Society},
  doi = {10.1103/PhysRevLett.132.246301},
  url = {https://link.aps.org/doi/10.1103/PhysRevLett.132.246301}
}

@article{XJ-Liu_2009,
  title = {{Effect of Induced Spin-Orbit Coupling for Atoms via Laser Fields}},
  author = {Liu, Xiong-Jun and Borunda, Mario F. and Liu, Xin and Sinova, Jairo},
  journal = {Phys. Rev. Lett.},
  volume = {102},
  issue = {4},
  pages = {046402},
  numpages = {4},
  year = {2009},
  month = {Jan},
  publisher = {American Physical Society},
  doi = {10.1103/PhysRevLett.102.046402},
  url = {https://link.aps.org/doi/10.1103/PhysRevLett.102.046402}
}

@article{Muller_2005,
doi = {10.1088/0305-4470/38/36/002},
url = {https://doi.org/10.1088/0305-4470/38/36/002},
year = {2005},
month = {aug},
publisher = {},
volume = {38},
number = {36},
pages = {7807},
author = {Müller, C A and Miniatura, C and Akkermans, E and Montambaux, G},
title = {{Mesoscopic scattering of spin $s$ particles}},
journal = {J. Phys. A: Math. Gen.},
}

@Article{Galitski_2013_review,
author={Galitski, Victor
and Spielman, Ian B.},
title={{Spin--orbit coupling in quantum gases}},
journal={Nature},
year={2013},
month={Feb},
day={01},
volume={494},
number={7435},
pages={49-54},
issn={1476-4687},
doi={10.1038/nature11841},
url={https://doi.org/10.1038/nature11841}
}

@article{Cheuk_2012,
  title = {{Spin-Injection Spectroscopy of a Spin-Orbit Coupled Fermi Gas}},
  author = {Cheuk, Lawrence W. and Sommer, Ariel T. and Hadzibabic, Zoran and Yefsah, Tarik and Bakr, Waseem S. and Zwierlein, Martin W.},
  journal = {\emph{ibid.}},
  volume = {109},
  issue = {9},
  pages = {095302},
  numpages = {5},
  year = {2012},
  month = {Aug},
  publisher = {American Physical Society},
  doi = {10.1103/PhysRevLett.109.095302},
  url = {https://link.aps.org/doi/10.1103/PhysRevLett.109.095302}
}

@article{Hasan_2022,
  title = {{Wave Packet Dynamics in Synthetic Non-Abelian Gauge Fields}},
  author = {Hasan, Mehedi and Madasu, Chetan Sriram and Rathod, Ketan D. and Kwong, Chang Chi and Miniatura, Christian and Chevy, Fr\'ed\'eric and Wilkowski, David},
  journal = {Phys. Rev. Lett.},
  volume = {129},
  issue = {13},
  pages = {130402},
  numpages = {6},
  year = {2022},
  month = {Sep},
  publisher = {American Physical Society},
  doi = {10.1103/PhysRevLett.129.130402},
  url = {https://link.aps.org/doi/10.1103/PhysRevLett.129.130402}
}

@Article{X-Hou_2026,
author={Hou, Xiangrui and Wu, Zhaoxin and Wang, Fangyu and Zhu, Shiyao and Yan, Bo and Yang, Zhaoju},
title={{Quantum boomerang effect of light}},
journal={Nat. Commun.},
year={2026},
month={Jan},
day={16},
volume={17},
number={1},
pages={1579},
issn={2041-1723},
doi={10.1038/s41467-026-68293-8},
url={https://doi.org/10.1038/s41467-026-68293-8}
}

@article{Huang_2016,
   author = {Huang, L. and Meng, Z. and Wang, P. and Peng, P. and Zhang, S.-L. and Chen, L. C. and Li, D. and Zhou, Q. and Zhang, J.},
   title = {{Experimental realization of two-dimensional synthetic spin-orbit coupling in ultracold Fermi gases}},
   journal = {Nat. Phys.},
   volume = {12},
   number = {6},
   pages = {540-+},
   ISSN = {1745-2473},
   DOI = {10.1038/nphys3672},
   url = {<Go to ISI>://WOS:000377475700010},
   year = {2016},
   type = {Journal Article}
}

@article{Kohda_2012,
  title = {{Gate-controlled persistent spin helix state in (In,Ga)As quantum wells}},
  author = {Kohda, M. and Lechner, V. and Kunihashi, Y. and Dollinger, T. and Olbrich, P. and Sch\"onhuber, C. and Caspers, I. and Bel'kov, V. V. and Golub, L. E. and Weiss, D. and Richter, K. and Nitta, J. and Ganichev, S. D.},
  journal = {Phys. Rev. B},
  volume = {86},
  issue = {8},
  pages = {081306(R)},
  numpages = {5},
  year = {2012},
  month = {Aug},
  publisher = {American Physical Society},
  doi = {10.1103/PhysRevB.86.081306},
  url = {https://link.aps.org/doi/10.1103/PhysRevB.86.081306}
}

@Article{Leroux_2018,
   author={Leroux, F. and Pandey, K. and Rehbi, R. and Chevy, F. and Miniatura, C. and Gr{\'e}maud, B. and Wilkowski, D.},
   title={{Non-Abelian adiabatic geometric transformations in a cold strontium gas}},
   journal={Nat. Commun.},
   year={2018},
   month={Sep},
   day={04},
   volume={9},
   number={1},
   pages={3580},
   issn={2041-1723},
   doi={10.1038/s41467-018-05865-3},
   url={https://doi.org/10.1038/s41467-018-05865-3}
}

@Article{Q-Liang_2024,
   author={Liang, Qian and Dong, Zhaoli and Pan, Jian-Song and Wang, Hongru and Li, Hang and Yang, Zhaoju and Yi, Wei and Yan, Bo},
   title={{Chiral dynamics of ultracold atoms under a tunable SU(2) synthetic gauge field}},
   journal={Nat. Phys.},
   year={2024},
   month={Nov},
   day={01},
   volume={20},
   number={11SN  - 1745-2481},
   pages={1738-1743},
   doi={10.1038/s41567-024-02644-4},
   url={https://doi.org/10.1038/s41567-024-02644-4}
}

@article{YJ-Lin_2011,
   author = {Lin, Y.-J. and Jim\'{e}nez-Garc\'{i}a, K. and Spielman, I. B.},
   title = {{Spin-orbit-coupled Bose-Einstein condensates}},
   journal = {Nature},
   volume = {471},
   number = {7336},
   pages = {83-U99},
   ISSN = {0028-0836},
   DOI = {10.1038/nature09887},
   url = {<Go to ISI>://WOS:000287924100038},
   year = {2011},
   type = {Journal Article}
}

@Article{Madasu_2025,
author={Madasu, Chetan S. and Mitra, Chirantan and Gabardos, Lucas and Rathod, Ketan D. and Zanon-Willette, Thomas and Miniatura, Christian and Chevy, Fr{\'e}d{\'e}ric and Kwong, Chang Chi and Wilkowski, David},
title={{Experimental realization of a SU(3) color-orbit coupling in an ultracold gas}},
journal={Nat. Commun.},
year={2025},
month={Sep},
day={26},
volume={16},
number={1},
pages={8448},
issn={2041-1723},
doi={10.1038/s41467-025-63142-6},
url={https://doi.org/10.1038/s41467-025-63142-6}
}

@article{Suzuki_2025,
  title = {{Disorder-Induced Slow Relaxation of Phonon Polarization}},
  author = {Suzuki, Yuta and Murakami, Shuichi},
  journal = {Phys. Rev. Lett.},
  volume = {135},
  issue = {4},
  pages = {046301},
  numpages = {8},
  year = {2025},
  month = {Jul},
  publisher = {American Physical Society},
  doi = {10.1103/z8wj-f384},
  url = {https://link.aps.org/doi/10.1103/z8wj-f384}
}

@article{Z-Wu_2016,
   author = {Wu, Z. and Zhang, L. and Sun, W. and Xu, X.-T. and Wang, B.-Z. and Ji, S.-C. and Deng, Y. and Chen, S. and Liu, X.-J. and Pan, J.-W.},
   title = {{Realization of two-dimensional spin-orbit coupling for Bose-Einstein condensates}},
   journal = {Science},
   volume = {354},
   number = {6308},
   pages = {83-88},
   ISSN = {0036-8075},
   DOI = {10.1126/science.aaf6689},
   url = {<Go to ISI>://WOS:000387777900034},
   year = {2016},
   type = {Journal Article}
}

@article{P-Wang_2012,
  title = {{Spin-Orbit Coupled Degenerate Fermi Gases}},
  author = {Wang, Pengjun and Yu, Zeng-Qiang and Fu, Zhengkun and Miao, Jiao and Huang, Lianghui and Chai, Shijie and Zhai, Hui and Zhang, Jing},
  journal = {Phys. Rev. Lett.},
  volume = {109},
  issue = {9},
  pages = {095301},
  numpages = {5},
  year = {2012},
  month = {Aug},
  publisher = {American Physical Society},
  doi = {10.1103/PhysRevLett.109.095301},
  url = {https://link.aps.org/doi/10.1103/PhysRevLett.109.095301}
}

@article{ZY-Wang_2021,
   author = {Wang, Z.-Y. and Cheng, X.-C. and Wang, B.-Z. and Zhang, J.-Y. and Lu, Y.-H. and Yi, C.-R. and Niu, S. and Deng, Y. and Liu, X.-J. and Chen, S. and Pan, J.-W.},
   title = {{Realization of an ideal Weyl semimetal band in a quantum gas with 3D spin-orbit coupling}},
   journal = {Science},
   volume = {372},
   number = {6539},
   pages = {271-+},
   ISSN = {0036-8075},
   DOI = {10.1126/science.abc0105},
   url = {<Go to ISI>://WOS:000641286700036},
   year = {2021},
   type = {Journal Article}
}

@Article{Walser_2012,
author={Walser, M. P. and Reichl, C. and Wegscheider, W. and Salis, G.},
title={{Direct mapping of the formation of a persistent spin helix}},
journal={Nat. Phys.},
year={2012},
month={Oct},
day={01},
volume={8},
number={10},
pages={757-762},
issn={1745-2481},
doi={10.1038/nphys2383},
url={https://doi.org/10.1038/nphys2383}
}

@Article{Kilic_2025,
author={Kilic, Berkay and Alvarruiz, Sergio and Barts, Evgenii and van Dijk, Bertjan and Barone, Paolo and S{\l}awi{\'{n}}ska, Jagoda},
title={{Universal symmetry-protected persistent spin textures in noncentrosymmetric crystals}},
journal={Nat. Commun.},
year={2025},
month={Aug},
day={27},
volume={16},
number={1},
pages={7999},
issn={2041-1723},
doi={10.1038/s41467-025-63136-4},
url={https://doi.org/10.1038/s41467-025-63136-4}
}

@Article{Tao_2018,
author={Tao, L. L. and Tsymbal, Evgeny Y.},
title={{Persistent spin texture enforced by symmetry}},
journal={Nat. Commun.},
year={2018},
month={Jul},
day={17},
volume={9},
number={1},
pages={2763},
issn={2041-1723},
doi={10.1038/s41467-018-05137-0},
url={https://doi.org/10.1038/s41467-018-05137-0}
}

@article{X-Liu_2012,
  title = {{Unified theory of spin dynamics in a two-dimensional electron gas with arbitrary spin-orbit coupling strength at finite temperature}},
  author = {Liu, Xin and Sinova, Jairo},
  journal = {Phys. Rev. B},
  volume = {86},
  issue = {17},
  pages = {174301},
  numpages = {14},
  year = {2012},
  month = {Nov},
  publisher = {American Physical Society},
  doi = {10.1103/PhysRevB.86.174301},
  url = {https://link.aps.org/doi/10.1103/PhysRevB.86.174301}
}

@article{X-Liu_2011,
  title = {{Spin dynamics in the strong spin-orbit coupling regime}},
  author = {Liu, Xin and Liu, Xiong-Jun and Sinova, Jairo},
  journal = {Phys. Rev. B},
  volume = {84},
  issue = {3},
  pages = {035318},
  numpages = {8},
  year = {2011},
  month = {Jul},
  publisher = {American Physical Society},
  doi = {10.1103/PhysRevB.84.035318},
  url = {https://link.aps.org/doi/10.1103/PhysRevB.84.035318}
}

@article{Zhao_2020,
  title = {{Purely Cubic Spin Splittings with Persistent Spin Textures}},
  author = {Zhao, Hong Jian and Nakamura, Hiro and Arras, R\'emi and Paillard, Charles and Chen, Peng and Gosteau, Julien and Li, Xu and Yang, Yurong and Bellaiche, Laurent},
  journal = {Phys. Rev. Lett.},
  volume = {125},
  issue = {21},
  pages = {216405},
  numpages = {6},
  year = {2020},
  month = {Nov},
  publisher = {American Physical Society},
  doi = {10.1103/PhysRevLett.125.216405},
  url = {https://link.aps.org/doi/10.1103/PhysRevLett.125.216405}
}

@article{XZ-Lu_2020,
title = {{Discovery Principles and Materials for Symmetry-Protected Persistent Spin Textures with Long Spin Lifetimes}},
journal = {Matter},
volume = {3},
number = {4},
pages = {1211-1225},
year = {2020},
issn = {2590-2385},
doi = {https://doi.org/10.1016/j.matt.2020.08.028},
url = {https://www.sciencedirect.com/science/article/pii/S2590238520304550},
author = {Xue-Zeng Lu and James M. Rondinelli},
}

@article{J-Ji_2022,
  title = {{Symmetry-protected full-space persistent spin texture in two-dimensional materials}},
  author = {Ji, Junyi and Lou, Feng and Yu, Rui and Feng, J. S. and Xiang, H. J.},
  journal = {Phys. Rev. B},
  volume = {105},
  issue = {4},
  pages = {L041404},
  numpages = {6},
  year = {2022},
  month = {Jan},
  publisher = {American Physical Society},
  doi = {10.1103/PhysRevB.105.L041404},
  url = {https://link.aps.org/doi/10.1103/PhysRevB.105.L041404}
}

@article{Muszynski_2022,
  title = {{Realizing Persistent-Spin-Helix Lasing in the Regime of Rashba-Dresselhaus Spin-Orbit Coupling in a Dye-Filled Liquid-Crystal Optical Microcavity}},
  author = {Muszy\'{n}ski, Marcin and Kr\'ol, Mateusz and Rechci\'{n}ska, Katarzyna and Oliwa, Przemys\l{}aw and Kedziora, Mateusz and \L{}empicka-Mirek, Karolina and Mazur, Rafa\l{} and Morawiak, Przemys\l{}aw and Piecek, Wiktor and Kula, Przemys\l{}aw and Lagoudakis, Pavlos G. and Pietka, Barbara and Szczytko, Jacek},
  journal = {Phys. Rev. Appl.},
  volume = {17},
  issue = {1},
  pages = {014041},
  numpages = {6},
  year = {2022},
  month = {Jan},
  publisher = {American Physical Society},
  doi = {10.1103/PhysRevApplied.17.014041},
  url = {https://link.aps.org/doi/10.1103/PhysRevApplied.17.014041}
}

@Article{Bliokh_2015_review,
author={Bliokh, K. Y. and Rodr{\'i}guez-Fortu{\~{n}}o, F. J. and Nori, F. and Zayats, A. V.},
title={Spin--orbit interactions of light},
journal={Nat. Photon.},
year={2015},
month={Dec},
day={01},
volume={9},
number={12},
pages={796-808},
issn={1749-4893},
doi={10.1038/nphoton.2015.201},
url={https://doi.org/10.1038/nphoton.2015.201}
}

@article{Y-Yang_2024_review,
author = {Yi Yang  and Biao Yang  and Guancong Ma  and Jensen Li  and Shuang Zhang  and C. T. Chan },
title = {{Non-Abelian physics in light and sound}},
journal = {Science},
volume = {383},
number = {6685},
pages = {eadf9621},
year = {2024},
doi = {10.1126/science.adf9621},
URL = {https://www.science.org/doi/abs/10.1126/science.adf9621},
}

@Article{Y-Chen_2019,
author={Chen, Yuntian and Zhang, Ruo-Yang and Xiong, Zhongfei and Hang, Zhi Hong and Li, Jensen and Shen, Jian Qi and Chan, C. T.},
title={{Non-Abelian gauge field optics}},
journal={Nat. Commun.},
year={2019},
month={Jul},
day={16},
volume={10},
number={1},
pages={3125},
issn={2041-1723},
doi={10.1038/s41467-019-10974-8},
url={https://doi.org/10.1038/s41467-019-10974-8}
}

@Article{Y-Li_2022,
   author={Li, Yao and Ma, Xuekai and Zhai, Xiaokun and Gao, Meini and Dai, Haitao and Schumacher, Stefan and Gao, Tingge},
   title={{Manipulating polariton condensates by Rashba-Dresselhaus coupling at room temperature}},
   journal={Nat. Commun.},
   year={2022},
   month={Jul},
   day={01},
   volume={13},
   number={1},
   pages={3785},
   issn={2041-1723},
   doi={10.1038/s41467-022-31529-4},
   url={https://doi.org/10.1038/s41467-022-31529-4}
}

@Article{Ma_2016,
author={Ma, L. B. and Li, S. L. and Fomin, V. M. and Hentschel, M. and G{\"o}tte, J. B. and Yin, Y. and Jorgensen, M. R. and Schmidt, O. G.},
title={{Spin--orbit coupling of light in asymmetric microcavities}},
journal={Nat. Commun.},
year={2016},
month={Mar},
day={18},
volume={7},
number={1},
pages={10983},
issn={2041-1723},
doi={10.1038/ncomms10983},
url={https://doi.org/10.1038/ncomms10983}
}

@article{Polimeno_2021,
author = {L. Polimeno and A. Fieramosca and G. Lerario and L. De Marco and M. De Giorgi and D. Ballarini and L. Dominici and V. Ardizzone and M. Pugliese and C. T. Prontera and V. Maiorano and G. Gigli and C. Leblanc and G. Malpuech and D. D. Solnyshkov and D. Sanvitto},
journal = {Optica},
number = {11},
pages = {1442--1447},
publisher = {Optica Publishing Group},
title = {{Experimental investigation of a non-Abelian gauge field in 2D perovskite photonic platform}},
volume = {8},
month = {Nov},
year = {2021},
url = {https://opg.optica.org/optica/abstract.cfm?URI=optica-8-11-1442},
doi = {10.1364/OPTICA.427088},
}

@article{Rechcinska_2019,
author = {Katarzyna Rechcińska  and Mateusz Król  and Rafał Mazur  and Przemysław Morawiak  and Rafał Mirek  and Karolina Łempicka  and Witold Bardyszewski  and Michał Matuszewski  and Przemysław Kula  and Wiktor Piecek  and Pavlos G. Lagoudakis  and Barbara Pi\c{e}tka  and Jacek Szczytko},
title = {{Engineering spin-orbit synthetic Hamiltonians in liquid-crystal optical cavities}},
journal = {Science},
volume = {366},
number = {6466},
pages = {727-730},
year = {2019},
doi = {10.1126/science.aay4182},
URL = {https://www.science.org/doi/abs/10.1126/science.aay4182},
}

@Article{J-Wu2022,
author={Wu, Jiexiong and Wang, Zhu and Biao, Yuanchuan and Fei, Fucong and Zhang, Shuai and Yin, Zepeng and Hu, Yejian and Song, Ziyin and Wu, Tianyu and Song, Fengqi and Yu, Rui},
title={{Non-Abelian gauge fields in circuit systems}},
journal={Nat. Electron.},
year={2022},
month={Oct},
day={01},
volume={5},
number={10},
pages={635-642},
issn={2520-1131},
doi={10.1038/s41928-022-00833-8},
url={https://doi.org/10.1038/s41928-022-00833-8}
}

\end{document}